\documentclass[lettersize,journal]{IEEEtran}
\usepackage{amsmath,amsfonts}
\usepackage{algorithmic}
\usepackage{algorithm}
\usepackage{array}
\usepackage{multirow}
\usepackage{tabularx}
\usepackage{calc}
\usepackage{etoolbox}

\usepackage{booktabs}
\usepackage{makecell}
\usepackage{colortbl}
\usepackage[table]{xcolor}
\usepackage{textcomp}
\usepackage{stfloats}
\usepackage{url}
\usepackage{verbatim}
\usepackage{graphicx}
\usepackage{cite}
\usepackage{xcolor}
\usepackage{subfigure}
\usepackage{pifont}
\usepackage{array}
\usepackage{caption}
\usepackage{subcaption}
\definecolor{red}{rgb}{1,0,0} 
\definecolor{green}{rgb}{0,0.7,0}  
\definecolor{myblue}{RGB}{199,217,236}
\definecolor{mygrey}{RGB}{236,236,237}

\usepackage{xurl}

\ifodd 1
\else

\fi

\begin{document}

\title{Mobile Edge Intelligence for Large Language Models: A Contemporary Survey}
\author{Guanqiao Qu,~\IEEEmembership{Graduate Student Member,~IEEE}, Qiyuan Chen, Wei Wei,~\IEEEmembership{Graduate Student Member,~IEEE},\\Zheng Lin,~\IEEEmembership{Graduate Student Member,~IEEE}, Xianhao Chen,~\IEEEmembership{Member,~IEEE}, and Kaibin Huang,~\IEEEmembership{Fellow,~IEEE}
\thanks{The work was supported in part by the Research Grants Council of Hong Kong under Grant 27213824, in part by HKU-SCF FinTech Academy R\&D Funding, and in part by HKU IDS Research Seed Fund under Grant IDS-RSF2023-0012. The work of K. Huang described in this paper was supported in part by the Research Grants Council of the Hong Kong Special Administrative Region, China under a fellowship award (HKU RFS2122-7S04), the Areas of Excellence scheme grant (AoE/E-601/22-R), Collaborative Research Fund (C1009-22G), and the Grant 17212423. Part of the described research work is conducted in the JC STEM Lab of Robotics for Soft Materials funded by The Hong Kong Jockey Club Charities Trust.

Guanqiao Qu, Qiyuan Chen, Wei Wei, Zheng Lin, Xianhao Chen, and Kaibin Huang are with the Department of Electrical and Electronic Engineering, University of Hong Kong, Pok Fu Lam, Hong Kong SAR, China. Xianhao Chen is also with HKU Musketeers Foundation Institute of Data Science, University of Hong Kong, Pok Fu Lam, Hong Kong SAR, China. (e-mail: gqqu@eee.hku.hk; qiyuanchen@connect.hku.hk; weiwei@eee.hku.hk; linzheng@eee.hku.hk; xchen@eee.hku.hk; huangkb@eee.hku.hk).
\textit{(Corresponding author: Xianhao Chen.)}}}

%
%

\markboth{}%
{Shell \MakeLowercase{\textit{et al.}}: A Sample Article Using IEEEtran.cls for IEEE Journals}


\maketitle

\begin{abstract}
On-device large language models (LLMs), referring to running LLMs on edge devices, have raised considerable interest since they are more cost-effective, latency-efficient, and privacy-preserving compared with the cloud paradigm. Nonetheless, the performance of on-device LLMs is intrinsically constrained by resource limitations on edge devices. Sitting between cloud and on-device AI, mobile edge intelligence (MEI) presents a viable solution by provisioning AI capabilities at the edge of mobile networks. This article provides a contemporary survey on harnessing MEI for LLMs. We begin by illustrating several killer applications to demonstrate the urgent need for deploying LLMs at the network edge. Next, we present the preliminaries of LLMs and MEI, followed by resource-efficient LLM techniques. We then present an architectural overview of MEI for LLMs (MEI4LLM), outlining its core components and how it supports the deployment of LLMs. Subsequently, we delve into various aspects of MEI4LLM, extensively covering edge LLM caching and delivery, edge LLM training, and edge LLM inference. Finally, we identify future research opportunities. We hope this article inspires researchers in the field to leverage mobile edge computing to facilitate LLM deployment, thereby unleashing the potential of LLMs across various privacy- and delay-sensitive applications.
\end{abstract}

\begin{IEEEkeywords}
Large language models, foundation models, mobile edge computing, edge intelligence, 6G, split learning.
\end{IEEEkeywords}
\section{Introduction\label{introduction}}
\subsection{Background\label{background}}
The recent advent of large language models (LLMs) has been a milestone in artificial intelligence (AI) technology to enable general-purpose intelligence. LLMs excel in various domains, i.e., not only in the task they are built for, i.e., generating text responses, but also in tasks such as multimodal content analysis, summarization, and generalization. For instance, the GPT-4 multimodal model accepts image and text inputs, which produces text outputs exhibiting human-level performance on various professional and academic benchmarks. Apart from these general-purpose models, sometimes called foundation models, LLMs can be fine-tuned to downstream tasks, catering to specific industries and application scenarios. For example, medical LLM, such as Med-PaLM M~\cite{medpalm}, was designed by Google to provide high-quality answers based on rich data modalities spanning text, imaging, genomics, and more. Google DeepMind also developed Robotics Transformer 2 (RT-2)~\cite{RT2}, a vision-language-action AI model for controlling robots. The broad spectrum of use cases highlights the profound impact of LLMs on our everyday lives.

Due to extensive resource demands, most commercial LLMs are confined to cloud data centers for service provisioning. Regrettably, cloud-based LLM provisioning brings inherent drawbacks, including data privacy breaches, high bandwidth costs, and long service latency. Specifically, users must upload their data to cloud centers to access LLM services, resulting in significant communication delays. Moreover, uploading private data poses a serious risk to user privacy, especially in privacy-sensitive applications like smart health. During the transmissions and storage of data on cloud servers, unauthorized data access can expose sensitive personal information, such as medical records and biometric data, to malicious actors \cite{li2024personal,rahman2023survey,sun2024efficient}, leading to the misuse of data for malicious purposes and severely compromising user confidentiality. Given these concerns, there is a continuing development trend in on-device LLM deployment, sparking a competitive race among major industry players. For instance, Google has launched Gemini Nano on Pixel 8 Pro smartphones with 1.8-billion and 3.25-billion parameters, respectively~\cite{Google2023Gemini}. Qualcomm plans to launch Llama 2 support on Snapdragon-powered flagship smartphones and personal computers~\cite{Qualcomm2023Llama}. On-device LLM deployment enables local processing of sensitive personal data and provides low response time, which is crucial for delay-sensitive applications such as robot planning and autonomous driving. 


\subsection{Motivation: From Cloud LLMs to On-device LLMs to MEI LLMs\label{motivation}}
Although on-device LLM is becoming a fast-growing field, the widespread deployment of on-device LLMs faces severe limitations. In particular, the scarcity of computing, memory, and storage resources on edge devices substantially limits the scale of on-device LLM. On the one hand, the existing industrial efforts focus on sub-10B (10 billion parameters) LLMs due to the extensive resource requirements for on-device deployment \cite{liu2024mobilellm,qualcomm2023}. For instance, Google's Gemini Nano, which relies on 4-bit models with 1.8B and 3.25B parameters, can only support relatively ``basic'' functionalities like summarizing text, suggesting smart replies in context, and checking grammar~\cite{Google2023Gemini}. However, as the desired functionalities become more complex, deploying larger-scale LLMs on the device becomes necessary. On the other hand, while on-device fine-tuning paves the way for personalized and context-aware AI, serving as a fundamental block for superior AI performance, the existing on-device LLM products generally do not incorporate on-device training (fine-tuning) functionalities due to the extensive training cost, which is much more resource-intensive than supporting AI inference alone.

To address the aforementioned dilemma, mobile edge computing offers a promising solution. The 6G mobile network aims to deliver low-latency AI inference and training services for a wide range of mobile devices by leveraging the network-endowed computing capabilities, say, on base stations. This leads to a paradigm called ``mobile edge intelligence (MEI)''. Specifically, MEI sits between on-device AI and cloud-based AI, featuring a modest scale of computing resources located close to users, which is more capable than edge devices yet less powerful than cloud centers. Benefiting from the short distance between edge devices and edge servers, large-scale LLMs can be supported with lower service latency and bandwidth costs. Meanwhile, the 6G edge can continuously fine-tune LLMs for adapting to ever-evolving environments by exploiting the memory, energy, and computing power on edge servers, which is hard to achieve via edge devices alone. As such, the 6G mobile edge is expected to be essential in democratizing LLMs to edge devices, and providing comprehensive discussions and review on this trend is the focus of our survey paper.


\subsection{Comparisons with Prior Surveys and Our Contributions}
\begin{table*}[!t]
\centering
\caption{Summary of the related surveys/articles.}
\label{table_comp}
\renewcommand{\arraystretch}{1.4}
\setlength{\tabcolsep}{2mm}
\begin{tabular}{|c|m{0.22\textwidth}|c|c|c|c|c|c|c|c|}
\hline
\makecell[c]{\multirow{4}*{\textbf{Ref.}}} & \makecell[c]{\multirow{4}*{\textbf{Description}}} & \multicolumn{4}{c|}{\centering\arraybackslash \textbf{Scenarios}} & \multicolumn{3}{c|}{\centering\arraybackslash \textbf{Perspectives}} \\ \cline{3-9}
  &  & 
  \makecell[c]{\textbf{Efficient}\\\textbf{(on-device) LLM}\\\textbf{techniques}} & \makecell[c]{\textbf{Edge LLM caching}\\\textbf{and delivery}} & \makecell[c]{\textbf{Edge LLM}\\\textbf{training}} & \makecell[c]{\textbf{Edge LLM}\\\textbf{inference}} & \makecell[c]{\textbf{Storage}\\\textbf{eff.}} & \makecell[c]{\textbf{Comp.}\\\textbf{eff.}} & \makecell[c]{\textbf{Comm.}\\\textbf{eff.}} \\ \hline
\cite{zhao2023survey} 
    &  Reviews recent progress of LLM pre-training, fine-tuning, usage, and capacity evaluation, along with the public resources for deploying LLM.
    & {\color{green}\ding{52}}  & {\color{red}\ding{55}}  & {\color{red}\ding{55}}   & {\color{red}\ding{55}}   
    & {\color{red}\ding{55}}  & {\color{green}\ding{52}}   & {\color{red}\ding{55}} \\ \hline
\cite{xu2024survey} 
    &  Introduces resource-efficient approaches to deploying LLMs, including model architectures, training and inference algorithms, and practical system designs.
    & {\color{green}\ding{52}}  & {\color{red}\ding{55}}  & {\color{green}\ding{52}}   & {\color{green}\ding{52}}    
    & {\color{green}\ding{52}}  & {\color{green}\ding{52}}  & {\color{red}\ding{55}} \\ \hline 
\cite{wan2024efficient} 
    &  Explores efficient LLM deployment via model-centric methods (model compression, training, inference, and architecture design), data-centric approaches (data selection and prompt engineering), and framework-centric strategies (specific training, inference, and serving frameworks).
    & {\color{green}\ding{52}}  & {\color{red}\ding{55}}  & {\color{red}\ding{55}}   & {\color{red}\ding{55}}
    & {\color{red}\ding{55}}  & {\color{green}\ding{52}}  & {\color{red}\ding{55}}\\ \hline
\cite{xu2023unleashing} 
    &  Overviews the deployment of AI-generated content applications in mobile networks, including mobile devices, edge servers, and cloud centers. 
    & {\color{green}\ding{52}}  & {\color{red}\ding{55}}  & {\color{green}\ding{52}}   & {\color{green}\ding{52}}    
    & {\color{red}\ding{55}}  & {\color{green}\ding{52}}  & {\color{green}\ding{52}} \\ \hline
\cite{kachris2024survey} 
    &  Surveys the current hardware acceleration methods for energy-efficient on-device LLM training and inference.
    & {\color{green}\ding{52}}  & {\color{red}\ding{55}}  & {\color{red}\ding{55}}   & {\color{red}\ding{55}}    
    & {\color{red}\ding{55}}  & {\color{green}\ding{52}}  & {\color{red}\ding{55}} \\ \hline
\cite{bai2024efficiency} 
    &  Reviews resource-efficient techniques for LLM deployment, covering computational, memory, energy, economic, and network resources based on their applicability across architecture design, pre-training, fine-tuning, inference, and system design. 
    & {\color{green}\ding{52}}  & {\color{red}\ding{55}}  & {\color{red}\ding{55}}   & {\color{green}\ding{52}}    
    & {\color{green}\ding{52}}  & {\color{green}\ding{52}} & {\color{green}\ding{52}} \\ \hline
\cite{yuan2024llm} 
    &  Summarizes current research on efficient LLM inference, including compression, fast decoding, and optimization for compiler/system/hardware. 
    & {\color{green}\ding{52}}  & {\color{red}\ding{55}}  & {\color{red}\ding{55}}   & {\color{red}\ding{55}}    
    & {\color{green}\ding{52}}  & {\color{green}\ding{52}}  & {\color{red}\ding{55}} \\ \hline
\cite{liu2024understanding} 
    & Reviews the evolution of low-cost on-device LLM training and inference techniques. 
    & {\color{green}\ding{52}}  & {\color{red}\ding{55}}  & {\color{red}\ding{55}}   & {\color{red}\ding{55}}    
    & {\color{green}\ding{52}}  & {\color{green}\ding{52}}  & {\color{red}\ding{55}}   \\ \hline
\cite{han2024parameter} 
    &  Overviews PEFT algorithms for LLMs, reviews computing-efficient applications and techniques, and introduces system design for PEFT.
    & {\color{green}\ding{52}}  & {\color{red}\ding{55}}  & {\color{red}\ding{55}}   & {\color{red}\ding{55}}    
    & {\color{green}\ding{52}}  & {\color{green}\ding{52}}  & {\color{red}\ding{55}}  \\ \hline
Ours 
    &  Reviews the state-of-the-art approaches to edge LLM training, inference, caching, and delivery in MEI, with an emphasis on enhancing storage, computing, and communication efficiency of LLM deployment at the network edge.
    & {\color{green}\ding{52}}  & {\color{green}\ding{52}}  & {\color{green}\ding{52}}   & {\color{green}\ding{52}}    
    & {\color{green}\ding{52}}  & {\color{green}\ding{52}}   & {\color{green}\ding{52}} \\ \hline
\end{tabular}
\end{table*}

While MEI provisions powerful resources, supporting LLM services is still a non-trivial research task. The deployment of LLMs is much more resource-intensive than conventional deep neural networks (DNNs), such as convolutional neural networks (CNNs), which is the main hurdle in bringing LLMs to the network edge. Apart from communication-computing resource requirements \cite{zhou2023opportunities,chien2023reducing,wang2023overview}, improving the resource and energy efficiency of LLMs is also crucial for successfully deploying them at the network edge. This survey paper aims to provide a contemporary survey of this converging trend, i.e., MEI and LLMs, \textit{primarily from the perspective of resource-efficient deployment of LLMs with MEI}, including storage efficiency, computing efficiency, and communication efficiency at the network edge. This paper differs from the prior survey papers on efficient LLM training/fine-tuning and inference, such as \cite{zhu2023survey,wan2024efficient,kachris2024survey,liu2024understanding,han2024parameter,zhao2023survey,miao2023efficient,yuan2024llm}. While these papers focus on improving computing efficiency, including efficient LLM compression methods \cite{wan2024efficient}, hardware optimization \cite{kachris2024survey}, efficient fine-tuning \cite{han2024parameter}, and fast decoding \cite{yuan2024llm}, for efficient LLM training and inference, they overlook the impact of communications on LLM training, inference, and caching and delivery, which is a significant bottleneck in mobile edge networks. This paper also differs from existing surveys/articles on LLM edge deployment, such as \cite{fang2023large,xu2023unleashing,bai2024efficiency,xu2024survey}, 
which explore LLM-empowered AI service provisioning with cloud-edge synergy, including generative AI services \cite{xu2023unleashing}, remote LLM inference \cite{fang2023large}, and LLM-driven sensing, digital twins, and task-oriented communications \cite{10648594} in cloud-edge networks. Unfortunately, they do not discuss \textit{resource-efficient deployments}, such as parameter-efficient fine-tuning, split inference/learning, efficient LLM caching and delivery, and \textit{their interplay with wireless edge networks}. Lastly, it is noted that this survey paper is fundamentally different from these papers on ``LLMs for networks"~\cite{shen2024large,huang2023large,zhou2024large}, where the design objective is to employ LLMs to optimize edge networks \cite{shen2024large,huang2023large} or telecommunication networks \cite{zhou2024large} rather than leveraging edge computing to support LLMs. Our paper, instead, explores the coordination of communication-computing resources in mobile edge networks to support LLM application provisioning, including caching, delivery, training, and inference of LLMs services at the network edge. The comparisons with some relevant surveys/papers are provided in Table \ref{table_comp}. The major contributions of this paper are summarized as follows. 


\begin{itemize}
\item We present the application scenarios that motivate the deployment of LLMs at the network edge and the supporting MEI framework for LLM service provisioning. Although the use cases of LLMs have been extensively discussed in other places, we will emphasize the necessity or benefits of provisioning these applications at the mobile edge based on their service requirements, including latency, bandwidth, and privacy needs.
\item We provide the first comprehensive survey on edge LLM caching and delivery, edge LLM training, and edge LLM inference within 6G edge networks. We will particularly concentrate on the resource-efficient deployment of LLMs to improve the storage, communication, and computing efficiency of LLMs at the network edge.
\item We identify several crucial future research directions for the integration of LLMs and MEI, including green edge AI, secure edge AI, and quality-aware edge training for LLMs.
\end{itemize}

As outlined in Fig. \ref{fig_survey_outiline}, the survey is organized as follows. Section \ref{application} illustrates four key applications that demonstrate the necessity of deploying LLMs at the network edge. Section \ref{challenges} presents an overview of LLMs and MEI, and Section \ref{sec:on-device} introduces the state-of-the-art resource-efficient LLM techniques. In Section \ref{sec:architecture}, we present the MEI for LLM (MEI4LLM) framework, which supports the deployment of LLMs at the network edge. This framework consists of AI-native architecture, parameter-sharing LLM caching and delivery, distributed LLM training/fine-tuning, and distributed LLM inference. Sections \ref{sec:caching}, \ref{sec:learning}, and \ref{sec:inference} delve into efficient techniques for edge LLM caching and delivery, edge LLM training, and edge LLM inference, respectively, considering storage efficiency, computing efficiency, and communication efficiency. Finally, we outline the roadmap for future research opportunities in Section \ref{sec:further} and present our conclusions in Section \ref{conclusion}.

\begin{figure*}[t!]
\centering
\includegraphics[width=0.9\textwidth]{surveyoutline.pdf}
\vspace{-0.25cm}
\centering\caption{The outline of this survey.}
\label{fig_survey_outiline}
\end{figure*}

\section{Application Scenarios\label{application}}
\begin{figure*}[ht]
    \centering
    \begin{minipage}{0.6\textwidth} 
        \centering
        \includegraphics[width=\textwidth]{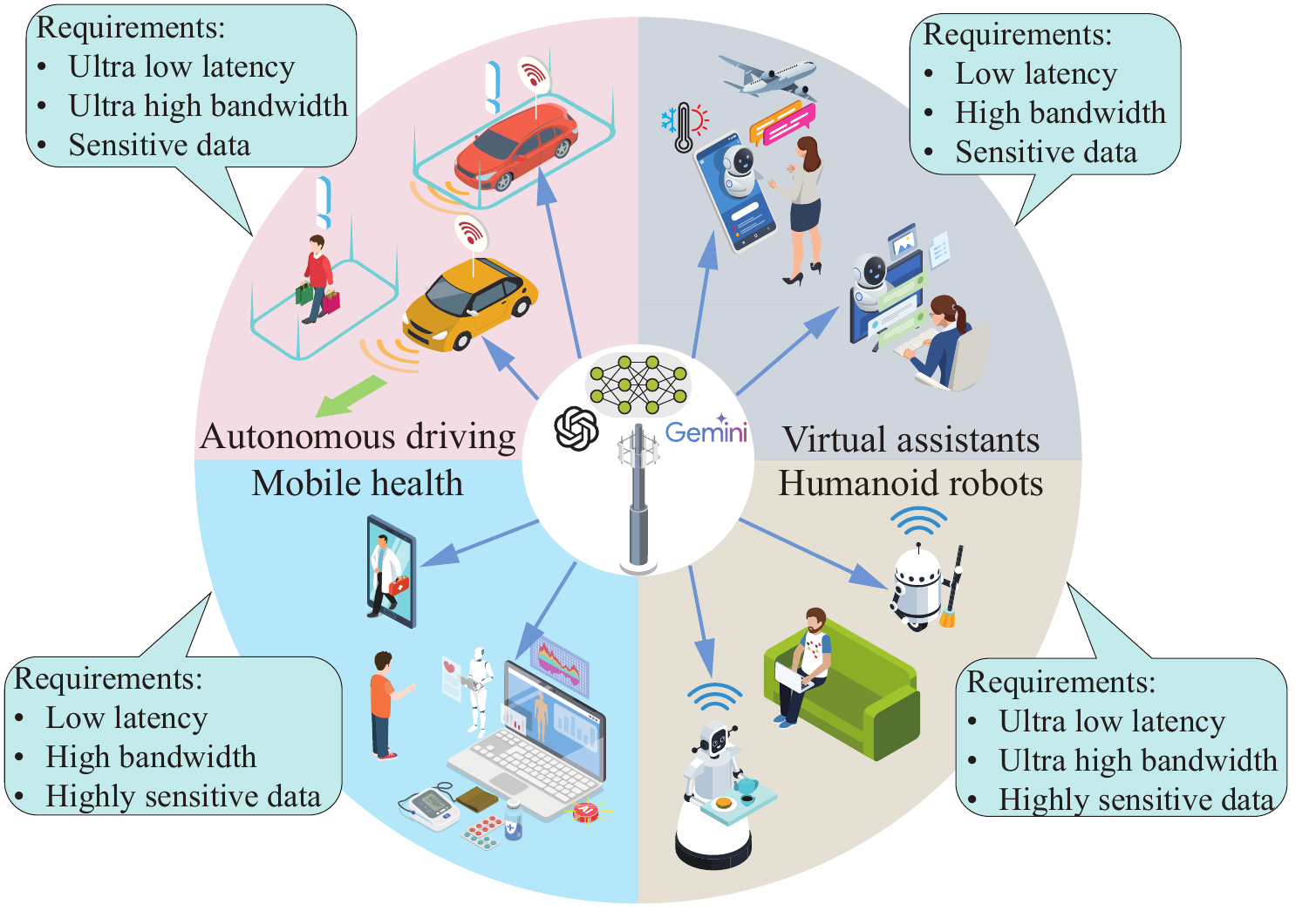}
    \end{minipage}
    \\
    \vspace{0.3cm}
    \begin{minipage}{\textwidth}
        \centering
        \renewcommand{\arraystretch}{1.4}
        \begin{tabular}{|>{\columncolor{myblue}}p{2.5cm}|>{\columncolor{mygrey}}p{2.8cm}|>{\columncolor{myblue}}p{3.5cm}|>{\columncolor{mygrey}}p{2.5cm}|>{\columncolor{myblue}}p{3.5cm}|}
            \hline
             & \makecell[c]{Mobile health} &  \makecell[c]{Humanoid robots} & \makecell[c]{Virtual assistants} & \makecell[c]{Autonomous driving}  \\ \hline
            \makecell[c]{Latency reqs.} & \makecell[c]{$\le$400 ms \cite{skorin2010analysis}} & \makecell[c]{10-100 ms \cite{3gpp.22.874}} & \makecell[c]{200 ms \cite{nvidia2021creating}} & \makecell[c]{10 ms \cite{3gpp.22.874}} \\ \hline
            \makecell[c]{Bandwidth reqs.} & \makecell[c]{10-50 Mbps \cite{qureshi2022communication}} & \makecell[c]{80 Mbps-12 Gbps \cite{3gpp.22.874}} & \makecell[c]{144 Mbps \cite{3gpp.22.261}} & \makecell[c]{80 Mbps-12 Gbps \cite{3gpp.22.874}} \\ \hline
        \end{tabular}
    \end{minipage}
    \caption{Killer LLM-empowered applications demonstrating the need for deploying LLMs at the network edge and the corresponding latency and bandwidth requirements on practical cases.}
    \label{fig_application}
\end{figure*}

Although LLMs can be applied to a broad range of tasks, we focus on the application scenarios that motivate the deployment of LLMs at the network edge. As illustrated in Fig. \ref{fig_application}, we will present four killer LLM-powered applications: mobile health, humanoid robots, virtual assistants, and autonomous driving. We focus on their service requirements, including latency, bandwidth, and privacy requirements, to underscore the need for deploying the LLMs at the network edge.

\textbf{Mobile health:} Healthcare is one of the most promising applications for LLMs. Google's Med-PaLM 2, for example, is an LLM fine-tuned on medical datasets, capable of delivering answers to medical inquiries~\cite{singhal2023towards}. Recently, Fitbit and Google Research have joined forces to build an LLM to support personalized health and wellness features, which aims to help people get summarization and recommendations from the data generated from their mobile devices~\cite {Google2024New}. Specifically, this model can deliver personalized coaching capabilities, like actionable messages and guidance, based on personal fitness goals. Apart from this, healthcare LLMs can also assist in medical question answering, diagnosis, treatment, and medical report generation~\cite{he2023survey}. With these exciting applications, mobile health has the following service requirements, making it better to deploy LLMs at the mobile edge: 

1) Latency requirements: Some LLM-empowered mobile health applications demand prompt warning messages to avert undesirable health consequences. For example, the state-of-the-art fall detection algorithms can achieve a latency of 37 ms~\cite{yu2023practical}. Moreover, the tolerant audio/video conferencing latency for an emergency accident and a routine checkup varies from 0 to 150 ms and 150 to 400 ms, respectively~\cite{skorin2010analysis}. LLMs should also achieve low latency to support the aforementioned applications, i.e., triggering warnings and generating advice upon fall detection. Since these scenarios require analytics of high-dimensional data/features to trigger warnings, uploading the data to cloud centers will experience long latency and high delay jitter through the backbone network.

2) Bandwidth costs: Medical LLMs often possess multimodal processing capabilities. For instance, Google's Med-PaLM 2 has a multimodal version, called Med-PaLM M, which processes rich data modalities spanning text, imaging, genomics, and more, to interpret the biometrics of subjects. Emerging 5G-enabled mobile health applications also incorporate medical augmented reality/virtual reality (AR/VR), where the requirement ranges from 10–50 Mbps for 360-degree 4K video \cite{qureshi2022communication}. Centralizing such data on the cloud for training or inference will consume significant network bandwidth, which is expensive for consumers and service providers.

3) Privacy requirements: Health information is among the most sensitive categories of data as defined by numerous laws. For example. Article 9 of the General Data Protection Regulation (GDPR) classifies health data as special personal data~\cite{GDPR}, the collection/processing of which is subject to the data subject's explicit consent. Given the strict regulations and growing public awareness of privacy, mobile health applications demand deploying LLMs at the network edge to enable localized data processing.

\textbf{Humanoid robots:} By harnessing LLMs, there is great potential for instilling human-like intelligence into humanoid robots. This is a version previously in science fiction. One example is Optimus, a general-purpose robotic humanoid under development by Tesla. The recent version, i.e., Optimus Generation 2, demonstrates smooth action about dancing and poaching an egg. Also, by combining LLMs and humanoid robots, NVIDIA has introduced Project GR00T, an initiative that utilizes a general-purpose foundation model for humanoid robot learning, which takes multimodal instructions and past interactions as inputs to generate robotic actions. With the power of LLMs, humanoid robots can efficiently perform numerous tasks, from assisting in warehouses to executing rescue missions to offering support in hospitals, elderly communities, and homes. However, humanoid robot applications also face strict latency and data privacy requirements:

1) Latency requirements: Robotic applications have very stringent latency requirements to enable robots to act swiftly in ever-changing environments and respond immediately to human instructions. According to 3GPP, 5G remote-controlled robotics requires 10-100 ms E2E latency and 2 ms intermediate data uploading latency~\cite{3gpp.22.874}, which is usually unattainable with cloud computing that incurs round-trip latency often larger than 100 ms\cite{satyanarayanan2009case}.

2) Bandwidth costs: Robots are equipped with multimodal sensors, thereby involving intensive data communications to servers, including vision data and high dimensional features. According to 3GPP, considering split inference, the required upload data rate varies from 80 Mbps to 12 Gbps, depending on the neural network architecture~\cite{3gpp.22.874}, leading to significant bandwidth costs if deployed in the cloud.

3) Privacy requirements: A primary concern of robotic applications, especially in smart home environments, is data privacy. Robots gather personal data daily by monitoring and engaging with people, involving highly sensitive data about the owners' daily activities. This necessitates the localized processing of LLMs.

\textbf{Virtual assistants:} LLMs can facilitate the everyday lives of human beings based on virtual assistants deployed on our smartphones or personal computers. While earlier virtual assistants like Siri and Google Assistant focused on basic functions such as audio recognition and Internet searches, the advent of LLMs has transformed the landscape. The advanced virtual assistants can serve as general-purpose agents, fundamentally altering how we interact with computers and phones, manage business tasks, and navigate our daily lives. For example, Microsoft Copilot, powered by OpenAI's GPT-4, can aid users in drafting documents, emails, presentations, and much more~\cite{Spataro2023Introducing}. Virtual assistants also require real-time responses to enhance user experience and localized data processing to preserve data ownership:

1) Latency requirements: A virtual assistant must be able to respond with an accurate answer in almost real time. According to NVIDIA, a 200 ms E2E latency can cause human beings to perceptively experience the delay and hamper user experience~\cite{nvidia2021creating}. It is important to note that the communication latency requirement would be even stricter as it only constitutes part of the E2E latency. Moreover, data transmissions across backbone networks often experience significant delay jitter, which can degrade service quality. These factors highlight the importance of deploying LLMs at the mobile edge.

2) Bandwidth costs: Future virtual assistants process text, audio, images, and videos. For instance, GPT-4 with vision, sometimes referred to as GPT-4V, allows image input to generate answers, such as instructing users on fixing a bike based on an input image. According to 3GPP, split AI image recognition may require an uplink data rate of 144 Mbps \cite{3gpp.22.261}. Given the potential for millions or billions of users, decentralizing this data for processing at the edge can considerably reduce bandwidth consumption costs.

3) Privacy requirements: Virtual assistants assist people in their everyday lives, which inevitably involve personal information such as email exchanges, Internet search records, and location information. Moreover, using LLMs to create emails, documents, or presentations may expose proprietary company information. These concerns underscore the importance of processing data on edge devices to maintain privacy and protect data ownership.

\textbf{Autonomous driving:} Existing autonomous driving solutions mostly rely on a modular approach, which divides driving into separate components, such as perception, prediction, and planning. The modular design inherently possesses limited capabilities for tasks requiring complex and human-like reasoning, where LLMs excel \cite{hu2024agentscodriver}. For example, when construction workers are at an intersection, LLMs have shown the ability to reason and make informed decisions about the right route. In the auto industry, Ghost Autonomy secured a \$5 million investment from OpenAI, the company owning GPT models, to bring large-scale and multi-modal LLMs to autonomous driving to handle the long tail of rare and complex driving scenarios~\cite{Ghost2023Ghost}. In China, Geely develops an LLM for autonomous driving offering services like vehicle-to-outside voice interactions and entertainment functions~\cite{TechNode2023Geely}. Although the relevant development is still in the early stages, it can be anticipated that LLMs will provide in-car and autonomous driving services to consumers soon. Undoubtedly, autonomous driving relies on in-car and edge computing to support real-time requirements and privacy preservation.

1) Latency requirements: Autonomous driving is one of the most mission-critical and delay-sensitive applications \cite{fang2024pacp}. 
According to 3GPP, autonomous driving cases may have a 10 ms E2E latency requirement~\cite{3gpp.22.874}.
To generate real-time responses and actions to rapidly changing vehicular environments, deploying LLMs in the cloud is ill-suited, making it essential to move LLMs to the network edge.

2) Bandwidth costs: Autonomous vehicles come equipped with multi-modal sensors, including multiple cameras and LiDAR, which generate up to 4TB of data per day~\cite{chen2024vehicle,hu2024collaborative}. Centralizing such data from a vast number of vehicles could easily overload the backbone network and the cloud. Moreover, according to 3GPP, in the split inference scenario, the uplink data rate ranges from 80 Mbps to 12 Gbps in different model architectures \cite{3gpp.22.874}.

3) Privacy requirements: Uploading vehicle data to cloud centers inevitably results in location privacy leakage, which is widely recognized as sensitive personal data. Processing data on vehicles or at the edge can enhance the data privacy of end users.

\textbf{Lessons Learned}: The stringent QoS requirements of LLM-empowered applications highlight the necessity of deploying LLMs in edge networks. First, applications demanding ultra-low latency cannot be supported by cloud computing due to the significant transmission delays over mobile backhaul and backbone networks. Second, ultra-high bandwidth costs resulting from large volumes of multimodal data not only raise the costs for both users and application developers (which need to pay the Internet service providers) but also result in network congestion. Third, with the sensitivity of the user input data, raw data sometimes must remain locally to prevent privacy leakage. These factors underscore the importance of deploying LLMs at the edge to provide timely, efficient, and privacy-preserving AI services to edge devices.

\section{Preliminaries I: An Overview of LLMs and MEI\label{challenges}}
As discussed in Section \ref{application}, the emergence of LLMs has unlocked new avenues for enhancing various edge applications. To fully leverage MEI to support LLMs within edge networks, it is essential to comprehend the key concepts behind both LLMs and MEI. Towards this direction, Section \ref{challenges}  provides preliminaries of LLMs and MEI. Section \ref{sec:on-device} reviews resource-efficient LLM techniques, which are indispensable for the edge deployment of LLMs. By laying out the foundational knowledge, these two sections prepare for the subsequent discussions on the integration of LLMs and MEI.
\subsection{Large Language Models}
\subsubsection{The Transformer architecture} LLMs are mostly built with Transformer-based architecture. Transformer~\cite{vaswani2017attention} have sparked a significant paradigm shift in the domain of natural language processing (NLP), demonstrating exceptional performance on a broad range of language tasks, including text classification~\cite{soyalp2021improving}, machine translation~\cite{ott2018scaling} and question answering~\cite{guan2022block}. For example, Bidirectional Encoder Representations from Transformers (BERT)~\cite{devlin2018bert} have achieved state-of-the-art performance in question-answering tasks, showcasing superiority in capturing contextual information efficiently. The breakthroughs from Transformers have extended beyond NLP by achieving tremendous success in computer vision. Transformer models and their variants have been widely employed in various image processing tasks such as image recognition~\cite{dosovitskiy2020image}, object detection~\cite{carion2020end}, and image segmentation~\cite{ye2019cross}. For instance, the Vision Transformer (ViT)~\cite{dosovitskiy2020image} segments images into non-overlapping patches and utilizes Transformer encoders to extract features, yielding superior detection accuracy over traditional CNNs~\cite{qiu2024ifvit}.

\begin{figure}[ht]
\centering
\includegraphics[width=0.45\textwidth]{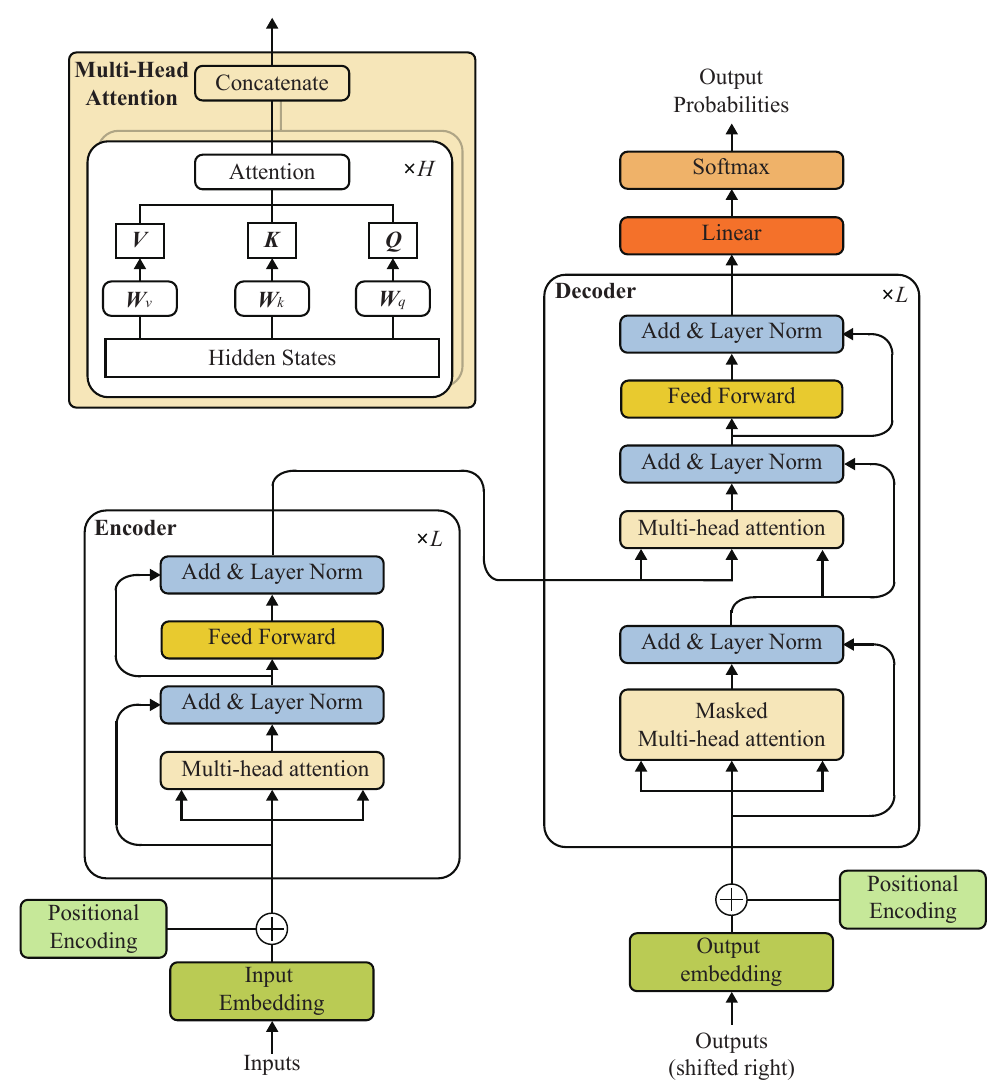}
\vspace{-0.25cm}
\caption{The Transformer architecture, adapted from \cite{vaswani2017attention}.}
\label{fig:transformer}
\end{figure}

A representative Transformer architecture is illustrated in Fig. \ref{fig:transformer}. Unlike the recursive connections for short-term context and sequential processing in Recurrent Neural Networks (RNNs), Transformers employ self-attention mechanisms to comprehensively capture intricate dependencies between sequence elements to learn long-range relationships. The core of the Transformer architecture design lies in the encoder-decoder architecture, consisting of stacked layers with multi-head self-attention mechanisms. These mechanisms prioritize processing different elements in the input sequence, enhancing the model's ability to generate output tokens effectively. 

Transformers typically operate as follows \cite{xu2024survey,zhao2023survey,vaswani2017attention}. First, input data, such as words or sentences, are segmented into token sequences through a process called tokenization, using tokenizers like WordPiece \cite{Gage1994ANA,sennrich-etal-2016-neural} or Byte-Pair Encoding \cite{6289079,devlin2018bert}. These transformed tokens are then converted into vectors by an embedding layer, which also incorporates position encoding to capture the input’s sequential information. Transformers then leverage the self-attention mechanism to evaluate the relationships between words within the inputs using their intermediate representations, i.e. query Q, key K, and value V, obtained by multiplying the input vectors with corresponding weight matrices. Afterward, layer normalization and feed-forward networks (FFNs) are applied. By using the self-attention mechanism and stacking multiple layers that integrate self-attention, layer normalization, and FFNs, the encoder converts the input sequence into context-rich representations, while the decoder employs these representations to generate the output sequence, considering the input and previously generated tokens. Finally, the output of the last Transformer layer is passed through a linear layer to generate the final result.
Moreover, Transformers typically utilize autoregressive decoding to produce results. Each new token is predicted based on the input token sequence and then appended to the input sequence, forming a longer sequence for subsequent inference steps. To avoid redundant computation of previously processed tokens in earlier steps, a key-value (KV) cache can be employed, where the intermediate states, key, and value are stored after each inference step, allowing for more efficient processing.

Self-attention lies in the heart of Transformers. The self-attention mechanisms embedded within Transformers overcome the limitation of short-term context inherent in RNNs, comprehensively grasping long-range dependencies and enhancing their ability to capture intricate relationships in sequences. Although attention modules have been widely used in feed-forward and recurrent networks~\cite{chaudhari2021attentive,de2022attention}, Transformers exclusively rely on attention mechanisms and employ a unique implementation (i.e., multi-head attention (MHA)) for parallelization optimization, facilitating scalability on high-complexity models and large-scale datasets. Other alternatives, such as hard attention~\cite{vinyals2015show}, are inherently stochastic, which necessitates Monte Carlo sampling for attention position sampling. Moreover, in contrast to convolutional or recursive counterparts~\cite{goodfellow2016deep,lecun2015deep,graves2012long}, Transformer requires minimal prior knowledge of problem structure. This characteristic renders it suitable for model pre-training via pretext tasks on large-scale unlabeled datasets~\cite{devlin2018bert,vaswani2017attention}, enabling the encoding of highly expressive and generalizable representations. These representations effectively capture the relationships among entities in a given dataset, laying the groundwork for subsequent supervised fine-tuning in downstream tasks.

\subsubsection{Unimodal LLMs} 
LLMs mainly refer to advanced transformer-based language models consisting of billions of parameters. These models are extensively pre-trained on massive datasets using deep learning techniques \cite{minaee2024large,zhao2023survey}. LLMs excel in tasks related to language comprehension, reasoning, and text generation, making them capable of solving complex tasks by producing coherent and contextually appropriate content. Notable examples include Meta’s LLaMA \cite{Qualcomm2023Llama} and OpenAI’s GPT-4 \cite{achiam2023gpt}. 
Presently, major players in the AI industry are dedicated to crafting their LLMs and applying them across various domains. For instance, OpenAI has developed the highly-regarded chat LLM, GPT-3~\cite{dale2021gpt}, demonstrating exceptional performance across various NLP tasks, such as text generation and machine translation. Google has introduced the medical LLM, Med-PaLM~\cite{singhal2023towards}, capable of offering expert-level medical guidance and diagnoses. Facebook proposed an innovative image classification LLM, DEiT~\cite{touvron2021training}, which integrates self-supervised learning with the Transformer architecture to achieve race-level image classification performance with limited annotated data. These LLMs are trained on extensive and varied datasets available on the Internet~\cite{lin2024splitlora}. 

\begin{figure}[ht]
\centering
\includegraphics[width=0.3\textwidth]{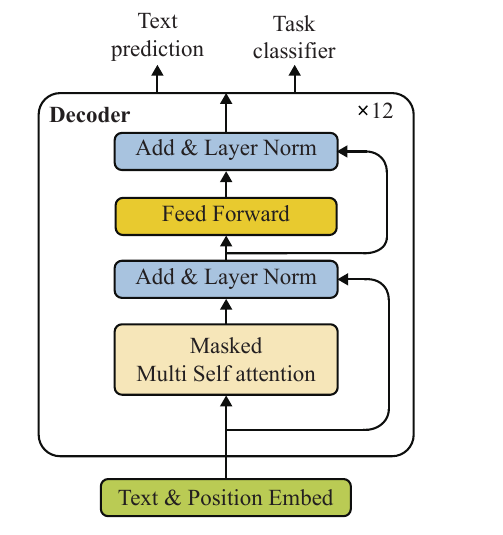}
\vspace{-0.25cm}
\caption{The decoder-only LLM architecture, which is adopted by GPT models \cite{radford2018improving}.}
\label{fig:decoder}
\end{figure}

A variety of LLMs have been built and evolved based on the Transformer architecture. LLM architectures can be classified into three categories: encoder-only LLMs, encoder-decoder LLMs, and decoder-only LLMs \cite{zhou2024large,minaee2024large,xu2024survey}. An overview of some popular LLMs with these three architectures is illustrated in Table \ref{table_overview_llm}.
\textbf{Encoder-only LLMs}, e.g. BERT \cite{devlin2018bert} and ALBERT \cite{Lan2020ALBERT}, exclusively consist of encoder components. The encoder is responsible for processing input sequences to generate contextualized representations for each token. Despite lacking a decoder to produce output sequences, encoder-only LLMs still exhibit exceptional performance in various NLP tasks such as text classification, sentence similarity computation, and language understanding due to their efficient feature extraction capabilities and adaptable representations. 
\textbf{Encoder-decoder LLMs}, exemplified by models like T5~\cite{raffel2020exploring}, represent a pivotal advancement in the NLP domain, integrating both encoder and decoder components within their architectures. The encoder processes input sequences to generate contextualized representations, while the decoder utilizes these representations to generate output sequences, typically in a sequence-to-sequence manner. Encoder-decoder LLMs find widespread application in tasks such as machine translation, text summarization, and question answering, owing to their ability to capture complex linguistic structures and contextual dependencies. 
\textbf{Decoder-only LLMs}, epitomized by the well-known GPT series~\cite{dale2021gpt,sanderson2023gpt}, constitute a significant branch of LLMs, only consisting of a decoder network. This type of LLMs constitutes a significant branch of LLMs. An example of the decoder-only LLM architecture is illustrated in Fig. \ref{fig:decoder}. Decoder-only LLMs adopt autoregressive decoding, which is widely used in both decoder-only and encoder-decoder LLMs, to generate output sequences based on previous tokens in the sequence. This architectural design makes them particularly well-suited for tasks where the model generates text sequentially, such as language generation, text completion, and dialogue response generation.

\begin{table}[!t]
\centering
\caption{An overview of some popular LLMs.}
\label{table_overview_llm}
\renewcommand{\arraystretch}{1.4}
\setlength{\tabcolsep}{2mm}
\begin{tabular}{|c|c|c|c|}
\hline
\textbf{LLM arch.} & \textbf{Model name} & \textbf{Release Year} & \textbf{Parameters} \\ 
\hline
\multirow{4}{*}{\makecell[c]{Encoder-\\only}}
    & BERT \cite{devlin2018bert} & 2018 & 110-340M \\ \cline{2-4}
    & ALBERT \cite{Lan2020ALBERT} & 2019 & 12-235M \\ \cline{2-4}
    & RoBERTa \cite{liu2019roberta} & 2019 & 355M \\ \cline{2-4}
    & XLNet \cite{NEURIPS2019_dc6a7e65}& 2019 & 110-340M \\ \cline{1-4}
\multirow{7}{*}{\makecell[c]{Decoder-\\only}}
    & GPT-1 \cite{radford2018improving} & 2018 & 117M \\ \cline{2-4}
    & GPT-2 \cite{radford2019language} & 2019 & 1.5B \\ \cline{2-4}
    & GPT-3 \cite{brown2020language} & 2020 & 175B \\ \cline{2-4}
    & PaLM \cite{10.5555/3648699.3648939} & 2022 & 8-540B \\ \cline{2-4}
    & Gemini Nano-2 \cite{team2023gemini} & 2023 & 3.25B \\ \cline{2-4}
    & LLaMA \cite{touvron2023llama} & 2023 & 7-65B \\ \cline{2-4}
    & GPT-4 \cite{achiam2023gpt} & 2023 & Close-sourced \\ \cline{1-4}
\multirow{5}{*}{\makecell[c]{Encoder-\\decoder}}
    & T5 \cite{raffel2020exploring} & 2019 & 60M-11B \\ \cline{2-4}
    & BART \cite{lewis-etal-2020-bart} & 2019 & 140M \\ \cline{2-4}
    & mT5 \cite{xue-etal-2021-mt5} & 2020 & 300M-13B \\ \cline{2-4}
    & HuBERT \cite{10.1109/TASLP.2021.3122291} & 2021 & 90M-1B \\ \cline{2-4}
    & Claude 3 \cite{claude2023} & 2024 & Close-sourced \\ \hline
\end{tabular}
\end{table}



\begin{figure}[t]
\centering
\includegraphics[width=0.45\textwidth]{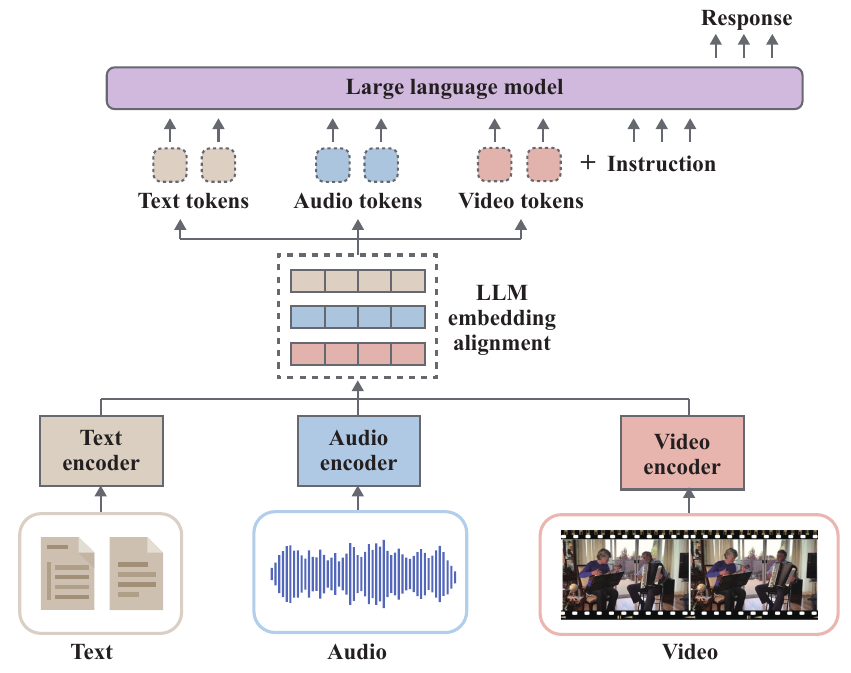}
\vspace{-0.25cm}
\caption{The structure of multimodal LLM. }
\label{fig:multi_modal_LLM}
\end{figure}

\subsubsection{Multimodal LLMs} Since traditional LLMs~\cite{chang2023survey,brown2020language,xu2024large} are mainly applied to textual data, the unimodal model training for LLMs limits their ability to comprehend other data types beyond text. For instance, traditional LLMs like GPT-3 and BERT~\cite{devlin2018bert} only rely on textual inputs. However, in numerous real-world scenarios, language comprehension is not limited to textual context but also visual cues, auditory signals, and contextual sensing information from diverse sensors.

To address the above issue, academia and industry extensively delve into the paradigm of multimodal LLMs shown in Fig.~\ref{fig:multi_modal_LLM}, amalgamating various modalities such as text, images, and audio, into a unified framework, unlocking the potential for handling diverse data types. For instance, GPT-4~\cite{sanderson2023gpt} excels at simultaneously processing both image and text inputs, exhibiting human-comparable performance across various benchmark tests. In image description tasks, GPT-4 utilizes both images and associated textual data to generate more precise and vivid descriptions, while in speech recognition tasks, it merges speech signals with textual information to improve speech comprehension and conversion. Multimodal perception plays a pivotal role in the pursuit of general AI, driven by the imperative to process complex real-world data \cite{yang2023modal}. This necessitates AI models capable of cross-modal information fusion and interactive learning, boosting training performance across multiple perceptual domains.

Multimodal LLMs inherit powerful learning capabilities of LLMs to empower diverse and complex multimodal tasks by integrating foundation models of various modalities. The LLMs provide robust language generation, zero-shot transfer capabilities, and in-context learning, whereas foundation models of other modalities offer informative representations from other data types~\cite{wu2023multimodal,cui2024survey}. Since foundation models of varied modalities are individually pre-trained, the primary challenge in constructing multimodal LLMs lies in how to connect these models to achieve high-performance collaborative training/inference. The predominant research in this domain focuses on refining modality alignment via multimodal pre-training~\cite{huo2021wenlan,qi2020imagebert} and multimodal instruction-tuning~\cite{peng2023instruction,zhang2023instruction}. Multimodal pre-training learns a common representation across modalities by training the model with multimodal datasets, such as XText \cite{Xtext2023}. During training, the model learns to correlate information from diverse modalities by optimizing predefined objectives, thus achieving alignment across modalities. This alignment enhances the model's understanding of inter-modality relationships, leading to superior performance across various cross-modal tasks. Multimodal instruction-tuning is a method of fine-tuning based on pre-trained models, aimed at improving model performance on specific tasks. It combines the models with one or more modality-related tasks, then fine-tunes the model using modality-labeled data to improve its alignment with modality-specific tasks. Multimodal instruction-tuning enables models to learn to empower unseen tasks by following new instructions, thereby improving the model's zero-shot performance and generalization capability.

\subsubsection{Practical applications: Generative/Interactive AI\label{generative}}
The rapid development of LLMs has a profound impact on various applications, particularly in generative AI (GAI) and interactive AI (IAI). GAI focuses on creating a wide range of content, including images, text, music, and video \cite{zhou2024survey}, collectively known as AI Generated Content (AIGC). By utilizing multimodal LLMs trained on high-quality datasets, GAI can effectively create superior AIGC based on input text~\cite{jeong2023study}. 
On the other hand, IAI can be regarded as the next phase of GAI. IAI responds to user queries in applications like chatbots and virtual assistants while enabling AI agents to adapt through user interaction, thereby continually improving the accuracy~\cite{Dirox,dureinforcement}. By leveraging powerful LLMs and the content generation strengths of GAI, IAI enables AI agents to emulate human interaction and generate meaningful and dynamic conversations with users~\cite{du2023usercentric,zhang2024interactivegenerative}. In this regard, LLMs are also regarded as the cornerstone of IAI because they facilitate complex interactive dialogues.


To enable AI agents to generate more accurate and up-to-date responses, retrieval augmented generation (RAG) can be integrated into LLMs to empower both IAI and GAI~\cite{zhang2024interactive}. Specifically, LLMs use the input sequence to retrieve relevant data from the external knowledge sources when generating responses, thus improving content generation performance \cite{nvidia_rag,lewis2020retrieval}. For example, Google combines RAG with Gemini to enhance LLMs’ ability to generate more accurate and contextually relevant responses for specific tasks~\cite{google_rag}. The main advantages of integrating RAG into LLMs are twofold. First, by connecting to knowledge sources rich in the latest information, RAG grounds LLMs on the most factual, accurate, and up-to-date content, reducing the likelihood of ``hallucinations'' in generated outputs and eliminating the need for frequently adapting LLMs. Second, RAG enables users to verify the sources of the models’ responses for improved trustworthiness \cite{ibm_rag}. 

\subsubsection{Industrial progress for LLMs}
LLMs have seen significant advancements in the industry, driven by the maturation of deep learning algorithms~\cite{wu2020liveness,yuan2023graph,wu2021toward}, increased computing capabilities, and the availability of large-scale datasets. Major technology companies, including OpenAI, Google, Microsoft, and Meta, have made substantial investments in LLM research and development, leading to the creation of prominent models like GPT series~\cite{dale2021gpt,sanderson2023gpt} and BERT~\cite{devlin2018bert}. These models have demonstrated exceptional performance across a spectrum of NLP tasks, including language translation, text generation, question answering, and sentiment analysis. Furthermore, multimodal LLMs have expanded beyond their initial domain of NLP and are shining in diverse sectors, such as healthcare, autonomous driving, and smart cities. For instance, in healthcare, Med-PaLM~\cite{singhal2023towards} is devised for medical image analysis, clinical document processing, and patient diagnosis, facilitating accurate diagnoses and treatment decisions by healthcare professionals. In the realm of autonomous driving, DriveMLM~\cite{wang2023drivemlm} bridges the gap between language decisions and vehicle control commands, enabling closed-loop autonomous driving in realistic simulators. As can be seen, the proliferation of LLMs offers substantial value across multiple industries.  

Recent advancements in on-device LLMs have garnered attention from the industry. For instance, Meta proposed an on-device LLM, named MobileLLM, utilizing deep and thin architectures, embedding sharing, and grouped-query attention mechanisms~\cite{liu2024mobilellm}. Google introduced a new instruction-tuning approach for building a mobile-centric text rewriting LLM~\cite{zhu2023towards}. Nevertheless, on-device LLMs often underperform compared with powerful LLMs of larger model sizes. For example, Google's Gemini Nano-1, designed for on-device deployment, contains only 1.8 billion parameters in a 4-bit format, which is distilled from larger Gemini models \cite{team2023gemini}. Due to its compact size, when the capabilities of such a small LLM are insufficient for the requirements of edge devices, these devices may still need to upload data to access large-scale LLMs, i.e., on edge servers.

\subsection{Mobile Edge Intelligence}
MEI has emerged as a promising paradigm integrating AI with mobile edge computing, revolutionizing the landscape of mobile services and applications \cite{wang2020thirty,wu2020caiauth,yuan2024satsense,peng2024sums}. As the precursor to MEI, mobile edge computing allows connected edge servers in mobile edge networks, such as base stations and wireless access points, to provide computing services to resource-limited edge devices. This significantly reduces the computing burden on edge devices and avoids the need to directly transfer massive amounts of data to the cloud, which would otherwise incur long transmission latency \cite{9442308,9979702}. MEI can be viewed as the confluence of AI and mobile edge computing. By leveraging mobile edge computing technology, MEI enables connected edge servers to execute AI algorithms with end-edge-cloud synergy, facilitating data collection, caching, processing, and analysis for data captured by edge devices in mobile AI applications. This greatly enhances both the efficiency and speed of data processing while enhancing data privacy \cite{9052677,9596610,zhang2019openei}. Specifically, MEI surmounts the limitations of traditional cloud-centric architectures by coordinating computing resources across a hierarchical structure that includes edge devices, edge servers, and cloud centers. This coordination can substantially optimize the performance of localized AI services, such as AI model caching, training, and inference \cite{8736011,9596610}, thereby facilitating the deployment of intelligent applications at the network edge.


By integrating AI and communications, the MEI framework enables mobile networks to provide services beyond communications, establishing a strong foundation for the Intelligence of Everything. Along this line, the usage case ``Integrated AI and Communication'' has been included in the IMT Framework Recommendation for 6G~\cite{HuaweiTech2023}. For standardization, the telecommunication standardization organizations 3GPP and ITU have depicted the prospects of edge intelligence in their white papers, respectively. ITU-3172~\cite{ITU-3172} elucidates the necessity of hosting machine learning (ML) functionalities closer to the network edge based on the latency-sensitive requirements of ML applications. In 3GPP release 18 for 5G standardization, MEI aims to support distributed learning algorithms, split AI/ML, and efficient AI model distribution \cite{3gpp.22.874}. The details are elaborated next. 
First, edge learning, such as federated learning, will be fully supported in edge networks, which enables edge servers to aggregate model updates and knowledge from multiple distributed edge devices, thereby improving the performance of AI/ML models. 
Second, the split AI/ML over 5G edge networks can facilitate the deployment of AI applications on devices with conflicting requirements, such as computation-intensive, energy-intensive, privacy-sensitive, and delay-sensitive requirements. For example, in the edge split inference, an AI model is partitioned into sub-models, and the computation-intensive and energy-intensive sub-models are offloaded to 5G edge servers (e.g., base stations). The edge server can execute the inference with the edge-side sub-model and the uploaded intermediate data from edge devices. 
At last, efficient AI model downloading ensures that an AI model can be delivered to edge devices with low latency when edge devices need to adapt to new AI tasks and environments. For example, autonomous driving vehicles need to download a new AI model within 1 second from a 5G edge server when the driving environments change. To integrate network-based AI algorithms into the 5G networks, the MEI framework needs to cater to the request for high-speed and stable data links between edge servers and edge devices. These links can enable high and constant uplink data rates for continuously uploading intermediate data/model updates to the edge server and high downlink data rates in a burst for downloading AI models to edge devices in a timely fashion. Moreover, the core of MEI is to capitalize on the proximity of data sources to edge computing devices (e.g., smartphones, laptops, and wearable gadgets) to enable intelligent decision-making closer to the data source. This distributed computing paradigm exhibits numerous advantages over conventional centralized architectures, including latency reduction, improved bandwidth utilization, data privacy preservation, and enhanced resilience to network failures.

On the application side, MEI carries substantial implications across various domains, such as smart healthcare, autonomous driving, and smart cities \cite{chen2024survey}. For instance, in the healthcare sector, MEI empowers real-time monitoring of patient health data and facilitates prompt interventions during emergencies. Likewise, in smart cities, MEI contributes to intelligent traffic management, environmental monitoring, and energy optimization, thereby fostering sustainability and enhancing the quality of life. Edge Intelligence has also witnessed notable industrial advancements, particularly with the proliferation of edge computing technologies and the advent of 5G networks. Leading enterprises, such as Microsoft, Google, Amazon, and NVIDIA, have developed edge AI platforms to support real-time AI services. For edge AI-empowered Internet of Things (IoT) applications, Microsoft ``Azure IoT Edge'', Google ``Cloud IoT'', Amazon ``Web Services IoT'', and NVIDIA ``EGX'' provide edge AI platforms to bring real-time AI services across a wide range of applications, spanning from live video analytics~\cite{ananthanarayanan2017real}, smart home~\cite{cao2017edgeos_h}, and industrial IoT~\cite{li2018deep}.

\subsection{Lessons Learned}

\begin{figure}[t]
\centering
\includegraphics[width=0.45\textwidth]{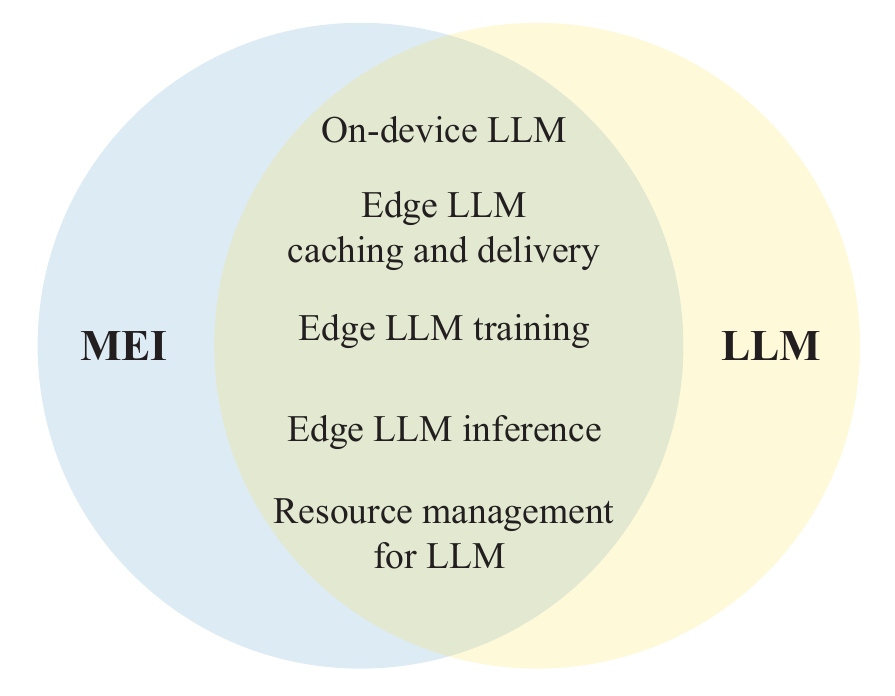}
\vspace{-0.25cm}
\caption{The convergence of LLMs and MEI.}
\label{fig:convergence}
\end{figure}

In the 6G era, LLMs can be deployed distributively across edge devices and servers within mobile edge networks. However, this deployment poses challenges due to the significant communication overhead and computing demands of LLMs. Therefore, it is necessary to advance MEI to efficiently support LLMs while maintaining superior performance, thereby expanding the applicability of LLMs to mobile AI applications within edge environments. As illustrated in Fig. \ref{fig:convergence}, integrating LLMs into mobile edge networks leads to numerous research problems, such as model caching and delivery, training, inference, and resource management.

\section{Preliminaries II: Resource-efficient LLM Techniques\label{sec:on-device}}
As we move forward to discuss the problems outlined in Fig. \ref{fig:convergence}, we first elaborate on on-device LLM, i.e., deploying LLMs on a resource-constrained device. On-device LLM deployment, including on-device inference and training, presents the following challenges.

\begin{itemize}
\item Excessive computing overhead: The computing overhead of on-device LLM deployment can be excessive. For instance, a LLAMA2-7B \cite{touvron2023llama} model with 7 billion parameters requires about 1700 giga floating point operations \cite{calflops} to perform a single forward pass on a sequence of 128 tokens. The Hexagon NPU, which is an advanced neural engine equipped on the Qualcomm Snapdragon X ELITE, can only provide a computing power of 45 tera operations per second. Furthermore, the backward propagation typically requires more computational resources than a forward pass~\cite{calflops}, implying that training would be even more challenging than inference.


\item Huge storage/memory demands: On the one hand, caching LLMs on edge devices consumes significant storage resources. LLMs specifically designed for on-device deployment even have billions of parameters, e.g., Google's on-device Gemini Nano-2 has 3.25 billion parameters. On the other hand, optimizing the model with the commonly used Adam optimizer during training usually requires 12 times the memory resources needed for inference \cite{malladi2024fine}, which can be unacceptable for mobile devices with limited memory. These facts indicate that deploying LLMs on edge devices for both training and inference poses stringent requirements on the storage and memory resources of edge devices.

\item High energy costs: The limited volume of battery in edge devices hinders the deployment of LLMs on edge devices. For instance, running an LLM quantized to INT4 with 13 billion parameters using llama.cpp (one of the most lightweight on-device LLM engines) on a Xiaomi 11 smartphone results in an energy consumption of about 56 J/token \cite{xu2023llmcad}. This implies that a smartphone with a battery capacity of 3000 mAh and an output voltage of 3.7 V can generate only approximately 700 tokens if the LLM is deployed on the smartphone. 
\end{itemize}

\begin{figure*}[ht]
\centering
\hspace{-2pt}\includegraphics[width=17 cm]{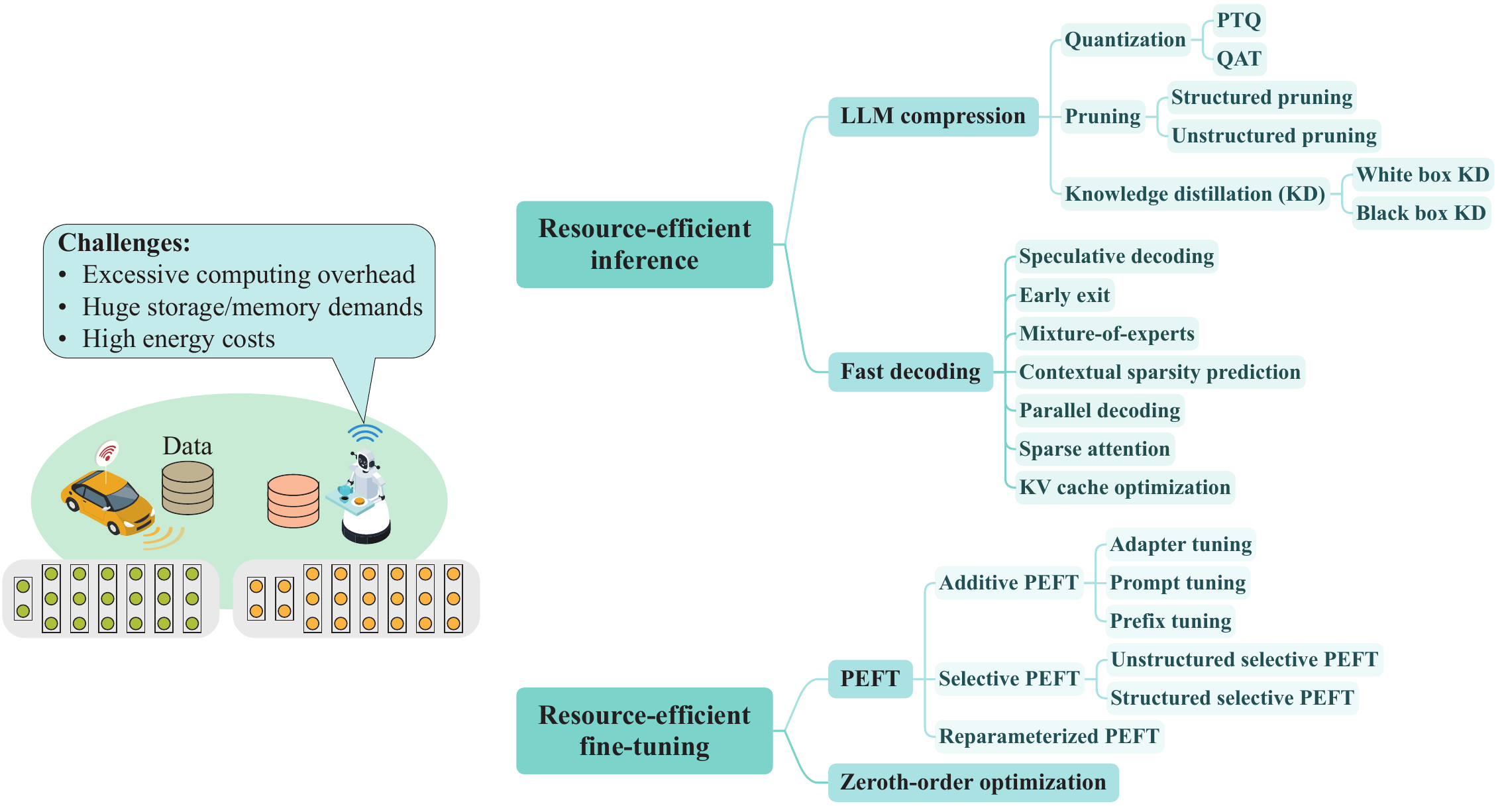}
\caption{
An overview of resource-efficient LLM techniques.}
\label{fig_ondevice}
\end{figure*}

To mitigate the above challenges, in this section, we will review the enabling techniques tailored for resource-efficient LLM deployment, focusing on resource-efficient inference and fine-tuning. These techniques are summarized in Fig. \ref{fig_ondevice}, and the comparison of related works is shown in Table \ref{table_on_device}. It is worth noting that the methods discussed in this section can decrease the complexity of LLM deployment on edge devices, edge servers, or in device-server collaboration. Consequently, these key techniques serve as the foundation for MEI4LLM and all subsequent sections.

\begin{table*}[!t]
\centering
\caption{Summary of related works on resource-efficient LLM techniques.}
\label{table_on_device}
\renewcommand{\arraystretch}{1.4}
\setlength{\tabcolsep}{2mm}
\begin{tabular}{|c|c|c|m{0.64\textwidth}|}
\hline
\textbf{Scenarios} & \textbf{Techniques} & \textbf{Ref.} & \makecell[c]{\textbf{Objectives}} \\ 
\hline
\multirow{35}{*}{\makecell[c]{Resource-efficient\\ inference}}
               & \multirow{10}{*}{LLM compression} 
                   & \cite{cheng2023optimize} & Focuses on weight-only quantization for LLMs with post-training quantization.  \\ \cline{3-4}
                   &  & \cite{lin2023awq} & Adopts uneven weight quantization for LLMs to quantize the critical weights in high precision for preserving inference accuracy. \\ \cline{3-4}
                   &  & \cite{ma2023llm} & Proposes LLM-Pruner to make pruning decisions for coupled structures in LLMs via structured pruning. \\ \cline{3-4}
                   &  & \cite{frantar2023sparsegpt} & Proposes SparseGPT for unstructured pruning and transforms the unstructured pruning problem into large-scale sparse regression problems. \\ \cline{3-4}
                   &  & \cite{gu2023minillm} & The proposed MiniLLM targets to minimize the reversed Kullback-Leibler divergence between the student and teacher model output distributions to enhance the performance of the student model via white box knowledge distillation. \\ \cline{3-4}
                   &  & \cite{wu2023lamini} & Uses the black box knowledge distillation to distillate knowledge from the teacher model through theGPT-3.5 Turbo APIs and generated instructions.\\ \cline{2-4}
               & \multirow{25}{*}{Fast decoding} 
                   & \cite{leviathan2023fast} & Proposes speculative decoding, where a lightweight language model first generates a sequence autoregressively, and a more powerful LLM then verifies and corrects the sequence. \\ \cline{3-4}
                   &  & \cite{soldaini2020cascade} & Inserts early exit modules in Transformer blocks for obtaining final outputs at exit points to skip the computation in subsequent blocks. \\ \cline{3-4}
                   &  & \cite{schuster2022confident} & Forwards the hidden states of the current token from the early exit layer to later layers for key-value cache computation.  \\ \cline{3-4}
                   &  & \cite{yi2023edgemoe} & Proposes EdgeMoE for efficient inference with mixture-of-experts-based LLMs, where the nox-expert weights that occupy less storage but need more computing resources are kept in memory all the time, and the required expert weights are only transferred between disks and memory as needed. \\ \cline{3-4}
                   &  & \cite{liu2023deja} & Proposes Deja Vu and inserts lightweight predictors after multi-layer perceptron and attention blocks to predict contextual sparsity of the next block, which can enable edge devices to activate and load only a small set of parameters of the next block for efficient inference. \\ \cline{3-4}
                   &  & \cite{cai2024medusa} & Proposes Medusa and introduces extra decoding heads on top of the last hidden states of an LLM to predict multiple subsequent tokens in parallel. After verifying all candidates, only one reasonable candidate is accepted for the next decoding phase.\\ \cline{3-4}
                   &  & \cite{fu2024break} & Introduces Lookahead Decoding and formulates autoregressive decoding as a non-linear system, which is solved by fixed-point Jacobi iteration. The Lookahead Decoding generates disjoint n-grams in parallel and verifies promising n-grams in parallel, ensuring continuity by matching the last token of the ongoing sequence.\\ \cline{3-4}
                   &  & \cite{lee2024hip} & Demonstrates that only the top-k key tokens need to be attended to for each query, which can generate a sparse attention mask, accelerate the inference process, and reduce computing resources without the significant performance trade-off.\\ \cline{3-4}
                   &  & \cite{liu2024kivi} & Develops a 2-bit quantization algorithm to quantize the key cache per channel and the value cache per token, respectively.\\ \cline{3-4}
                   &  & \cite{liu2024minicache} & The proposed MiniCache merges the key and value caches with high similarity at the same position across consecutive layers into a single cache, respectively, while those with substantial semantic meanings remain unmerged.\\ \cline{3-4}
                   &  & \cite{NEURIPS2023_a452a7c6} & The proposed Scissorhands maintains KV cache memory usage within a fixed budget. Non-influential tokens are dropped from the cache when the buffer is full.\\ \cline{3-4}
                   &  & \cite{zhang2024pyramidkv} & Dynamically allocate different KV cache budgets across various layers and strategically select important KV vectors to cache.\\ \hline
\multirow{12}{*}{\makecell[c]{Resource-efficient\\ fine-tuning}}
               & \multirow{9}{*}{\makecell[c]{Parameter-efficient\\ fine-tuning (PEFT)}} 
                   & \cite{houlsby2019parameter} & Inserts adapter modules into the Transformers and freezes the other parameters during fine-tuning via adapter tuning.   \\ \cline{3-4}
                   &  & \cite{lester2021power} & Uses prompt tuning to add soft prompt tokens at the beginning of the input tokens for fine-tuning.   \\ \cline{3-4}
                   &  & \cite{li2021prefix} & Adds trainable prefix parameters to the keys and values of the MHA in Transformer layers for fine-tuning via prefix tuning.   \\ \cline{3-4}
                   &  & \cite{fu2023effectiveness} & Uses a second-order approximation method to solve the reformulated trainable parameter selection in PEFT to choose the sparse trainable parameter for unstructured selective PEFT.  \\ \cline{3-4}
                   &  & \cite{guo2020parameter} & Proposes Structured Diff Pruning to partition parameters into groups and remove some groups before fine-tuning to realize structured selective PEFT.  \\ \cline{3-4}
                   &  & \cite{hu2021lora} & Proposes LoRA and introduces two additional trainable matrices with ranks much smaller than that of the pre-trained weight matrix. Only the newly introduced low-rank matrices are trainable during fine-tuning.  \\ \cline{2-4}
               & \multirow{1}{*}{\makecell[c]{Zeroth-order\\optimization}} 
                   & \cite{malladi2024fine} & Proposes MeZO to estimate model gradients only through forward propagation, and the estimated gradients are used for model parameter updating.\\ \hline
\end{tabular}
\end{table*}

\subsection{Resource-efficient Inference\label{sec:on-deviceinference}}
On-device LLM inference eliminates privacy leakage and the need for Internet connections. However, it requires substantial computing, memory, and energy resources. In what follows, we briefly present how to mitigate these problems by introducing efficient on-device LLM inference techniques.

\subsubsection{LLM compression}
LLM compression enables the deployment of compressed LLMs on edge devices. The design objective is to compress LLMs without substantially compromising inference accuracy. Although compression has been extensively studied in the field of traditional DNNs, the unique architectures and properties of LLMs necessitate the redesign of existing compression strategies. The compression techniques tailored for LLMs are detailed below. 

\textbf{Quantization:}
Quantization converts LLM parameters from high-precision floating-point numbers (e.g., FP16) to low-precision numbers (e.g., INT4), thereby reducing storage usage, computing latency, and energy footprint during inference. Classic model quantization can be divided into two categories: post-training quantization (PTQ) and quantization-aware training (QAT). PTQ involves directly converting the parameters of trained models into lower precision~\cite{lin2023awq}, while QAT considers quantization errors in the training phase to improve the performance of the quantized models~\cite{liu2023llm}. Although QAT typically yields better performance, it requires model retraining, which is extremely resource-intensive for on-device LLMs. Therefore, most quantization methods for on-device LLMs rely on PTQ~\cite{shao2023omniquant}. In contrast to traditional quantization strategies that target both weights and activations, LLM quantization primarily focuses on weight-only quantization \cite{cheng2023optimize}. The reasons are as follows. First, quantizing activations leads to more significant performance degradation for LLMs~\cite{xiao2023smoothquant}. Second, the primary sources of latency and energy consumption when generating new tokens with LLMs are often due to loading model parameters from memory~\cite{xu2023llmcad}. Therefore, weight-only quantization allows for more efficient loading of quantized weights from memory, making inference more efficient without significantly degrading inference accuracy.

In LLM quantization, uneven weight quantization can be adopted to preserve the inference accuracy of quantized LLMs since not all weights in LLMs contribute equally to the final inference results~\cite{lin2023awq}. For example, the authors in \cite{lin2023awq} propose the activation-aware weight quantization for uneven weight quantization. This method quantizes most of the weights in LLMs from FP16 to INT3/INT4 while only retaining approximately 1\% of the critical weights in FP16 format.

\textbf{Pruning:}
Unlike quantization, which reduces the bitwidth of LLM parameters, pruning aims to shrink the size of LLMs by directly removing redundant or non-significant parameters, thereby effectively reducing computing workload and storage usage. Based on the granularity of pruning, pruning methods can be classified into structured pruning \cite{fang2023depgraph} and unstructured pruning \cite{dong2017learning}. 1) Structured pruning prunes structured patterns, such as sub-blocks in LLMs. For example, LLM-Pruner \cite{ma2023llm} utilizes gradient information and estimated Hessian matrices to make pruning decisions for coupled structures in LLMs, such as attention heads. Then, the pruned LLMs are fine-tuned with LoRA \cite{hu2021lora} to recover model performance. The compressed LLMs pruned via structural pruning can be directly deployed and executed on standard computational frameworks without additional adjustments since the structured pruning removes entire structures in LLMs.
2) Unstructured pruning aims to remove individual weights, realizing finer-grained pruning by setting unimportant weights to zero. This approach makes the weights sparse, allowing us to leverage the sparsity to accelerate the inference of LLMs. For example, SparseGPT \cite{frantar2023sparsegpt} transforms the pruning problem into a series of large-scale sparse regression problems and addresses them with an innovative approximate solver. The proposed SparseGPT enables up to 60\% sparsity for OPT-175B \cite{zhang2022opt} before significant accuracy loss occurs. Although this approach can better preserve the performance of pruned LLM for on-device inference, it results in sparse models that require specialized hardware or software platforms during deployment, which is the major limitation of LLM unstructured pruning.

\textbf{Knowledge distillation:}
Knowledge distillation (KD)\cite{hinton2015distilling} involves transferring knowledge from a large and complex teacher model to a smaller student model. This technique enables the small-size student model to learn the behavior of the teacher model, making it suitable for deploying on resource-constrained edge devices while ensuring a competitive performance. 1) When the teacher model is fully accessible, the KD process is known as white-box KD, which allows the student model to learn the output distributions, intermediate features, and activations of the teacher model \cite{yuan2024llm,xu2024surveyKD}. For example, MiniLLM \cite{gu2023minillm} addresses the limitations of traditional KD loss functions, which do not perform well for text-generation tasks, by minimizing the reversed Kullback-Leibler divergence between the student and teacher model output distributions, thereby enhancing the performance of the student model. Additionally, the authors derive an effective optimization strategy to update the student model. 
2) When the internal structure of the models is inaccessible, the form of KD is referred to as black-box KD \cite{xu2024surveyKD}. This approach is particularly suitable for closed-source LLMs since most LLM access is limited to outputs via APIs. One method of black-box KD for LLMs involves generating numerous prompt-response pairs by the teacher LLM and enabling the student LLM to learn from the outputs of the teacher LLM. For instance, the authors in \cite{wu2023lamini} generate a set of 2.58 million instructions and use GPT-3.5 Turbo APIs to produce responses. These instructions are subsequently used to fine-tune various student language models. The distilled language model can achieve comparable performance or even outperform the 7B LLaMA \cite{touvron2023llama}. 

\subsubsection{Fast decoding for LLMs}
Fast decoding methods can be used in on-device LLM deployment to save computing resources of edge devices in inference and facilitate on-device inference. 
As explained in Section \ref{challenges}, LLMs typically generate tokens in an autoregressive manner, producing one token at a time based on previously generated tokens. However, as the size of LLMs grows, the volume of LLM parameters and intermediate data involved during inference also increases, leading to higher computational demands and prolonged inference latency. Additionally, as the output length increases, the LLM must repeat the decoding phase multiple times while caching all intermediate data, placing a significant burden on computing memory. To address these challenges and enable timely inference on resource-limited devices, fast decoding techniques have been developed. These techniques include leveraging lightweight LLMs for initial token generation, minimizing the volume of activated parameters/blocks, generating multiple subsequent tokens in a single forward pass, and reducing the volume of intermediate data. By applying these methods, memory footprint and inference latency can be significantly reduced. Typical fast decoding methods are elaborated below. 

\textbf{Speculative decoding:} LLMs typically generate text outputs in an autoregressive manner, producing one token based on the previously generated tokens in each forward propagation. Consequently, the number of iterations required to generate a sequence is equal to the length of the output sequence, resulting in significant latency and heavy computing overhead. To tackle this issue, the authors in \cite{leviathan2023fast} propose speculative decoding. In this method, a lightweight language model generates a sequence autoregressively. Then, these output tokens are processed by a more powerful LLM for verification and correction in a single inference step. 

Speculative decoding can significantly reduce LLM inference latency. In the conventional LLM inference paradigm, when the model's memory requirements exceed the capacity of edge devices, each inference process necessitates dynamically releasing inferred parameters from memory and loading new parameters from disk to memory \cite{Gerganov2024}. This process can account for up to 90\% of the inference latency \cite{xu2023llmcad}. In contrast, in speculative decoding, the lightweight model can remain in memory for continuous token generation on edge devices, and the powerful LLM verifies and corrects the entire token sequence generated by the lightweight model in a single inference step. This process reduces memory loading operations, thereby decreasing inference latency. In \cite{xu2023llmcad}, the authors demonstrate that speculative decoding can halve both per-token generation latency and energy consumption while maintaining the quality of the generated content, outperforming traditional autoregressive methods in inference. 

\textbf{Early exit:}
To minimize computing latency, the early exit strategy can be employed to bypass certain subsequent layers that input tokens traverse, effectively accelerating inference time. In the case of Transformers, early exit modules may be incorporated following some early blocks \cite{soldaini2020cascade, xin2020deebert}, where the output from intermediate layers is converted into final outputs at exit points, provided that the desired confidence value is attained. However, the conventional early exit techniques may not be effective in LLM inference since the hidden states of the token, which output the results in an early exit layer, may be missing in the later layers~\cite{10.1145/3527155}. For example, upon early exit, the hidden states and the KV caches of the corresponding token in the subsequent layers will be missing, hindering the generation of future token sequences~\cite{chen2024eellm}. To address this challenge, the hidden states of the current token can be forwarded from the exiting layer to subsequent layers for KV cache computation~\cite{schuster2022confident,li2021accelerating} if the token outputs the result in the exiting layer. 

\textbf{Mixture-of-experts:} Mixture-of-experts (MoE) can effectively scale up the LLM capacity and increase performance across various downstream tasks~\cite{fedus2022switch,lepikhin2020gshard,fedus2022review}. Specifically, the original FFN in the Transformer can be replaced with an expert network, which consists of multiple FFN experts and a router. During inference, the router directs the given input token to the best-fit FFN expert or experts. In this context, to alleviate the burden of loading substantial parameters into memory when deploying LLMs on edge devices, the MoE architecture can be leveraged so that only part of the LLM is activated and loaded during inference. For example, in \cite{yi2023edgemoe}, the authors propose the EdgeMoE to enhance memory efficiency for inference with MoE-based LLMs on edge devices. The non-expert weights, which occupy less storage but require more computation, are kept in memory all the time. In contrast, the large-size expert weights, which consume less powerful computing resources, are saved on disks. Only the required expert weights for specific tasks are activated and transferred between disks and memory as needed. To further reduce memory consumption, EdgeMoE quantities the expert weights into different bitwidths according to accuracy degradation thresholds specified by users in downstream tasks. Compared with the method that loads all model weights into running memory, EdgeMoE can save between 2.6 to 3.2 times the running memory. 

\textbf{Contextual sparsity prediction:}
Contextual sparsity prediction involves predicting the small and input-dependent sets of attention heads and multi-layer perceptron (MLP) parameters needed for inference computation \cite{liu2023deja,song2023powerinfer}. For example, the authors in \cite{liu2023deja} propose the Deja Vu inference system, where the lightweight predictors are inserted after MLP and attention blocks. Given the input to the current block, the contextual sparsity of the next block is predicted. Using the predicted sparsity, only a small set of MLP parameters or attention heads in the next block are activated and loaded into running memory by edge devices for inference computation. This method reduces computing overhead and inference latency while maintaining approximately the same inference accuracy. For example, with the Deja Vu inference system, the average accuracy across tasks of the OPT-175B does not drop until 75\% sparsity. Additionally, compared with the FasterTransformer, the Deja Vu inference system can reduce the inference latency of the OPT-175B by about half at around 75\% sparsity. To further reduce inference latency, the frequently activated parameters across various tasks in LLMs can be predicted in advance using general datasets. Edge devices can preload these parameters into running memory first and keep the other parameters in the storage space, which can be loaded into memory as needed during the online inference stage. 

\textbf{Parallel decoding:}
Parallel decoding enables an LLM to generate multiple subsequent tokens in a single forward pass without relying on a lightweight language model used in speculative decoding \cite{yuan2024llm,xiao2024clover}. For example, the authors in \cite{cai2024medusa} propose the Medusa and introduce extra decoding heads on top of the last hidden states of an LLM. During inference, each extra decoding head can predict multiple subsequent tokens in parallel for its designated position. These predictions are assembled into candidates and then processed in parallel with the tree-based attention mechanism. After verifying all candidates, a reasonable candidate is accepted for the next decoding phase. Additionally, in \cite{fu2024break}, the authors propose the Lookahead Decoding, where the autoregressive decoding is formulated as a non-linear system and solved by the fixed-point Jacobi iteration method. In each inference step, the LLM generates several disjoint n-grams in parallel and verifies the promising n-grams (subsequent tokens) in parallel from the n-gram pool, which caches the historically generated n-grams. The promising n-gram starts with a token that exactly matches the last token of the current ongoing sequence. These two methods fully exploit the computing resources that the autoregressive decoding would otherwise leave idle. Compared with the autoregressive decoding, Medusa achieves 41 times the operational intensity in attention matrix multiplication with a batch size of 16 and Llama 33B, indicating better utilization of computing resources of edge devices and reduced inference latency. 

\textbf{Sparse attention:}
Sparse attention is designed to reduce the computing overhead and memory usage associated with the attention mechanism in Transformers~\cite{deng2024attention}. Traditionally, the attention mechanism computes the relationships between all query elements and key sequences, resulting in substantial computing delay. However, it is found that ignoring less important interactions between queries and keys can save computing resources without degrading inference performance significantly~\cite{zeng2024cap,ribar2023sparq,Song_2024_CVPR}. In \cite{lee2024hip}, the authors propose a hierarchically pruned attention, demonstrating that appending only the top-k key tokens to each query can accelerate the inference process and reduce memory usage with minimal or no performance degradation. Furthermore, the authors in \cite{deng2024attention} present a comprehensive theoretical analysis of sparsity in attention mechanisms, offering valuable insights into potential trade-offs between computational efficiency and model performance. Furthermore, sparse attention is also advantageous in LLM fine-tuning. For example, in \cite{gui2023spt}, Gui et al. introduce a sparse MHA module for LLM fine-tuning. By substituting the original MHA module in LLMs with the sparse MHA module, the output becomes a sparse matrix, effectively reducing peak memory consumption and accelerating fine-tuning processes.

\textbf{KV cache optimization:} During the autoregressive decoding phase, LLMs need to store the KV cache of previous tokens to generate future tokens. This process results in a continuously increasing size of the KV cache as more tokens are generated, leading to higher memory consumption and increased inference latency. To enable efficient on-device LLM inference on edge devices, it is crucial to reduce the KV cache size while maintaining inference performance \cite{wan2024efficient,xu2024survey}. 
The first approach to reducing KV cache size is through KV cache compression. For example, the authors in \cite{liu2024kivi} develop a 2-bit quantization algorithm to quantize the key cache per channel and the value cache per token, respectively. This algorithm can achieve nearly identical inference accuracy while reducing peak memory usage by 2.6 times. Additionally, to preserve the layer-specific information during compression, the authors in \cite{liu2024minicache} introduce MiniCache. It merges the key and value caches with high similarity at the same position across consecutive layers into a single cache, respectively, while those with substantial semantic meanings remain unmerged. With 4-bit quantization, this method can reduce memory usage by 41\% while ensuring near-lossless model performance. 
Another approach involves KV cache eviction, which employs an eviction policy to dynamically select KV cache \cite{zhang2023ho,xiao2024efficient}. For example, in \cite{NEURIPS2023_a452a7c6}, the authors propose Scissorhands, which maintains KV cache memory usage within a fixed budget. The proposed system first identifies pivotal tokens with higher attention scores, which significantly influence future generations. Then, non-influential tokens are dropped from the cache when the buffer is full. This method can reduce inference memory usage of the KV cache by 5 times without degrading performance. Similarly, the authors in \cite{zhang2024pyramidkv} dynamically allocate different KV cache budgets across various layers and strategically select important KV vectors to cache. The proposed method can maintain model performance with only 12\% of the KV cache compared to a full KV cache.

\textbf{Real-world implementations:} The above efficient LLM inference techniques have already been deployed in many commercial products for on-device LLM applications nowadays. For example, Google DeepMind has reported that their most efficient LLM, Gemini Nano, designed for on-device deployment, is trained by distilling knowledge from larger Gemini models to enable efficient inference while preserving inference accuracy~\cite{team2023gemini}. Besides, Gemini Nano is quantized with 4-bit for on-device deployment. Moreover, Apple's newly released Apple Intelligence, integrated into the iPhone, iPad, and Mac, brings powerful generative AI to users. To optimize model performance on edge devices, Apple Intelligence employs a range of technologies, including speculative decoding, to accelerate the inference process \cite{apple_intelligence}.

\subsection{Resource-efficient Fine-tuning}
Compared to inference, on-device LLM training demands significantly higher memory and computing resources. For example, computing the gradients of the LLM OPT-13B \cite{zhang2022opt} consumes 12 times the memory needed for inference~\cite{malladi2024fine}. However, since LLM fine-tuning requires much less computing resources than full-parameter training, LLM fine-tuning is widely adopted in on-device LLM deployment. In this subsection, we will explore techniques that can effectively fine-tune LLM parameters under limited resources. 

\subsubsection{Parameter-efficient fine-tuning} Parameter-efficient fine-tuning (PEFT) \cite{han2024parameter} has emerged as a prominent solution for LLM fine-tuning by updating only a small number of parameters in the fine-tuning process. Popular PEFT techniques can be categorized into three main types, i.e., additive PEFT, selective PEFT, and reparameterized PEFT, which are elaborated next.

\begin{figure*}[t]
\centering
\includegraphics[width=0.85\textwidth]{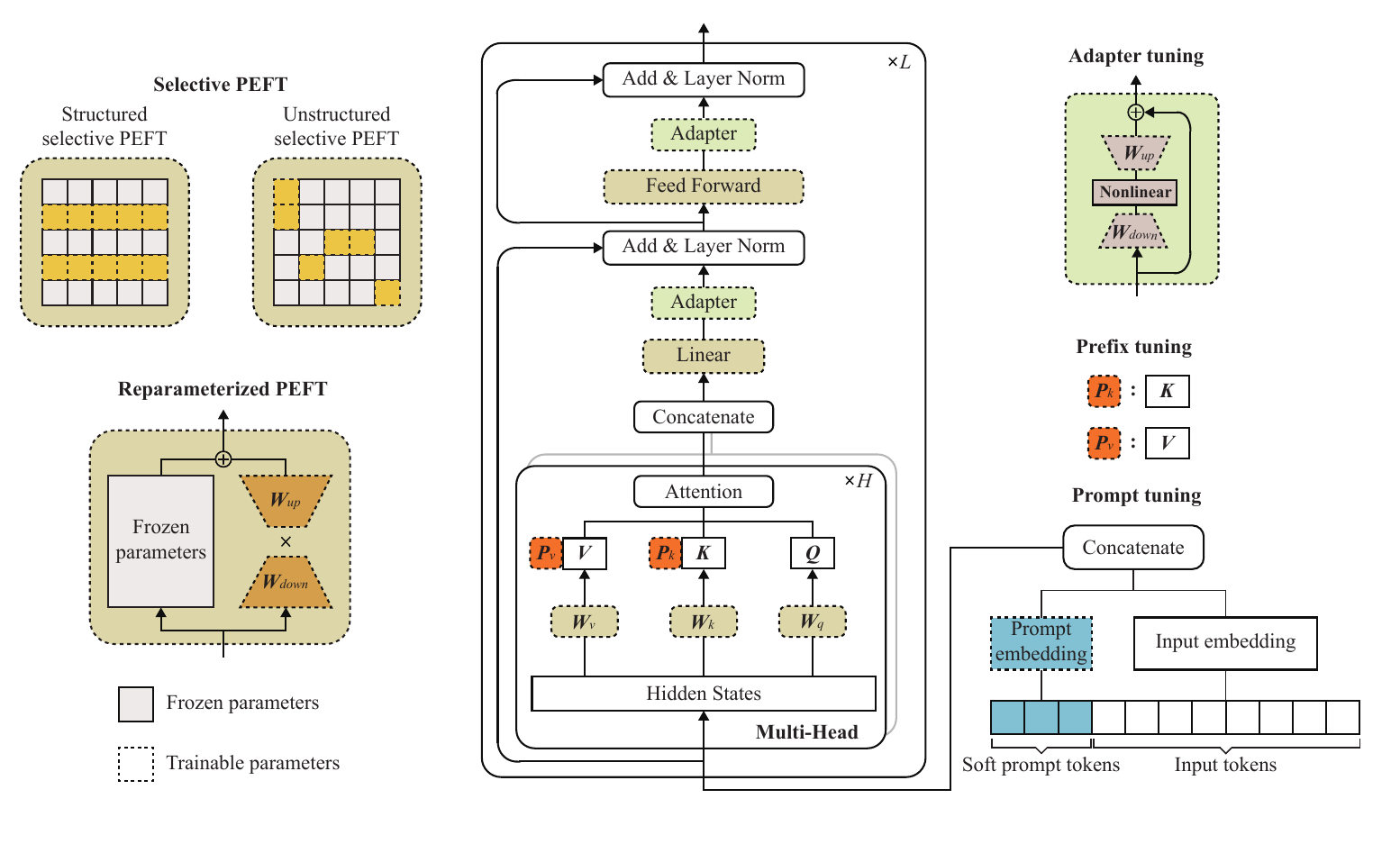}
\vspace{-0.25cm}
\caption{PEFT methods for resource-efficient fine-tuning.}
\label{fig:additive PEFT}
\end{figure*}

\textbf{Additive PEFT:}
To mitigate the computing burden of fine-tuning, the additive PEFT introduces trainable components with minimal parameters into LLMs while maintaining the pre-trained LLM parameters frozen. Additive PEFT can be further categorized into three types based on the locality of the introduced components, i.e., adapter tuning, prompt tuning, and prefix tuning, as shown in Fig.~\ref{fig:additive PEFT}. 1) As first introduced by \cite{houlsby2019parameter}, \textit{adapter tuning} inserts adapter modules into the Transformer layers and freezes the other parameters in the Transformer. In this approach, only the adapters are updated during fine-tuning. 2) \textit{Prompt tuning} adds soft prompt tokens at the beginning of the input tokens for fine-tuning \cite{lester2021power}. This method leverages the characteristic of LLMs that the encoding and token generation are based on the preceding tokens. 3) \textit{Prefix tuning} adds trainable prefix parameters to the keys and values of the MHA in each Transformer layer \cite{li2021prefix}. Although this approach increases the number of trainable parameters compared with prompt tuning, it allows for direct modification of representations within LLMs, enabling LLMs to respond more precisely to specific tasks.

\textbf{Selective PEFT:}
Though additive PEFT enables fine-tuning LLMs with fewer trainable parameters, it introduces additional inference latency by adding more parameters \cite{han2024parameter}. To address this problem, selective PEFT preserves the model architecture by freezing most parameters and updating only a smaller subset of parameters. Selective PEFT can be broadly classified into unstructured selective PEFT and structured selective PEFT, which are shown in Fig.~\ref{fig:additive PEFT}. 
1) \textit{Unstructured selective PEFT} determines the selection of trainable parameters individually, which can enhance the performance of fine-tuned models \cite{han2024parameter}. For example, the authors in \cite{fu2023effectiveness} reformulate the trainable parameter selection in PEFT as an optimization problem with a second-order approximation of the loss function and provide a second-order approximation method to solve this problem. By choosing the sparse trainable parameters, the sparse fine-tuned models outperform the fully fine-tuned models. However, the sparsity of the trainable parameters may not be well supported by existing deep learning frameworks and hardware, introducing additional workload for on-device LLM deployment during fine-tuning \cite{zhang2023increlora}. 
2) \textit{Structured selective PEFT} selects regular combinations of parameters, such as specific model layers, to enhance hardware computational efficiency in on-device LLM deployment \cite{zhang2023increlora}. 
For example, the authors in \cite{guo2020parameter} propose Structured Diff Pruning, which partitions parameters into groups based on matrix/bias vectors and strategically removes some groups. Then, only the parameters in the remaining groups are updated, thereby saving computing resources during LLM fine-tuning.

\textbf{Reparameterized PEFT:} Reparameterized PEFT techniques leverage low-rank matrices to reduce the number of trainable parameters during model fine-tuning. One of the most well-known methods in reparameterized PEFT for LLMs is LoRA \cite{hu2021lora}. For a pre-trained weight matrix in an LLM, LoRA introduces two additional trainable matrices with ranks much smaller than that of the pre-trained weight matrix, as illustrated in Fig.~\ref{fig:additive PEFT}. In the fine-tuning process, the pre-trained weight matrix is frozen, and only the newly introduced two low-rank matrices are trainable. This approach allows for efficient model fine-tuning because updating the low-rank matrices requires significantly less computational capability than updating the pre-trained weight matrix. Besides, LoRA does not add any additional inference latency since LoRA's fine-tuned weights are merged into the original weights of LLMs during inference \cite{han2024parameter}. Due to its salient advantages, LoRA has inspired numerous subsequent research works. For example, Quantized LoRA (QLoRA) \cite{dettmers2023qlora} aims to minimize memory consumption by combining quantization techniques with LoRA, enabling the fine-tuning of a language model with 65B parameters with a 48 GB GPU within 24 hours. The fine-tuned LLM achieves 99.3\% of the ChatGPT's performance on the evaluated tasks, demonstrating the effectiveness of QLoRA.

\subsubsection{Zeroth-order optimization}
Zeroth-order optimization \cite{baydin2022gradients} is a novel technique for model training, which estimates gradient updates only through the forward pass. This method considerably reduces the computational burden since the forward pass, equivalent to inference, demands far fewer computing resources than the backpropagation during training. Specifically, compared with prevalent first-order optimizers, such as Adam, zeroth-order optimizers do not need to store the intermediate results of the backpropagation during the training process, significantly reducing the memory consumption of LLM training. 
For example, the authors in \cite{malladi2024fine} propose the zeroth-order optimizer, MeZO, which adopts simultaneous perturbation stochastic approximation to estimate model gradients only through the forward propagation and uses the estimated gradients to update model parameters. 
Compared with the fine-tuning with Adam, the model fine-tuned with MeZO demonstrates competitive performance on 7 out of 11 tasks, using only 1/12 of the running memory while only causing less than a 1\% accuracy degradation. 
Moreover, to further enhance the efficiency of LLM fine-tuning, the zeroth-order optimization technique can be combined with PEFT methods, such as LoRA and prefix fine-tuning \cite{malladi2024fine}.

\subsection{Lessons Learned}
Resource-efficient LLM techniques enable efficient LLM training and inference on resource-limited edge devices and edge servers. Resource-efficient inference techniques focus on two main areas: model compression reduces computational demands by shrinking model sizes, whereas fast decoding improves inference speed and efficiency by optimizing data processing, reducing parameter use, and minimizing intermediate data generation. Resource-efficient fine-tuning techniques also have two primary focuses: they either shrink the number of parameters for updating or execute the forward pass only.

The overarching goal of these resource-efficient techniques is to minimize the computing workload and memory footprint, which is essential for deploying LLMs on devices with limited resources. However, there are still limitations to these approaches, as they tend to focus on single devices and do not fully exploit networked computing resources, thus placing less emphasis on communications between devices. By connecting intelligent devices, more effective model inference and training can be achieved through collaboration among edge devices and edge servers, which is the primary focus of MEI.

\section{An Overview of MEI4LLM\label{sec:architecture}}
\begin{figure*}[th]
\centering
\includegraphics[width=15 cm]{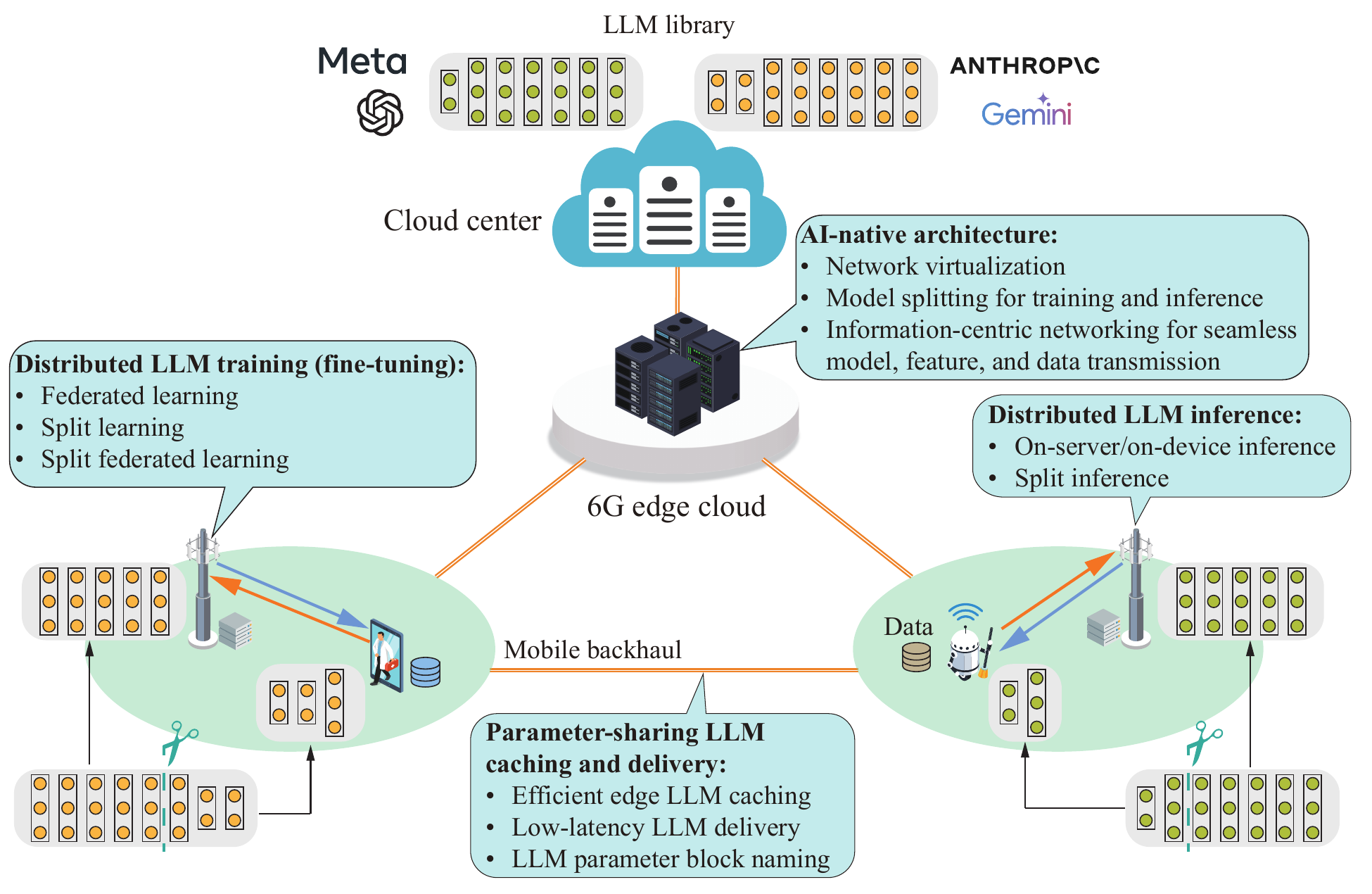}
\vspace{-0.2cm}
\caption{The MEI architecture for LLMs in 6G with four basic components, i.e., AI-native architecture, parameter-sharing LLM caching and delivery, distributed LLM training (fine-tuning), and distributed LLM inference.}
\label{fig_architecture}
\end{figure*}

As alluded to in Section \ref{sec:on-device}, while there are great efforts in pushing LLMs to the network edge, the aforementioned schemes focus on implementations on a single device, limiting its applicability and efficiency. By deploying LLMs in MEI, edge devices can collaborate with edge servers for collaborative learning and inference, which largely mitigates the resource scarcity of edge devices. In line with the ``NET4AI'' (network for AI) vision for the 6G era~\cite{huawei20216g}, this section presents an overview of the MEI framework that supports the deployment of LLMs, called MEI4LLM, as shown in Fig. \ref{fig_architecture}. The primary objective of MEI4LLM is to enable efficient collaboration among networked edge devices/servers with enhanced communication, computing, and storage efficiency. Clearly, this framework should also seamlessly integrate with security defense mechanisms to combat various attacks, such as data poisoning and model inversion attacks during training, evasion attacks during inference~\cite{wu2024wafbooster,li2024privacy}, and unauthorized access during model and data storage, to safeguard the security of the framework. However, since the security aspect is not our main focus, we will briefly discuss it in Section \ref{sec:further} for future research.


This section will begin by discussing the AI-native architecture specifically designed for MEI4LLM. Then, we will introduce the core components -- covering parameter-sharing LLM caching and delivery, distributed LLM training (fine-tuning), and distributed LLM inference -- that enable MEI4LLM to efficiently support the deployment and operations of LLMs within mobile edge networks. Further technical discussions of these core components will be elaborated in Sections \ref{sec:caching}, \ref{sec:learning}, and \ref{sec:inference}, respectively.

\subsection{The AI-native Architecture} 
6G is evolving into an AI-native architecture. To effectively support pervasive LLM applications, mobile networks are expected to have the following transformative aspects: 1) Adopting the ``task-oriented'' design principle. Instead of maximizing throughput or minimizing latency, the design goal can be to optimize the quality of the output, say, the cross entropy of output tokens from LLMs. 2) The mobile networks must support native model partitioning across edge servers and devices to facilitate real-time LLM inference/training. 3) information-centric networking protocols can be supported to enable fast LLM delivery over the edge networks with reduced latency.

Network virtualization is of paramount importance in achieving task-oriented design. Following the design principle of software-defined networking, the MEI4LLM needs central controllers that orchestrate network-wide computing resources and data transmissions, with the decoupled control and data plane. Moreover, since the design goal and optimization principles can vary across different LLM tasks, the protocols on the control plane must be fully programmable for specific LLM services. By collecting global network knowledge, such as the accuracy of LLMs, various quantization levels, user requirements on LLM services, channel conditions, battery status of users, and computing resource availability, the controller executes the customized algorithm to coordinate model training/inference and delivery across the distributed edge computing systems, with intermediate smashed data (i.e., intermediate activations and back-propagated gradients), model parameters, or user data exchanged across edge routers and servers. 

Further edge networks will evolve into the ``neural edge''~\cite{huawei20216g}, where neural network layers are distributed across edge nodes for collaborative computing. Analogous to cloud data centers with many GPUs to support large-scale LLMs, MET4LLM must feature flexible and model splitting for training and inference across distributed edge devices and servers. The optimal model splitting, placement, and data routing for large-scale models should be concertedly supported over edge networks. Moreover, both air interfaces and network designs should have native support for federated learning, split learning, and split inference for AI models, including LLMs. Since model training and inference are robust to packet errors, task-oriented wireless transmissions, say, for smashed data at the cut layer, can be carried out with appropriate error control to achieve the best efficiency-reliability trade-off. 

Lastly, since LLMs can be communication-intensive, information-centric networking can be implemented to ensure seamless model, feature, and data transmission across edge networks for efficient delivery of LLMs. In this respect, MEI4LLM should support LLM parameter block naming and name-based transmission protocol. By assigning names for popular LLM parameter blocks, the central controller in the MEI4LLM architecture can forward the parameter request to edge caches located nearby and support parameter block multicasting for users, thereby reducing latency and bandwidth consumption for delivering large-scale models across networks and to end users.


\subsection{Parameter-sharing LLM Caching and Delivery} Given the limited storage capacities of edge devices and real-time model fine-tuning in changing environments, LLMs must be delivered over mobile networks frequently to support subsequent usage. Since parameter blocks can be shared among various downstream LLMs~\cite{qu2024trimcachingICDCS,hu2021lora} and even reused within the same LLM \cite{Lan2020ALBERT}, LLM caching and delivery schemes must take advantage of this fact to reduce caching and delivery costs of shared LLM parameter blocks. Besides, to enable real-time model delivery, MEI4LLM can construct a lookup table, assigning names for LLM parameter blocks to facilitate content search and management, following the principle of information-centric networking \cite{6563278,8057300}. By doing so, the MEI4LLM paradigm places LLMs at appropriate sites, retrieves LLMs of interest from nearby edge servers, and enables routing/multicasting of LLM parameter blocks to mobile users. Details will be further discussed in Section \ref{sec:caching}.

\subsection{Distributed LLM Training (Fine-tuning)} It is foreseen that 6G MEI systems can efficiently fine-tune LLMs to adapt to local environments. Edge LLM fine-tuning can be triggered whenever the inference accuracy decreases or after a certain period when local environments have changed. For example, LLM-empowered virtual assistants should be fine-tuned regularly to better adapt to new trends in news media, top local restaurants, and popular attractions, resulting in improved decision-making and interactions with users \cite{10654534, ghimire2024opportunities}. LLM-empowered mobile health applications should be personalized to provide better predictions and health or fitness suggestions \cite{cosentino2024towards}.

In next-generation mobile networks, edge training for LLMs must answer two questions: 1) how to preserve user privacy and data ownership, and 2) how to support large-scale model training through the collaboration of edge nodes. To enhance the data privacy of users, federated learning (FL) and split learning (SL) serve as two promising distributed learning frameworks to implement at the network edge \cite{9923620}. Specifically, FL allows edge devices to train models locally while only sharing model parameters with an edge server for aggregation, thus harnessing collective intelligence without sharing personal data \cite{8737464}. Alternatively, SL and its variant split federated learning (SFL), can be implemented to enable device-server co-training without sharing local raw data, which is particularly suitable for large-scale LLM fine-tuning for edge devices~\cite{lin2023split}, as model splitting allows for workload balancing across diverse edge nodes. To effectively support the intensive training, various resource-efficient training techniques, as detailed in Section \ref{sec:on-device}, can be combined with FL or SL. These discussions will be provided in Section \ref{sec:learning}.

In addition, considering RAG \cite{fan2024survey,chen2024benchmarking}, external knowledge sources should also be cached at the network edge, ensuring timely acquisition of the data/knowledge for an LLM. However, RAG involves extra storage costs and delivery latency for retrieving external knowledge. In comparison, fine-tuning an LLM consumes extra computing power while eliminating the need for external databases. In this regard, an important research problem is determining data selection for either LLM training (fine-tuning) or RAG in MEI systems.



\subsection{Distributed LLM Inference} To accommodate for resource-intensive LLMs,
edge servers and edge devices must perform distributed inference for LLMs in a concerted manner, depending on the communication-computing workload and privacy requirements. There are different ways for edge inference. On-server inference requires users to upload raw data to the server. This approach eliminates the computing burden on edge devices while potentially violating users' privacy requirements. For instance, multimodal LLMs may collect sensitive audio and video data in home environments, where users are often reluctant to share. Conversely, on-device inference preserves privacy and eliminates communication costs while imposing an intensive computing workload on edge devices. Split inference sits in between, with edge devices and servers holding parts of the AI models, which has been mentioned in 3GPP 5G technical specifications. Split inference involves uploading features from edge devices to edge servers for co-inference. To facilitate LLM inference, MEI4LLM can tailor consumer services by appropriately selecting from these schemes based on communication-computing resource status and privacy requirements, as elaborated in Section \ref{sec:inference}.

\subsection{Lessons Learned for MEI4LLM}
Clearly, MEI4LLM is nothing but a special case of MEI. However, the need to train and deploy LLMs at the edge serves as a key stimulus for the evolution of MEI towards its next stage. Specifically, the extreme resource demands of LLMs necessitate a fundamental rethinking of traditional MEI. For this reason, MEI4LLM must feature the following characteristics: \textit{1) the native support of model splitting and parallel training/inference across interconnected edge nodes to facilitate the deployment of large-scale models 2) the integrated design of wireless communications and resource-efficient LLM training/inference techniques such as parameter-efficient fine-tuning and token (representation) reduction to make LLM deployment cost-effective.} The remainder of this survey paper will discuss these two critical aspects in detail.

\section{Edge Caching and Delivery for LLMs\label{sec:caching}}
Edge LLM caching and delivery play indispensable roles in both the training and inference of LLMs, serving as cornerstones of edge LLM deployment. For this reason, we begin with discussions on edge caching and delivery and then present edge training and inference in the subsequent sections. Compared with conventional edge service/content caching and delivery, \textit{the main distinction of edge LLM caching and delivery is the exploitation of parameter shareability}, which is commonly observed in LLMs, aiming to increase storage and communication efficiency over edge networks. While parameter shareability can exist in traditional DNNs, it is much more prevalent and significant in LLMs due to the widespread adoption of PEFT techniques, requiring our special design attention. In what follows, this section presents the techniques that leverage this characteristic of LLMs for optimizing edge caching and delivery. The related works for efficient edge LLM caching and delivery are outlined in Table \ref{table_caching} for readers' convenience.
\begin{table*}[!t]
\centering
\caption{Summary of related works for edge LLM caching and delivery.}
\label{table_caching}
\renewcommand{\arraystretch}{1.4}
\setlength{\tabcolsep}{2mm}
\begin{tabular}{|c|c|c|m{0.64\textwidth}|}
\hline
\textbf{Scenarios} & \textbf{Techniques} & \textbf{Ref.} & \makecell[c]{\textbf{Objectives}} \\ 
\hline
\makecell[c]{Edge LLM\\caching}
               & \makecell[c]{Parameter-sharing\\LLM caching} 
                   & \cite{qu2024trimcaching} & Proposes TrimCaching framework for LLMs, where shared parameter blocks of LLMs only need to cached once in an edge server for storage efficiency.  \\ \hline
\multirow{2.5}{*}{\makecell[c]{Edge LLM\\delivery}}
               & \multirow{2.5}{*}{\makecell[c]{Parameter-sharing\\wireless model\\downloading}}
                   & \cite{wu2023efficient} & Proposes a model multicasting and assembling framework, where shared parameter blocks of models requested by users are multicast to users, and the specific parameter blocks are unicast to each user separately.   \\ \cline{3-4}
                   & & \cite{10183793} & Compresses model weights in low bitwidth for fast model downloading.   \\ \hline
\end{tabular}
\end{table*}

\subsection{Edge LLM Caching}
Edge model caching can realize low model downloading latency by distributing AI models to wireless edge servers in advance. Distinct from service placement for computation offloading, edge model caching focuses on caching AI models for downloading from edge servers to end users~\cite{qu2024trimcachingICDCS,9522156}. This paradigm enables users to fetch AI models from edge servers directly instead of accessing remote cloud data centers, which incurs excessive downloading latency~\cite{10183793,qu2024trimcaching}. However, implementing edge LLM caching presents several challenges: 1) \textit{Limited storage capacity for LLM caching}: Service providers aim to place as many popular LLMs as possible on edge servers to enhance the cache hit ratio and decrease model downloading latency for users. Nevertheless, the immense size of LLMs presents a significant challenge for their storage on edge servers; 2) \textit{High LLM edge cache (re)placement costs}: Over time, previously cached LLMs may no longer align with the changing user requests. To address this, service providers may replace LLMs on edge servers to better accommodate up-to-date requests, imposing a substantial burden on mobile backhaul networks. In what follows, parameter-sharing model caching is presented to address the above challenges.

\subsubsection{Parameter-sharing LLM caching} 
\begin{figure}[!t]
\centerline{\includegraphics[width=0.35\textwidth]{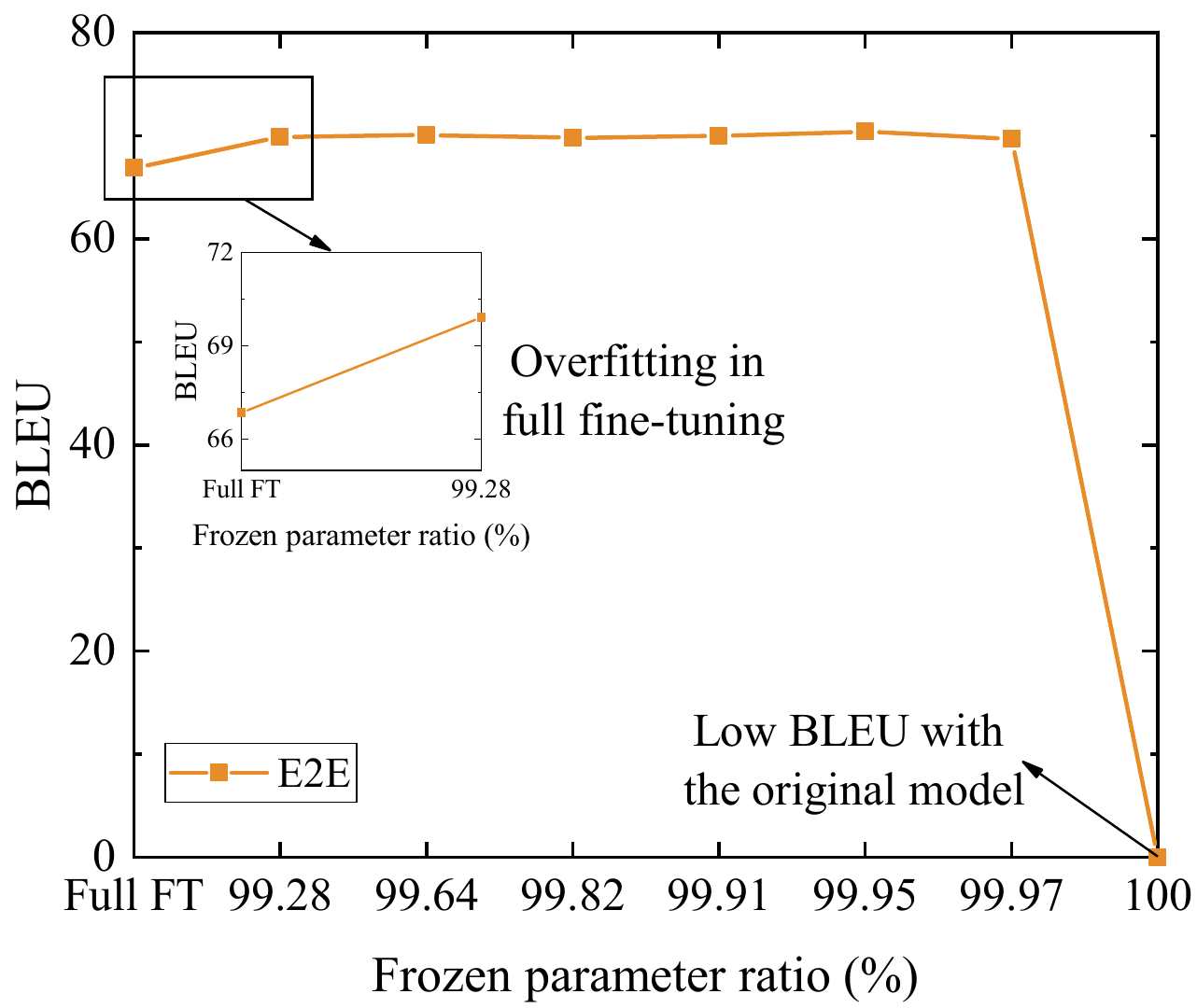}}
	\caption{BLEU v.s the frozen parameter ratio of the fine-tuned GPT-2 large. In this example, LoRA enables the integration of fine-tuned parameters ranging from hundreds of kilobytes to tens of megabytes into a pre-trained GPT-2 large model, which has a size of around 3.02 GB. This process creates multiple GPT-2 large models tailored for specific downstream tasks. The frozen parameter ratio refers to the ratio of the pre-trained GPT-2 large model size to the fine-tuned GPT-2 large model size. BLEU is used to evaluate machine-translated texts against high-quality reference translations. ``Full FT" in the figure indicates the full parameter fine-tuning.}
	\label{fig_gpt2}
\end{figure}

\begin{figure}[!t]
\centerline{\includegraphics[width=0.45\textwidth]{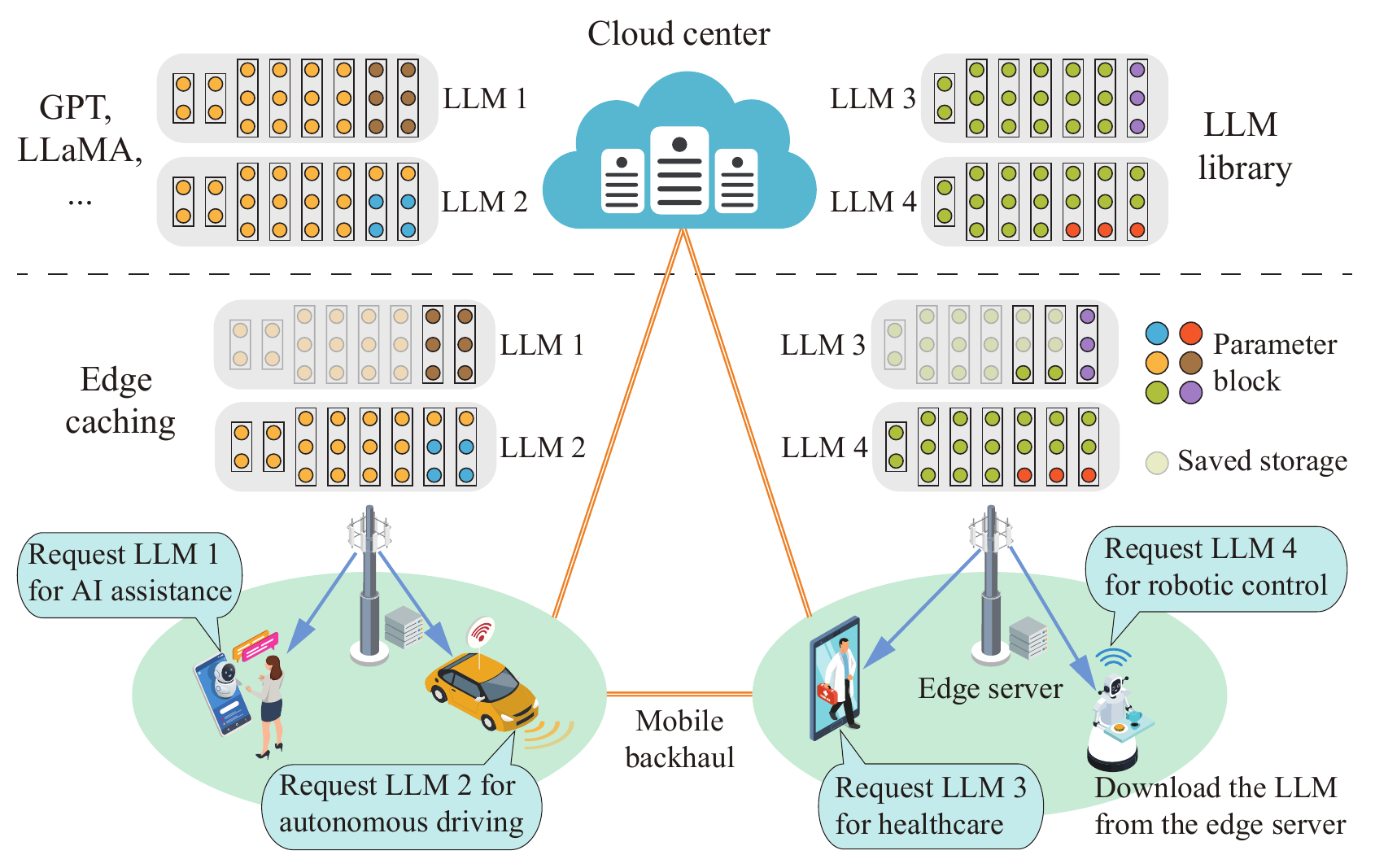}}
	\caption{The TrimCaching mechanism for caching LLMs in wireless edge networks. Popular LLMs are placed on edge servers, where users can download the requested LLMs from the edge network. To enhance storage efficiency, shared parameters across LLMs are cached only once on an edge server.}
	\label{fig_trim_caching}
\end{figure}

\begin{figure}[th]
\centerline{\includegraphics[width=0.35\textwidth]{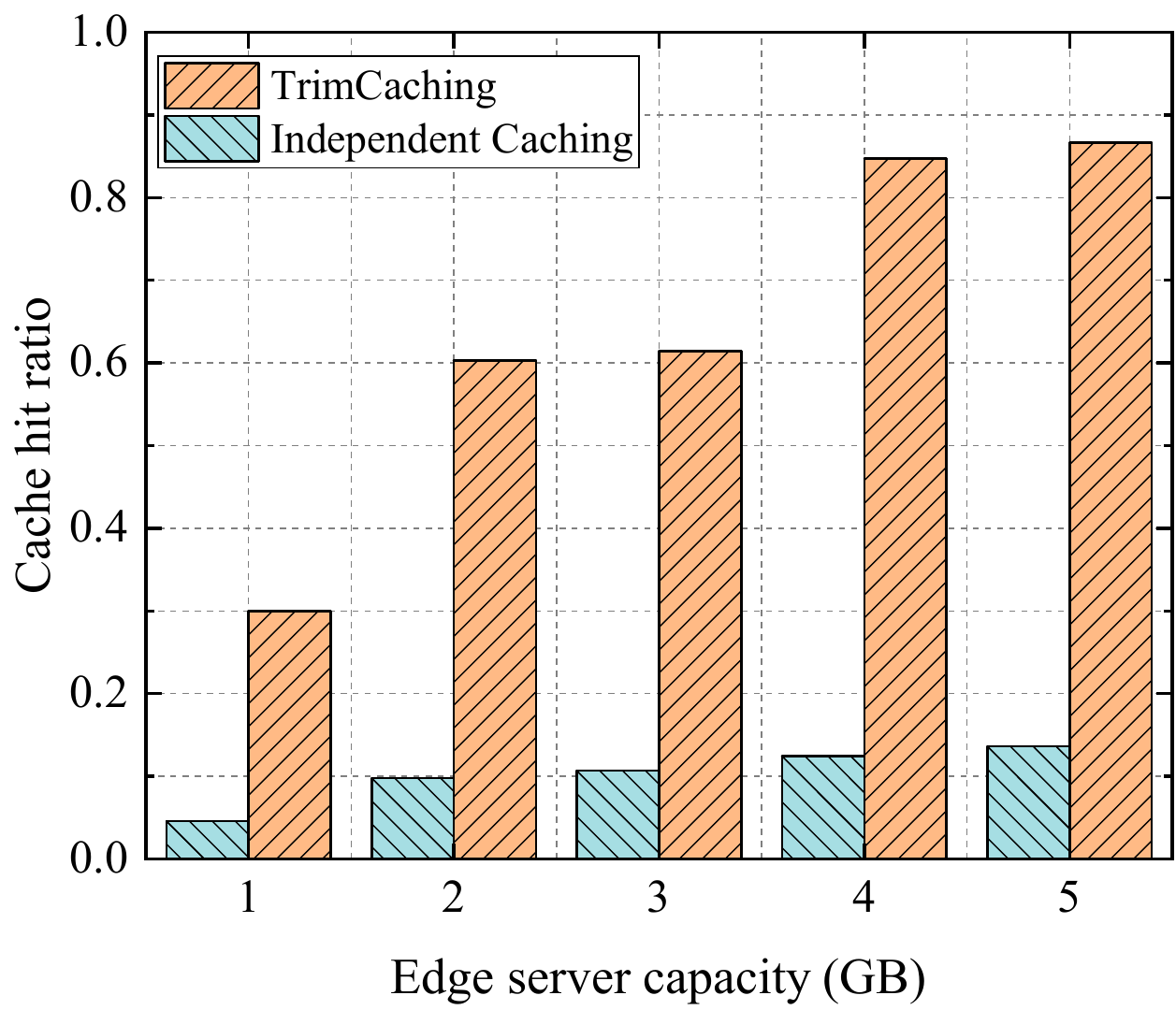}}
	\caption{Cache hit ratio v.s edge server capacity. A cache hit occurs if the requested GPT-2 model can be served by any edge server in the edge networks within the E2E latency requirements of end users. The GPT-2 models, including GPT-2 small, medium, and large, are fine-tuned with LoRA for parameter sharing in 100 downstream tasks. 40 users and 12 edge servers are located in a square area of 1 $\text{km}^{\text{2}}$. The coverage radius of edge servers is 275 m. The other parameter settings can be referred to \cite{qu2024trimcaching}.} 

	\label{fig_trim_gpt2}
\end{figure}

Parameter-sharing model caching can be adopted to improve storage and transmission efficiency at the network edge. As discussed in Section \ref{sec:on-device},  PEFT, such as LoRA, is widely adopted for adapting LLMs to downstream tasks. In LoRA, the pre-trained LLM parameters are frozen, and only the newly introduced parameters are trainable, typically accounting for less than 1\% of the original LLM parameters. Therefore, most parameters of various LLMs fine-tuned with LoRA for downstream tasks are shared from pre-trained LLMs, which should be leveraged to enhance caching efficiency significantly. Take LoRA and GPT-2 as an example. Fig. \ref{fig_gpt2} demonstrates that the inference performance almost remains unchanged even when 99.97\% of parameters in the GPT-2 large model are frozen parameters from the pre-trained GPT-2 large. Based on this observation, in \cite{qu2024trimcaching}, we propose an AI model placement strategy, called TrimCaching, to maximize the cache hit ratio under server storage capacity and service latency constraints by exploiting the parameter-sharing properties of AI models, particularly LLMs. In the TrimCaching framework, only one copy of the shared parameter blocks across LLMs is cached on one edge server, thereby improving storage efficiency, as illustrated in Fig. \ref{fig_trim_caching}. Compared with the Independent Caching strategies for edge LLM caching \cite{10436771,xu2024cached}, which do not consider parameter sharing across LLMs, the TrimCaching strategy can significantly improve the cache hit ratio as shown in Fig. \ref{fig_trim_gpt2}. In other words, even with storage-limited edge servers, a large number of popular LLMs can still be cached on network edge servers under the TrimCaching framework, thereby considerably reducing LLM downloading latency compared with fetching them from the cloud.

While the parameter-sharing model caching under multi-cell scenarios has been investigated in \cite{qu2024trimcaching}, this paradigm can be extended to consider many different scenarios in cellular networks, such as centralized-RAN (C-RAN) and heterogeneous networks (HetNets). Moreover, mobility-aware edge caching can also be developed by exploiting the knowledge of user mobility patterns. For instance, in \cite{poularakis2013exploiting}, a distributed approximation algorithm based on large deviation inequalities is developed for content placement based on the assumption that users move randomly based on a discrete-time Markov chain model. Similar algorithms can be developed to address the parameter-sharing model caching problem with high user mobility.

\subsubsection{Edge LLM cache replacement}
Since the popularity of models can evolve over time, another fundamental research problem in edge caching is LLM replacement. By replacing outdated content with new data, edge servers can continuously refresh their caches with new content to satisfy ever-changing user requests~\cite{8964499}. The two most classic replacement strategies are the recency-based and the frequency-based strategies, which remove the least recently used (LRU) objects and least frequently used (LFU) objects and then replace them with updated content. However, such naive strategies fail to exploit the cooperative caching among edge nodes and parameter shareability among LLMs. Two directions can be further explored to improve the performance. First, shared parameter blocks among LLMs and cooperation among neighboring edge servers can be considered to achieve parameter replacement with minimal backhaul transmission costs. Second, model replacement should be optimized with inference accuracy in mind, as LLMs are often more effective in general tasks than specific ones. This suggests that users may request an LLM that closely matches their needs, i.e., achieving required prediction accuracy rather than a specific LLM, thereby leading to less frequent replacement.

To implement the replacement policies, one direction is centralized proactive caching, say the scheme in \cite{qu2024trimcachingICDCS}, after a certain period. However, under high user mobility, such methods may involve high system complexity and communication costs for transferring LLMs. The other direction is to develop distributed algorithms, often based on the Markov Decision Process or reinforcement learning, to make replacement decisions without knowing the full information of other edge nodes~\cite{gu2014distributed}. Both of the directions are worth exploring further.

\subsubsection{Edge caching for RAG}
As mentioned in Section \ref{generative}, integrating LLMs and RAG, which enables LLMs to retrieve relevant data from external knowledge sources, is essential for generating reliable and up-to-date responses without retraining/fine-tuning. However, since retrieving the information from the remote cloud can be time-consuming, the most popular external knowledge should be cached at the network edge to enable LLMs to fetch the most factual, accurate, and up-to-date content. Intriguingly, this edge caching problem naturally differs from traditional ones as the caching should be optimized by considering the training status or internal knowledge of LLMs. Specifically, if LLMs can already memorize or easily infer certain content, such knowledge can be removed from the external knowledge sources to save storage space at the network edge, thereby enhancing storage efficiency for caching external knowledge sources. Besides, edge servers can cache specific external knowledge sources frequently requested by their associated users, which can significantly increase the QoS for responses to users in different regions. On the other hand, edge caching for RAG is tightly coupled with on-the-fly LLM fine-tuning, i.e., we need to choose which data to fine-tune outdated LLMs and which data to cache for RAG on edge servers. With this in mind, it is worth investigating a joint LLM fine-tuning and knowledge source caching problem under the context of RAG to enhance the reliability of LLMs under latency constraints, taking into account the data retrieval latency from external knowledge sources at the edge/cloud servers and the fine-tuning costs by feeding new data to LLMs.





\subsection{Edge LLM Delivery}
\begin{figure*}[!th]
\centerline{\includegraphics[width=0.8\textwidth]{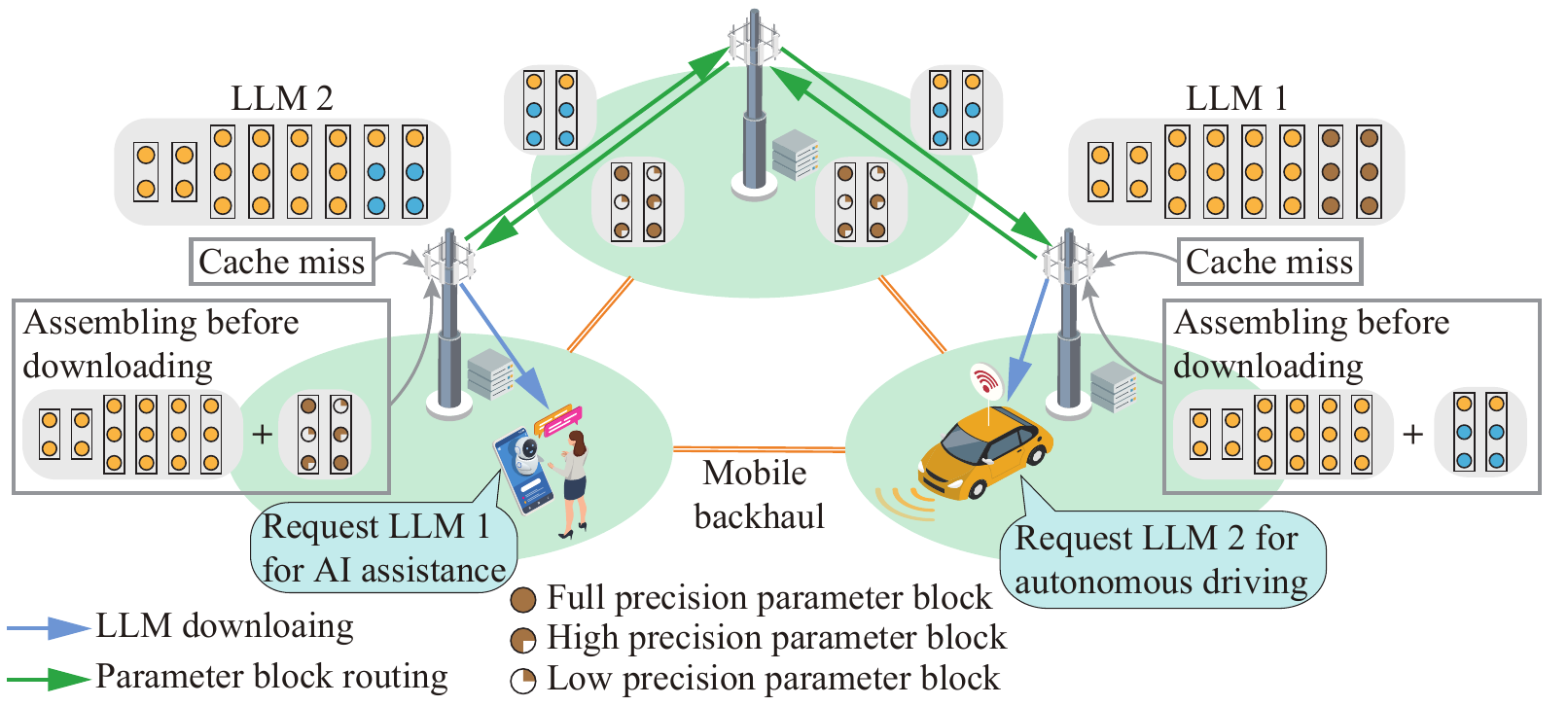}}
	\caption{The parameter-sharing backhaul/backbone LLM delivery framework. In this example, the AI assistant and autonomous driving applications request LLM 1 and LLM 2, respectively, which are not cached on the associated edge server. Therefore, LLM 1 and LLM 2 need to be delivered from another edge server. In this framework, to reduce communication overhead and latency, only the specific LLM parameter blocks need to be transmitted since the shared backbone/parameter blocks are already cached on the associated edge server. The entire LLM can be assembled before downloading, or on edge devices once all the needed parameter blocks have been received. To further reduce delivery latency, parameter-sharing LLM delivery can be combined with various compression techniques, where the compression ratio can be determined by jointly considering model performance and channel/backhaul conditions.}
	\label{model_delivery}
\end{figure*}

An essential step in fetching the models from where they are cached to end users is delay-efficient model delivery. This process encompasses both model routing within backhaul/backbone networks and model downloading via wireless access links, facing the following challenges: 1) \textit{Excessive backhaul/backbone delivery latency}: When the requesting LLMs are not cached on the associated edge server, the LLMs need to be routed within the edge network. However, compared with traditional AI models, LLMs have significantly larger model sizes. Therefore, LLM routing needs to be carried out among edge servers with moderate backhaul traffic; 2) \textit{Significant wireless downloading latency:} AI model downloading needs to be finished with low latency to fulfill the QoS requirements of end users. As envisioned by 3GPP, autonomous driving applications require AI model downloading within 1 second \cite{3gpp.22.874}. However, the large model size of LLMs hinders fast model downloading, making it extremely hard to meet the stringent service latency requirements. These call for the following solution approaches.



\subsubsection{Parameter-sharing backhaul/backbone model delivery} To reduce the model delivery costs within the backhaul/backbone networks, parameter-efficient model delivery can be developed by exploiting the parameter shareability across LLMs, as illustrated in Fig. \ref{model_delivery}. When an edge server does not cache an LLM but caches other LLMs with the shared parameter blocks, only the missing parameter blocks of the LLM need to be delivered, thereby reducing data delivery costs. For example, for LLMs fine-tuned with LoRA, only the specific LoRA parameters must be transmitted if the shared backbone has been cached on the edge server. Moreover, upon delivery of parameter blocks, the parameter blocks can be fetched from different serving edge servers. The entire LLM can be assembled as long as all the needed parameter blocks reach the destination (e.g., requesting server or user). For this reason, considering multi-hop backhaul/backbone communication networks, caching-aware data routing can be developed to exploit multicasting of overlapped parameter blocks to improve network throughput.

\begin{figure}[!t]
\centerline{\includegraphics[width=0.4\textwidth]{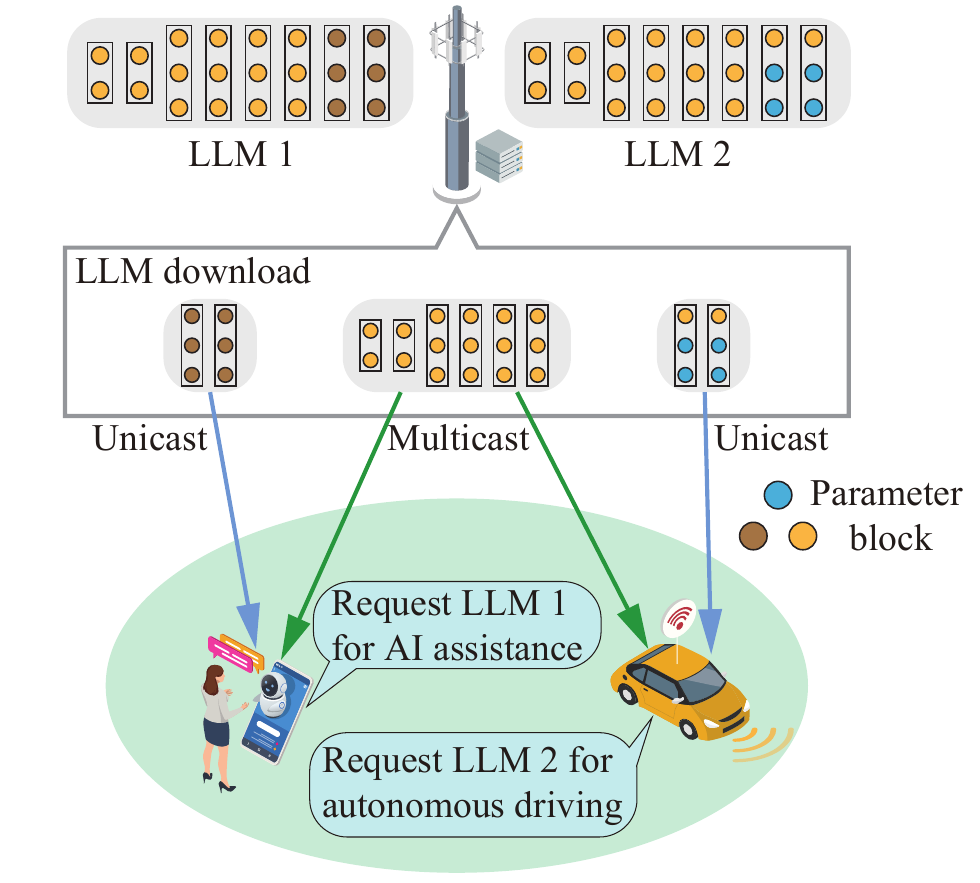}}
	\caption{Parameter-sharing wireless LLM downloading.}
	\label{fig_parameter_model_download}
\end{figure}

\subsubsection{Parameter-sharing wireless model downloading} To reduce the costs of wireless model downloading from base stations (edge servers) to users, it is essential to consider parameter-sharing wireless model downloading. As illustrated in Fig. \ref{fig_parameter_model_download}, to decrease downloading latency, the key idea is to multicast the reusable parameter blocks, thereby achieving timely downloading. In \cite{wu2023efficient}, the authors propose a model multicasting and assembling framework by exploiting the shareable parameters across AI models, i.e., Transformers as considered in the paper. This framework multicasts shared parameter blocks to multiple requesting users while unicasting specific parameter blocks to each user. A user then assembles the downloaded parameter blocks to obtain the desired LLM. There is a trade-off between parameter sharing and downloading latency. Although a model with more shared parameter blocks can be downloaded with lower latency, this will also degrade model performance in downstream tasks. To reduce the total model downloading latency and ensure the QoS requirements, the proposed framework aims to maximize the occurrence of the same parameter block across different models within the accuracy QoS requirement of downstream inference tasks.

Parameter-sharing LLM delivery/downloading can also be integrated with various LLM compression techniques. 
Take quantization as an example; by compressing LLM weights in low bitwidth, LLM downloading can be accomplished with low latency \cite{frantar2023gptq,10183793}. However, as illustrated in Section \ref{sec:on-deviceinference}, setting a uniform quantization bitwidth in LLMs degrades the inference performance significantly since not all weights contribute equally to the final outputs \cite{lin2023awq}. Therefore, weights in LLMs can be assigned with non-uniform bitwidths according to weight importance \cite{kim2024squeezellm}. For example, weights corresponding to larger activations or higher quantization errors can be quantized in high bitwidth before downloading, which can preserve the performance of the quantized LLMs~\cite{lin2023awq,dettmers2023spqr,kim2024squeezellm}. Moreover, quantization should also be optimized by considering the popularity/shareability of parameter blocks across various downstream LLMs. By integrating LLM compression methods, the design for LLM delivery/downloading by jointly considering wireless channel conditions, weight importance, parameter block shareability, and model performance is worth further investigation.

\subsubsection{Joint parameter-sharing model caching and delivery} Parameter-sharing model caching and delivery are tightly coupled over wired networks, wireless networks, or the hybrid of both. On the one hand, model placement considerably affects backhaul and wireless link traffic across edge networks. On the other hand, radio resource allocation and data routing influence the optimal decisions of model placement. In view of the shared parameter blocks across LLMs, this joint problem significantly differs from existing caching and delivery schemes because of the shared parameters of various contents and the fact that a model can be recovered by fetching various blocks from different source nodes. In multi-hop networks, the joint caching and data routing problem can be studied by considering the parameter shareability and multicasting of the shared model parts. In cache-aided cellular networks~\cite{tandon2016cloud}, one can jointly optimize the model placement, fronthaul/backhaul cost, and radio resource allocation to facilitate fast AI model downloading to edge devices.

\subsection{Lessons Learned}
Edge caching and delivery are highly challenging in the context of LLMs because of their enormous size. To mitigate these issues, the aforementioned techniques aim to 1) only transmit a small proportion of the model (i.e., task-specific parameter block), 2) store or multicast the shared parameter blocks across communication networks, and 3) jointly optimize LLM capabilities and external knowledge sources. \textit{All these features arise from exploiting ``reusable knowledge'' in LLMs to save bandwidth and storage resources as much as possible}. Clearly, these new features will generate a rich set of research problems, particularly considering the various scenarios in edge networks and multi-dimensional resource management therein. For instance, diverse network architectures such as multi-cell wireless networks, heterogeneous wireless networks, and a combination of wireless and wired networks can be explored, along with optimizing various sets of radio resources, including spectrum resources and transmit power. The guiding design principle is to optimize the usability of LLMs after caching and delivery while minimizing resource utilization based on the aforementioned principles regarding reusable knowledge.

\section{Edge Training for LLMs\label{sec:learning}}

\begin{figure*}[!ht]
	\centering
	\subfigure[Centralized edge learning.]{\includegraphics[width=3.2in]{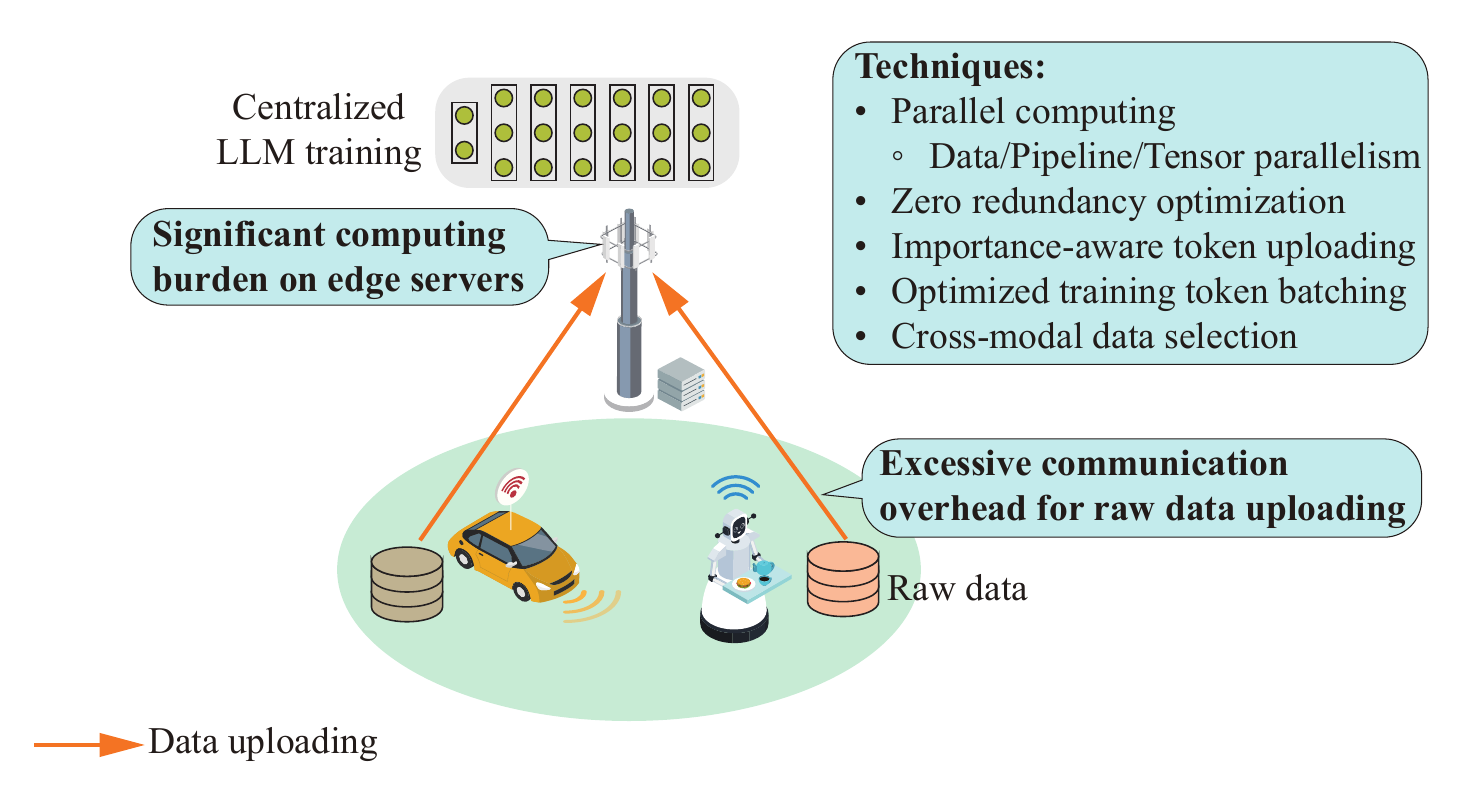}\label{fig_centralized_learning}}
	\quad
	\subfigure[Federated edge learning.]{\includegraphics[width=3.2in]{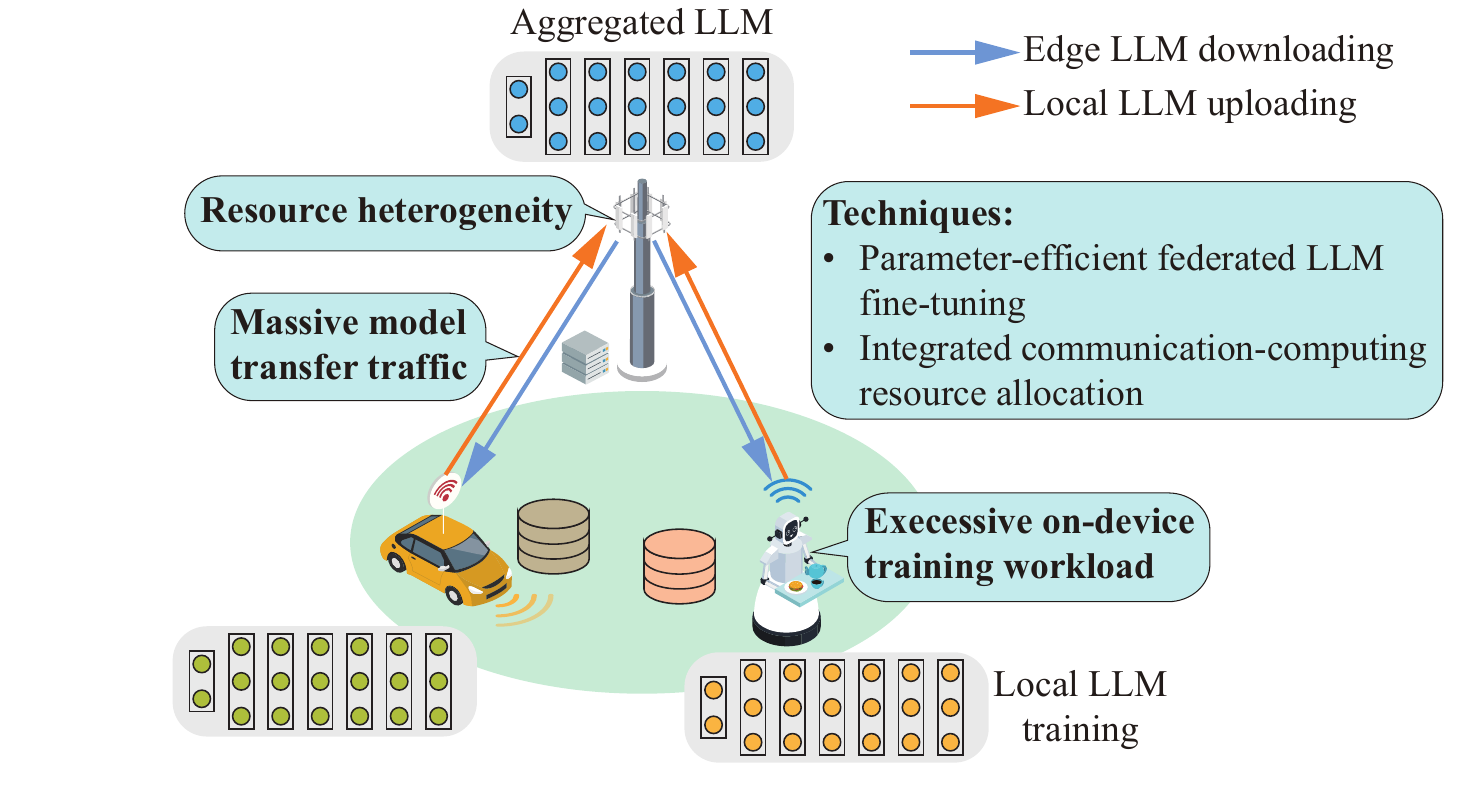}\label{fig_federated_learning}}
        \\ 
        \subfigure[Split learning.]{\includegraphics[width=3.2in]{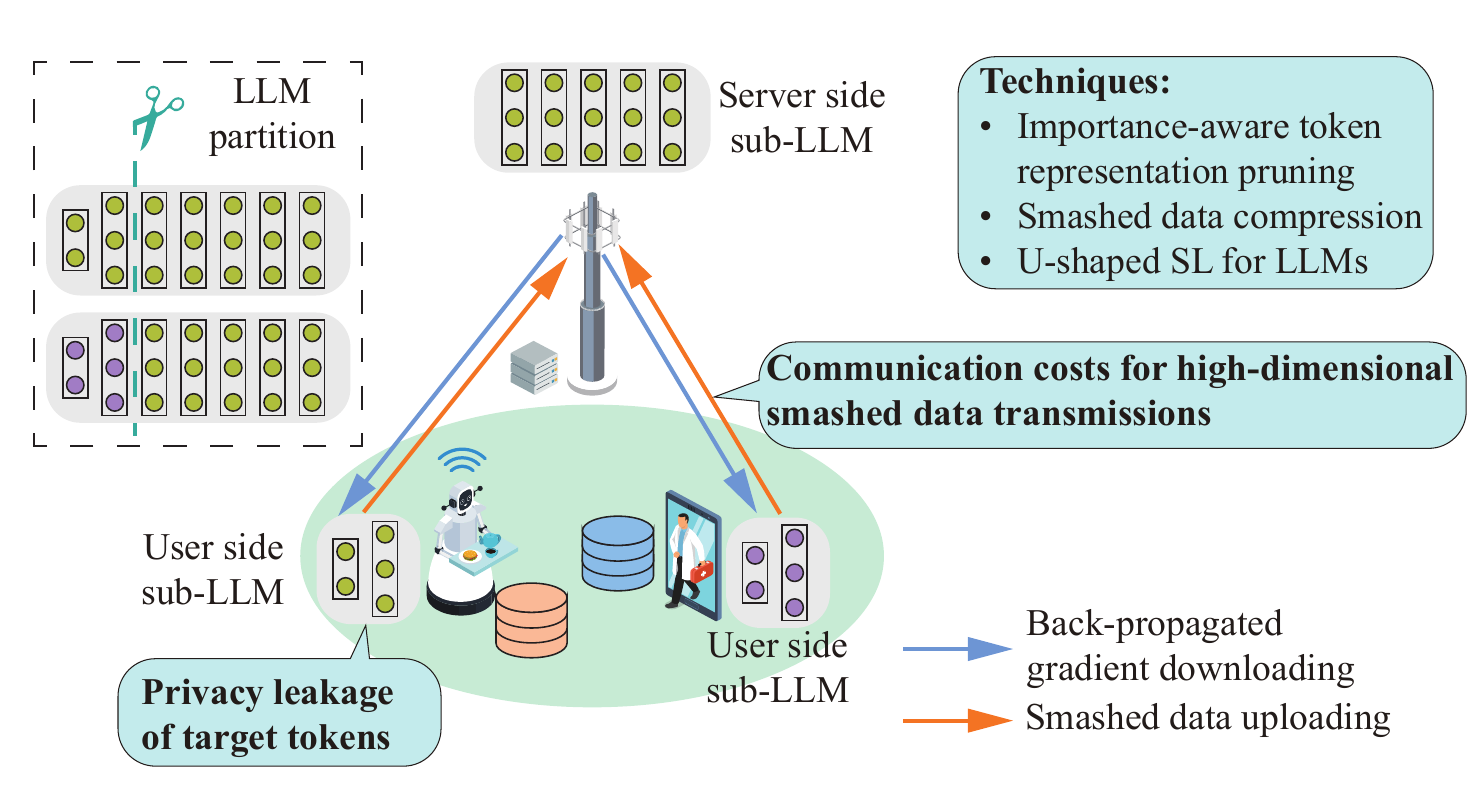}\label{fig_split_learning}}
	\quad
	\subfigure[Hierarchical collaborative learning.]{\includegraphics[width=3.2in]{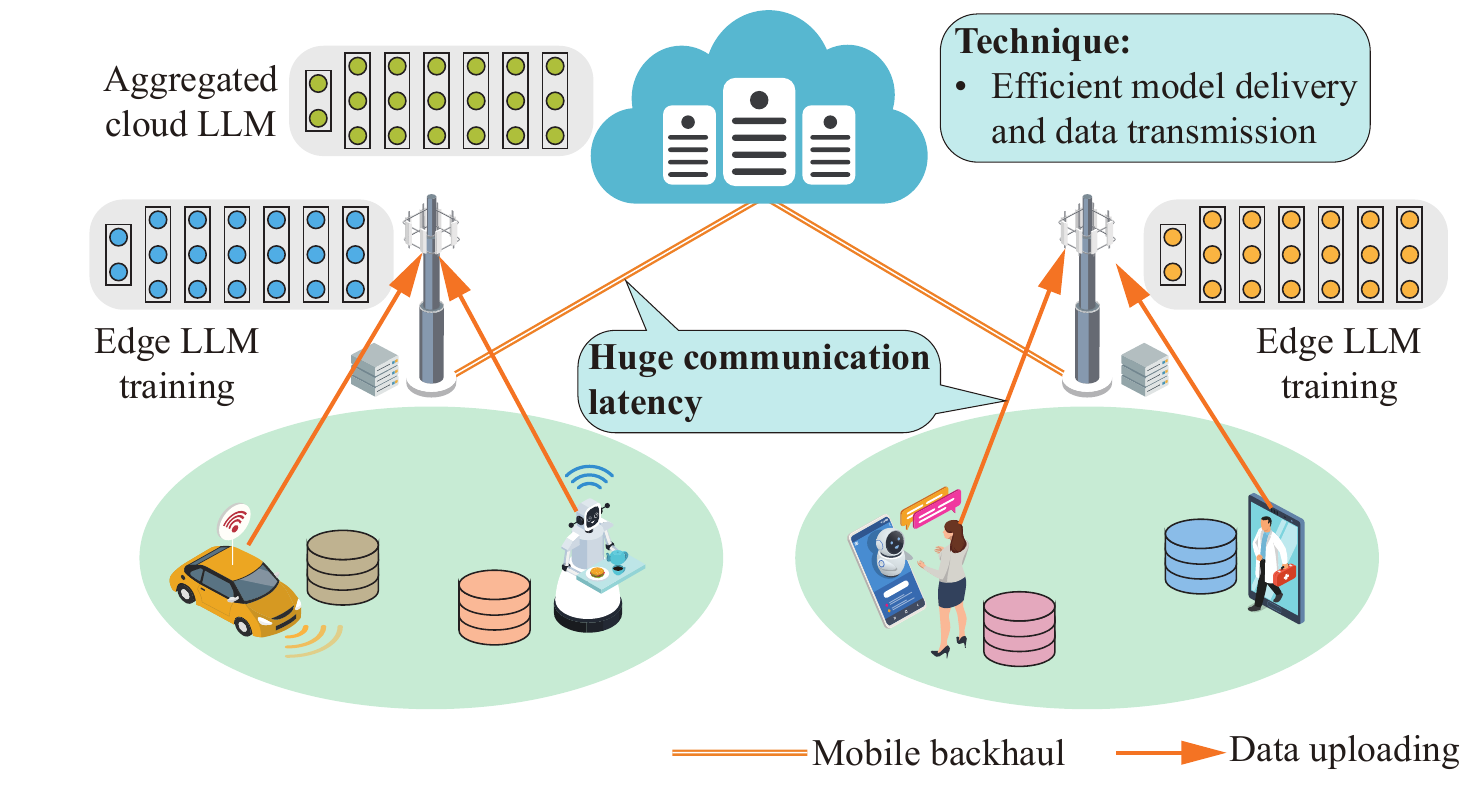}\label{fig_hierarchical_learning}}
	\caption{Edge training frameworks for LLMs. This figure presents the associated challenges and corresponding techniques to address them in each sub-figure. In Fig. \ref{fig_centralized_learning}, edge devices upload raw data to the edge server for global LLM training. In Fig. \ref{fig_federated_learning}, FL is adopted, where the edge server aggregates the locally trained LLMs from edge devices. In Fig. \ref{fig_split_learning}, an LLM is partitioned into two sub-models, with the split learning process executed by exchanging intermediate data between the edge server and the edge device. At last, hierarchical collaborative learning can be categorized into cloud-end, cloud-edge, and cloud-edge-end collaborative learning. We take cloud-edge collaborative learning as an example in Fig. \ref{fig_hierarchical_learning}, where edge devices upload the data to edge servers for training, and the cloud aggregates different edge-trained LLMs.}\label{fig:edge_learning}
\end{figure*}

Edge training performs model training at the network edge to extract intelligence from data sources. The main distinction of edge LLM training from traditional edge training lies in \textit{the large scale of AI models, which can be too large to fit into an edge server, as well as the optimization of PEFT over wireless networks.} As shown in Fig. \ref{fig:edge_learning}, this section elaborates on edge LLM training through four categories: centralized edge learning, federated edge learning, split learning, and hierarchical collaborative learning. For readers' convenience, the related works on edge LLM training are summarized in Table \ref{table_training}.
\begin{table*}[!t]
\centering
\caption{Summary of related works on edge LLM training.}
\label{table_training}
\renewcommand{\arraystretch}{1.4}
\setlength{\tabcolsep}{2mm}
\begin{tabular}{|c|c|c|m{0.64\textwidth}|}
\hline
\textbf{Scenarios} & \textbf{Techniques} & \textbf{Ref.} & \makecell[c]{\textbf{Objectives}} \\ 
\hline
\multirow{12}{*}{\makecell[c]{Centralized\\edge learning}}
               & \multirow{10}{*}{\makecell[c]{Scaling-up\\training}} 
                   & \cite{xu2020automatic} & Adopts data parallelism to replicate models and distribute datasets across different processors. The gradients from all processors are aggregated for model updating. \\ \cline{3-4}
                   &  & \cite{huang2019gpipe} & Leverages pipeline parallelism to divide models into sub-models and divide a mini-batch of training samples into micro-batches in each training iteration. Different processors can update the sub-models with different micro-batches at the same time. \\ \cline{3-4}
                   &  & \cite{shoeybi2020megatronlm} & Splits per attention head parameters in self-attention modules across multiple processors to realize the parallel processing for the matrix multiplication via tensor parallelism. \\ \cline{3-4}
                   &  & \cite{narayanan2021efficient} & Combines tensor, pipeline, and data parallelism to efficiently train LLMs. \\ \cline{3-4}
                   &  & \cite{rajbhandari2020zero} & Proposes ZeRO where each processor only holds a portion of optimizer states, gradients, and parameters during training, and the rest can be obtained from other data processes as needed. \\ \cline{3-4}
                   &  & \cite{ren2021zero} & Introduces ZeRO-Offload, which efficiently optimizes GPU memory by utilizing CPU memory. It partitions gradients and optimizer states across GPUs and offloads them to the CPU during training.\\ \cline{2-4}
               & \multirow{3}{*}{\makecell[c]{Importance-aware\\ token uploading}} 
                   & \cite{9098940} & Demonstrates that the convergence rate and model accuracy during training can be improved by allocating more radio resources to training data samples with higher importance. \\ \cline{3-4}
                   &  & \cite{9107235} & Illustrates that models can converge faster via appropriate user scheduling by considering the signal-to-noise ratio and data uncertainty metric.\\ \hline
\multirow{7}{*}{\makecell[c]{Federated edge\\learning}}
               & \multirow{7}{*}{\makecell[c]{Parameter-efficient\\federated LLM\\fine-tuning}} 
                   & \cite{jiang2023lowparameter} & Combines LoRA and FL, enabling edge devices to update and upload only the LoRA parameters to the edge server. \\ \cline{3-4}
                   &  & \cite{kuang2023federatedscopellm} & Introduces FederatedScope-LLM, which provides PEFT techniques and versatile programming interfaces to facilitate FL for LLMs. \\ \cline{3-4}
                   &  & \cite{chen2023prompt} & Integrates prompt tuning in FL for real-world meteorological forecasting tasks with distributed sensors. \\ \cline{3-4}
                   &  & \cite{che2024federated} & Proposes FedPepTAO, which focuses on updating parameters in proper prompt layers of LLMs. \\ \cline{3-4}
                   &  & \cite{yang2023efficient} & Proposes pFedPG, where clients update the client-specific prompts during FL for personalization. \\ \hline
\multirow{5}{*}{\makecell[c]{Split\\learning}}
               & \multirow{2}{*}{\makecell[c]{Token representation\\reduction in SL}} 
                   & \cite{10283579} & Edge devices can send the most informative smashed data to the server in SL for reducing communication overhead. \\ \cline{3-4}
                   &  & \cite{Zheng_2023} & Quantizes smashed data in SL/SFL/PSL with low precision for communication-efficient SL. \\ \cline{2-4}
                & \multirow{1}{*}{\makecell[c]{U-shaped SL}} 
                   & \cite{ohta2023lambdasplit} & Adopts U-shaped SL for LLMs. The framework places the computation-sensitive decoders on edge servers and places the bottom layers with the text input module and the top layers with the text output module on edge devices, which can protect the privacy of the ground-truth target tokens. \\ \hline
\multirow{10}{*}{\makecell[c]{Hierarchical\\collaborative\\learning}}
               & \multirow{4}{*}{\makecell[c]{Cloud-end\\collaboration}} 
                   & \cite{10283579} & Proposes DLoRA that dynamically identifies and fine-tunes the most relevant personal PEFT modules on edge devices while keeping the frozen parameters of LLMs in the cloud for cloud-end collaborative training. \\ \cline{3-4}
                   &  & \cite{wang2023clouddevice} & Places a larger multimodal LLM and a smaller multimodal LLM in the cloud and the edge device, respectively. Edge devices select and upload informative multimodal data to the cloud for training adapters, which are then downloaded to the edge devices. \\ \cline{2-4}
                & \multirow{4}{*}{\makecell[c]{Cloud-edge\\collaboration}} 
                   & \cite{xu2023unleashing} & Generative AI models are first pre-trained in the cloud and then fine-tuned on edge servers for customization. \\ \cline{3-4}
                   &  & \cite{tian2024edgecloud} & The cloud trains a large-scale model, which is combined by models of multiple edge servers for multi-task and multi-modal learning and then distributes task-specific lightweight models back to the edge servers for personalized fine-tuning. \\ \cline{2-4}
                & \multirow{2}{*}{\makecell[c]{Cloud-edge-end\\collaboration}} 
                   & \cite{liu2020client} &  Proposes a hierarchical FL framework, which enables edge servers to aggregate the locally trained models from end users and the cloud to aggregate models from different edge servers. \\ \cline{3-4}
                   &  & \cite{liu2022hierarchical} & Adopts model quantization in the hierarchical FL framework for efficient model aggregation. \\ \hline
\end{tabular}
\end{table*}
\subsection{Centralized Edge Learning}\label{sec:edgelearning_central}
The most straightforward model training approach in MEI systems is gathering data from edge devices and conducting model training on edge servers. While edge servers are generally more powerful than edge devices, centralized edge learning presents several key challenges: 1) \textit{Significant computing burden on edge servers}: LLM training or fine-tuning demands significant computing resources and storage/memory capacity. For instance, training LLama-2 7B in FP32~\cite{touvron2023llama} empirically demands 112 GB of GPU memory~\cite{llama_memory}, which can be challenging for an edge server provided that a powerful H100 GPU only possesses 80 GB memory. 2) \textit{Excessive communication overhead for raw data uploading}: Considering multimodal raw sensing data, a tremendous amount of data should be uploaded to a centralized data center. These two challenges can be overcome by the following solution approaches.

\subsubsection{Scaling-up LLM training} 
Scaling-up training is essential for LLM training in centralized edge training. Due to memory and computing constraints, training large-scale LLMs on a single GPU is highly challenging. Therefore, leveraging distributed computing and memory resources, such as multiple GPUs on edge servers, come to the rescue. Scaling-up LLM training can be categorized into two main approaches: parallel training and GPU memory optimization, which are elaborated below.

\textbf{Parallel training for LLMs: }Considering the extreme training workload of LLMs, parallel computing must be employed to leverage resources across edge nodes to split LLM training to reduce training latency and share required memory space. The three most prominent parallel computing strategies are data parallelism, pipeline parallelism, and tensor parallelism \cite{jiang2024megascale,zhou2024training}: 1) Data parallelism enables replicating models across different processors (processors can refer to multiple GPUs in one server or different edge servers), with the uploaded dataset being shuffled and distributed among the processors. After each processor finishes the forward and backward propagation with the corresponding model replica in parallel, all the gradients are aggregated for synchronization to update the model parameters \cite{xu2020automatic}. 2) In pipeline parallelism, models are divided into different sub-models; each processor owns one sub-model. A mini-batch of training samples is further divided into micro-batches in each training iteration. The pipeline training process can be executed by feeding micro-batches into the sub-model, and different processors can update the sub-models with different micro-batches at the same time \cite{huang2019gpipe}.
\begin{figure}[ht]
\centering
\includegraphics[width=0.25\textwidth]{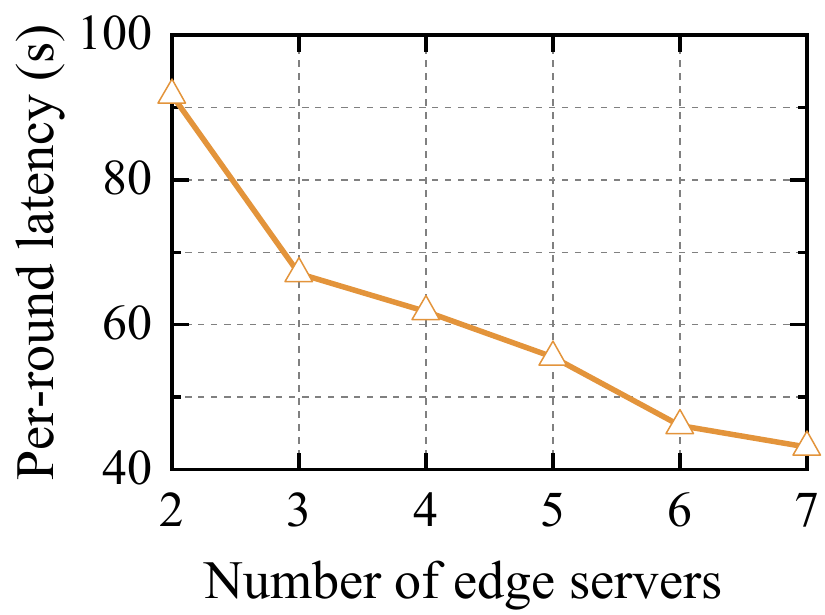}
\vspace{-0.25cm}
\caption{Per-round latency for split edge learning based on the number of edge servers participating in multi-hop split edge learning. We train a LLaMA-2 model on the E2E dataset \cite{DBLP:journals/corr/NovikovaDR17} to evaluate performance, assuming a wireless communication data rate of 200 Mbps between two edge servers. Per-round latency refers to the time taken to train one mini-batch of 128 samples.}
\label{fig:multi_hop}
\end{figure}
3) In tensor parallelism, the parameters and computations of each layer are divided across multiple processors for separate processing by multiple processors in parallel. For example, considering Transformers, the authors in \cite{shoeybi2020megatronlm} split per attention head parameters in self-attention modules across multiple processors such that the matrix multiplication corresponding to each attention head is stored locally on one processor. To leverage the advantages of these three parallel computing paradigms, in \cite{narayanan2021efficient}, Narayanan et al. combine tensor, pipeline, and data parallelism to efficiently train LLMs by harnessing a massive number of distributed GPUs. 

\textbf{GPU memory optimization: }Effective GPU memory optimization, such as Zero Redundancy Optimization (ZeRO), can be adopted to reduce memory usage in edge LLM training with distributed GPUs \cite{rajbhandari2021zero,ren2021zero,rajbhandari2020zero}. In \cite{rajbhandari2020zero}, Rajbhandari et al. develop ZeRO, where each processor only holds a portion of optimizer states, gradients, and parameters during training, and the rest can be obtained from other data processes as needed, thereby reducing the required GPU memory space for each processor. To further alleviate the pressure on GPU memory, the authors in \cite{ren2021zero} propose ZeRO-Offload, which effectively leverages the CPU memory. ZeRO-Offload partitions gradients and optimizer states across GPUs, offloading them to CPU memory for the entire training. During backward propagation, gradients are computed and averaged on the GPU, and each GPU offloads the averaged gradients belonging to its partition to the CPU memory. Once the gradients are available on the CPU, the optimizer states are updated directly on the CPU before being gathered back to the GPU.

\subsubsection{Parallel LLM training at the network edge} Parallel computing for LLM training can be extended to wireless edge networks. As illustrated by our simulations in Fig. \ref{fig:multi_hop}, pipeline parallelism significantly reduces end-to-end training latency by splitting the model into multiple sub-models and placing them on edge servers appropriately. Different from cloud-based approaches, parallel computing of LLMs encounters a more significant communication bottleneck at the network edge. Specifically, there exist heterogeneous communication and computing capabilities of edge servers. This situation, however, is normally oversimplified in large-scale GPU clusters. Since high-dimensional smashed data should be exchanged among edge servers via wired/wireless links, model splitting and placement should be optimized judiciously under computing-communication resource constraints. 

Similar problems have been considered for multi-hop split inference~\cite{hu2019dynamic,wang2021hivemind}, where the model splitting and/or placement is often mapped to a graph, which can be further mapped to shortest path problem~\cite{wang2021hivemind} or min-cut problem~\cite{hu2019dynamic}. Unfortunately, these problems are not directly applicable to parallel training of large-scale models, such as pipeline parallelism and tensor parallelism under edge networks. For instance, unlike multi-hop inference problems that often handle one input data at a time, pipeline parallelism for model training considers processing a batch of training data samples, which can be divided into multiple micro-batches as inputs to form the pipeline process. These characteristics set apart parallel training for large-scale models from inference problems in edge computing networks. Besides, by exploiting the special structure of LLMs, mostly relying on Transformers, the workload splitting problem might be simplified because it consists of repeated Transformer blocks with the same size of intermediate data output. Furthermore, model splitting, tensor splitting, and sub-model/layers placement can be jointly optimized with radio resource management problems, such as spectrum allocation and data routing~\cite{chen2022federated}, to mitigate network congestion with reduced transmission delay.


\subsubsection{Importance-aware token uploading} Another challenge in centralized edge training for LLMs lies in the communication bottleneck in input token uploading. Particularly, multimodal LLMs involve massive multimodal data uploading from edge devices, such as text, audio, high-dimensional images/videos, and LiDAR, which can cause network congestion and intolerable delay for delay-sensitive applications. To address this concern, existing works on importance-aware training data transmission schemes in edge learning can be extended to the context of LLMs. In \cite{9098940}, the authors demonstrate that the convergence rate and model accuracy during training can be improved by allocating more radio resources to training data samples with higher importance. Besides, it is shown in \cite{9107235} that a faster model convergence can be achieved via appropriate user scheduling by taking into account the signal-to-noise ratio and data uncertainty metric. 

In the context of LLMs, there are two additional characteristics that can be further exploited for efficient token uploading. First, compared with other types of training samples (i.e., images only), the input and target tokens in LLMs can vary in size significantly, implying that batching input/target tokens for training should consider their lengths. For instance, batching a long sequence with a short sequence for training may not be a wise choice because it can result in the idleness of GPU resources, considering that the short sequence can be finished shortly. It is noted that batching on the server side is tightly coupled with the scheduling of token uploading from edge devices. Second, the importance level of tokens can be obtained by exploiting the correlation across different modalities; for example, text data may reveal the important part of the image for training. As such, it is promising to explore the selection of important data across multiple modalities, such as text and images, from edge devices to improve the accuracy of LLMs while minimizing the overall data transmission overhead. 

\subsection{Federated Edge Learning\label{subsec:fed}}
Centralized edge learning allows edge servers to directly access the personal data of edge devices, which raises significant privacy concerns and potentially breaches data collection regulations. As such, there is a pressing need to employ FL for LLM training at the network edge \cite{fang2024automated}. In FL, clients cooperatively train a global model by sending their local model updates to an edge server for aggregation \cite{lin2023fedsn,zhang2024fedac}. When FL meets LLMs, there are some crucial challenges to address: 1) \textit{Excessive on-device training workload:} Compared with the memory bandwidth of deep learning accelerators (up to 7.8 TB/s), the memory bandwidth of embedded edge devices remains much lower (up to 0.2 TB/s)~\cite{woisetschlager2023federated}, leading to severe training time penalties. 2) \textit{Massive model transfer traffic:} On-device LLMs contain billions of parameters~\cite{Qualcomm2023Llama}. Uploading LLMs from massive edge devices to edge servers for model aggregation results in a heavy communication burden on telecommunication infrastructure, which can be extremely expensive for mobile subscribers. 3) \textit{ Resource heterogeneity:} The straggler issue is commonly observed in FL, where the training time is determined by the slowest client with the most scarce communication-computing capabilities. To mitigate this issue, integrated communication-computing resource allocation should be considered to fulfill the requirements of time-sensitive FL tasks.
As such, we present the following strategies to tackle the aforementioned challenges.

%


\subsubsection{Parameter-efficient federated LLM fine-tuning} To tackle the above challenges, parameter-efficient federated LLM fine-tuning can be adopted. Parameter-efficient federated LLM fine-tuning integrates PEFT into FL, enabling each client only to update and upload a small proportion of parameters, thereby reducing both communication and computing overhead \cite{jiang2023lowparameter,zhang2024building,zhang2023federated,kuang2023federatedscopellm,fan2023fatellm,babakniya2023slora}. In \cite{zhang2023federated}, Zhang et al. apply parameter-efficient tuning to FL for training LLMs. The authors demonstrate the effectiveness of parameter-efficient federated tuning methods in training LLMs and their capabilities to defend against data inference attacks. In \cite{jiang2023lowparameter}, Jiang et al. propose a low-parameter FL approach that operates under limited communication and computing resource constraints. By combining LoRA and FL, edge devices update and upload only the LoRA parameters to the edge server instead of the full parameters in LLMs. Some works push the practical deployment of FL for LLMs. For instance, FederatedScope-LLM \cite{kuang2023federatedscopellm} provides PEFT algorithms and versatile programming interfaces to facilitate FL for LLMs with low communication and computation costs without accessing the full model, making it well-suited for closed-source LLMs. Additionally, in \cite{fan2023fatellm}, an industrial-grade FL framework, called FATE-LLM, has been developed to support efficient FL for LLMs through PEFT methods, such as LoRA and P-Tuning-v2 \cite{liu2022ptuning}.

\textit{Federated LLM prompt tuning} is another parameter-efficient federated LLM fine-tuning technique, which adapts LLM by optimizing a small amount of task-specific prompt vectors on edge devices during the FL process. For example, MetePFL \cite{chen2023prompt} focuses on addressing communication and computation efficiency challenges in real-world meteorological forecasting tasks with distributed sensors via prompt tuning. From the communication perspective, prompt tuning adjusts only a small set of parameters for specific tasks, significantly reducing the size of parameters to be uploaded for aggregation. For example, if a model has hundreds of millions of parameters, a prompt might consist of only a few thousand parameters, i.e., 0.01\% to 0.1\% of the entire model size \cite{lester2021power}. From the computing perspective, prompt tuning requires updating only a small subset of LLM parameters, enabling resource-limited edge devices to participate. For instance, FedPepTAO \cite{che2024federated} devises parameter-efficient prompt tuning with adaptive optimization in FL. FedPepTAO focuses on updating parameters in proper prompt layers with small communication costs of LLMs, reducing the number of parameters transmitted and optimizing computing efficiency. Moreover, to leverage robust representations from LLMs while enabling efficient model personalization for clients in FL, the authors in \cite{yang2023efficient} propose pFedPG. The server-side personalized prompt generation module creates personalized prompts for all clients based on optimization direction information collected from clients. Then, clients update their client-specific prompts for personalization. Additionally, this approach ensures that only necessary and targeted parameters are updated, further optimizing communication and computing efficiency. 

\subsubsection{Resource management in FL for LLMs} Under the context of LLMs, the optimization of PEFT for FL over wireless networks can be a significant future direction, which is still in its nascent stage. The principle is to adapt the proportion of trainable parameters over time by considering both wireless channel conditions and model training status. Intuitively, a larger set of trainable parameters leads to better training performance while incurring more significant communication and computing latency. In \cite{chen2021communication}, based on the observation that model parameters gradually stabilize prior to model convergence, Chen et al. propose an adaptive parameter freezing scheme that freezes the non-synchronized stable parameters during the training process, eliminating the need for synchronizing the full models. However, this work does not consider the dynamic channel conditions in wireless networks. Considering PEFT techniques for LLMs such as LoRA, the rank of trainable matrices largely influences training accuracy and communication-computing latency in FL, as shown in \cite{fang2024automated}. Consequently, one essential research problem is how to jointly optimize the ranks in LoRA in LLMs and radio resource allocation in FL over wireless networks.

\subsection{Split Learning}\label{sec:edgelearning_split}
While FL can be combined with various efficient fine-tuning techniques for training LLMs, it is still extremely resource-intensive for lightweight edge devices. Specifically, models such as GPT-3 or BERT contain billions of parameters. It is challenging for edge devices, such as smartphones or IoT devices, to perform computation-intensive parameter updates locally even with PEFT~\cite{9978924}. To address these concerns, SL can be a promising LLM training paradigm in edge networks, which enables the co-training of large-scale models through the collaboration among edge servers and edge devices~\cite{lin2023pushing}. 

SL allows an edge server to take over the major training load from edge devices based on model splitting~\cite{gupta2018distributed}. Since its invention, SL has been applied to various scenarios, such as healthcare~\cite{vepakomma2018split,ha2021secure}. Unlike FL, SL only places a sub-model on edge devices for training, thereby considerably decreasing the workload of edge devices. The vanilla SL involves sequential interactions between an edge server and edge devices, which is a major bottleneck due to the waiting time of idle edge devices. The variants of SL, including parallel split learning (PSL) \cite{lin2023efficient, kim2022bargaining} and SFL \cite{thapa2022splitfed,lin2024adaptsfl}, can be applied to enable parallel training while utilizing the resources on multiple edge devices. Clearly, analogous to FL, SL can also be integrated with PEFT or other resource-efficient techniques in Section \ref{sec:on-device} to alleviate the workload on edge devices further. For instance, edge devices are only required to execute the forward pass by freezing client-side parameters, considerably reducing the computing workload and memory usage.

Although SL can facilitate LLM training by harnessing edge servers, several challenges still exist in employing SL for LLMs. 1) \textit{Communication costs for high-dimensional smashed data transmissions:} Although model partitioning leverages distributed computing resources and alleviates the computing load on edge devices, the communication overhead incurred by uploading cut-layer smashed data can be a major bottleneck. Considering GPT-3 Medium and an edge device with 100 data samples, each with 1024 tokens, the total smashed data volume at the cut layer can be approximately 400 MB for one training round \cite{brown2020language}. 2) \textit{Privacy leakage of target tokens:} While it is generally difficult to recover the raw training data of LLMs based on the received smashed data in SL~\cite{thapa2021advancements}, there is a privacy risk of target token leakage (i.e., label leakage). During the process of SL, one commonly used LLM splitting scheme is to place the sub-LLM with the input module on edge devices and the sub-LLM with the output module on edge servers. In such a case, edge devices need to upload the target tokens of the input data to edge servers for LLM training, which leads to target token leakage. As presented below, several methods can be employed to tackle the said challenges.


\subsubsection{Token representation reduction in SL} First, importance-aware token representation pruning can be used to eliminate the unimportant token representation at the cut layer from uploading. In \cite{10283579}, edge devices can selectively send the most informative smashed data to the server based on an online distillation method to perform SL. This approach improves performance while reducing communication costs by around 50\% compared with the benchmark without smashed data selection. Second, the smashed data compression can be adopted before transmissions. In this respect, quantization can be employed to efficiently reduce the communication overhead in SL/SFL/PSL \cite{Zheng_2023}. Specifically, quantization can convert float values (usually 32-bit) to low-bit representations, thereby enhancing the efficiency of smashed data transmission in SL. These methods can be adopted in the context of LLMs to shrink the data volume for uploading during the SL process.

\subsubsection{U-shaped SL for LLMs} In the SL framework, another common issue is the privacy leakage of ground-truth target tokens. In general, edge devices are required to transmit the target tokens of training samples to the server to calculate the loss function, leading to severe privacy concerns. For instance, the ground-truth target tokens of an LLM might be the disease type and health recommendations for a patient, which are considered sensitive personal data. To address this issue, U-shaped SL can be adopted for LLMs. U-shaped SL \cite{yang2022robust,lyu2023optimal} leaves both the head neural layers or transformer blocks with the input module and the tail layers/blocks with the output module on edge devices while only placing the intermediate layers/blocks on edge servers, thus effectively preserving label privacy. When applying this approach to LLM training, the computation-sensitive decoders can be placed on edge servers, and the bottom layers with the text input module and the top layers with the text output module can be stored on edge devices \cite{ohta2023lambdasplit}, leading to better privacy preservation.

\subsubsection{Resource management in SL for LLMs} To support SL for LLMs effectively and efficiently, the joint design of SL and radio resource allocation is necessary. Such joint problems of model splitting and/or radio resource allocation have been studied for SFL or PSL~\cite{lin2023efficient,lin2024adaptsfl}. In \cite{lin2023efficient}, we propose an efficient PSL approach by employing last-layer backpropagated gradient aggregation to reduce the computing, communication, and memory load on an edge server when serving multiple clients. Then, we design the joint model splitting and channel allocation strategies to mitigate the straggler issue in wireless PSL. However, it is shown that such an optimization problem is often integer programming due to the model splitting decisions, which can be NP-hard and compute-intensive when the number of edge devices is large~\cite{zhu2024esfl}.  With Transformer-based LLMs, the transformer blocks are generally repeated (unlike CNNs with varying output sizes for each layer) within an encoder or decoder, thus making the optimal optimization problem more solvable. This provides an opportunity to design the optimal radio resource allocation and model splitting strategy using a much more efficient algorithm. Particularly, when user-side computing capability is the limiting factor, a shallower model split point results in reduced communication-computing latency, as the communication costs remain consistent across different cutting blocks. However, it is important to note that a shallow split point may lead to increased privacy leakage of raw data; therefore, the split layer should not infringe on users' privacy requirements. In summary, the unique structure of Transformer-based LLMs can be harnessed to develop effective and efficient model splitting and resource allocation strategies under the SL framework.


 \subsection{Hierarchical Collaborative Learning}
The hierarchical collaborative learning paradigm can facilitate LLM training at scale\cite{xu2023unleashing,chen2024netgpt,gozalobrizuela2023chatgpt}. Compared with the previous edge-only paradigm, hierarchical collaborative learning provides improved flexibility to accommodate varied task complexity and resource availability by exploiting the synergy among clouds, edge servers, and edge devices. For instance, computation-intensive training tasks can be offloaded to cloud servers, whereas relatively easy ones can remain on the network edge to save communication latency/bandwidth. Besides, hierarchical collaborative learning is indispensable for LLMs to learn global knowledge across large geographical regions, where LLM updates can be synchronized at a central cloud. As presented below, hierarchical collaborative learning can be divided into three categories: cloud-end collaboration, cloud-edge collaboration, and cloud-edge-end collaboration.

\subsubsection{Cloud-end collaboration} The cloud and edge devices can collaboratively train LLMs together. To reduce the computing workload of edge devices and enhance user privacy, Gao et al. in \cite{gao2024dlora} propose a distributed PEFT framework called DLoRA. This framework maintains and fine-tunes personal PEFT modules on edge devices while storing the frozen parameters of LLMs in the cloud. By exchanging activations and gradients, edge devices and the cloud can collaboratively train the LLM. Besides, to reduce the communication overhead of edge devices, the authors adopt the Kill and Revive mechanism to dynamically identify and fine-tune the most relevant and significant PEFT module. 
In \cite{wang2023clouddevice}, Wang et al. propose a cloud-end collaborative learning framework for multimodal LLMs, in which a larger multimodal LLM is located in the cloud while a smaller multimodal LLM is allocated to the edge device. Edge devices, such as robots, upload multimodal data to the cloud to perform KD to train the adapters, which are then downloaded to the edge devices. To reduce the communication overhead in data uploading, the authors adopt the uncertainty-guided token sampling strategy, enabling edge devices to upload only the most informative tokens to the cloud.


\subsubsection{Cloud-edge collaboration} Cloud-edge collaborative LLM training enables the cloud and edge servers to collaboratively train LLMs. On the one hand, LLMs can initially undergo pre-training in the cloud, followed by additional fine-tuning on edge servers to enhance performance. In \cite{xu2023unleashing}, Xu et al. propose a cloud-edge collaborative training and fine-tuning framework for generative AI models, where models are pre-trained in the cloud to learn general features, after which the models are fine-tuned with context-aware data stored on edge servers for customization. On the other hand, edge servers can train LLMs locally and then send the model updates to the cloud for knowledge sharing, such as model aggregation in the FL framework, which can improve the generality of LLMs. In \cite{tian2024edgecloud}, the edge models of multiple edge servers are integrated using gating neural networks and linear projection connections to form a large-scale model suitable for multi-task and multi-modal learning. The cloud trains this large-scale model with the cloud common dataset and then distributes task-specific lightweight models back to the edge servers for personalized fine-tuning.

\subsubsection{Cloud-edge-end collaboration}
Cloud-edge-end collaborative learning, which has a three-tiered architecture, brings in more distributed computing resources and flexibility in resource management. In \cite{liu2020client}, the authors propose a hierarchical FL framework, where edge servers aggregate the locally trained models from their associated end users, and the cloud aggregates models from different edge servers. Furthermore, to reduce the communication cost in the hierarchical FL framework, the authors in \cite{liu2022hierarchical} adopt model quantization to compress model sizes for efficient model aggregation, where a convergence upper bound is provided to optimize the aggregation intervals. The hierarchical collaborative learning paradigm is suitable for LLM training. Since the raw data in some LLM applications, such as healthcare, is highly sensitive, users can join the training process with the help of both cloud and edge servers while keeping private data on their local devices~\cite{10491730}. However, despite these advantages, model delivery and data transmission among the cloud, edge servers, and edge devices must be appropriately handled since delivering model parameters of LLMs incurs significant communication latency, especially in links from the cloud to the edge server.

\subsubsection{Resource management in hierarchical collaborative learning for LLMs} In hierarchical collaborative training, model aggregation intervals (in FL or SFL) and model splitting must be carefully designed to maximize training accuracy with reduced latency and data traffic costs for LLMs based on the aforementioned techniques in Section \ref{subsec:fed} and \ref{sec:edgelearning_split}. For instance, the proportion of uploaded parameters in LLMs must be dynamically adjusted based on PEFT by taking into account the status of model training, wireless channel conditions, and the Internet delay in the backbone networks. Also, the model splitting and uploaded token representation pruning must be carefully devised across different tiers of the cloud-edge-end architecture. The increased number of network tiers adds more complexity to the optimization problem, giving rise to more challenges and opportunities in this direction.

\subsection{Lessons Learned}
Analogous to edge training for other AI models, edge LLM training aims to optimize training accuracy under limited communication-computing resources. Nevertheless, since LLM training is much more resource-demanding, the underpinning principle should be splitting the training and ``only updating a little''. In other words, the process must borrow the wisdom from large-scale model training, i.e., parallel training across multiple edge devices/servers, and PEFT over wireless networks. This implies that SL can be an important enabling edge learning framework in the era of LLM. To make the MEI tailored for LLMs, the key principle is to enable flexible model splitting based on the dynamic communication-computing resources within edge networks and determine the fine-tuning portion of LLMs by jointly considering communication-computing resources and the training status of LLMs. To ground the network optimization, a theoretical understanding of fine-tuning of LLMs, such as LoRA, is required, i.e., characterizing the relationship between training accuracy and the percentage of trainable parameters, to determine the optimal freezing ratio for fine-tuning over wireless networks.

\section{Edge Inference for LLMs\label{sec:inference}}
\begin{figure*}[!ht]
	\centering
	\subfigure[Centralized edge inference.]{\includegraphics[height=1.6in]{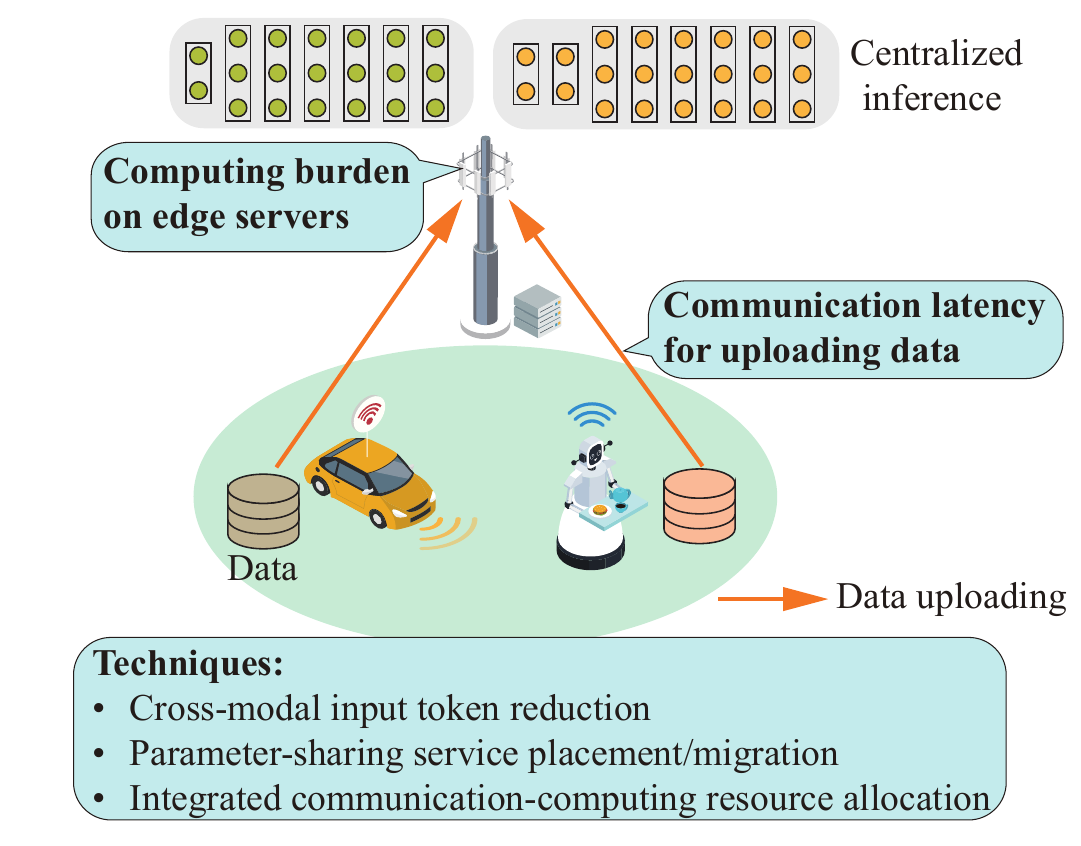}\label{fig_centralized_inference}}
	\quad
	\subfigure[Split inference.]{\includegraphics[height=1.6in]{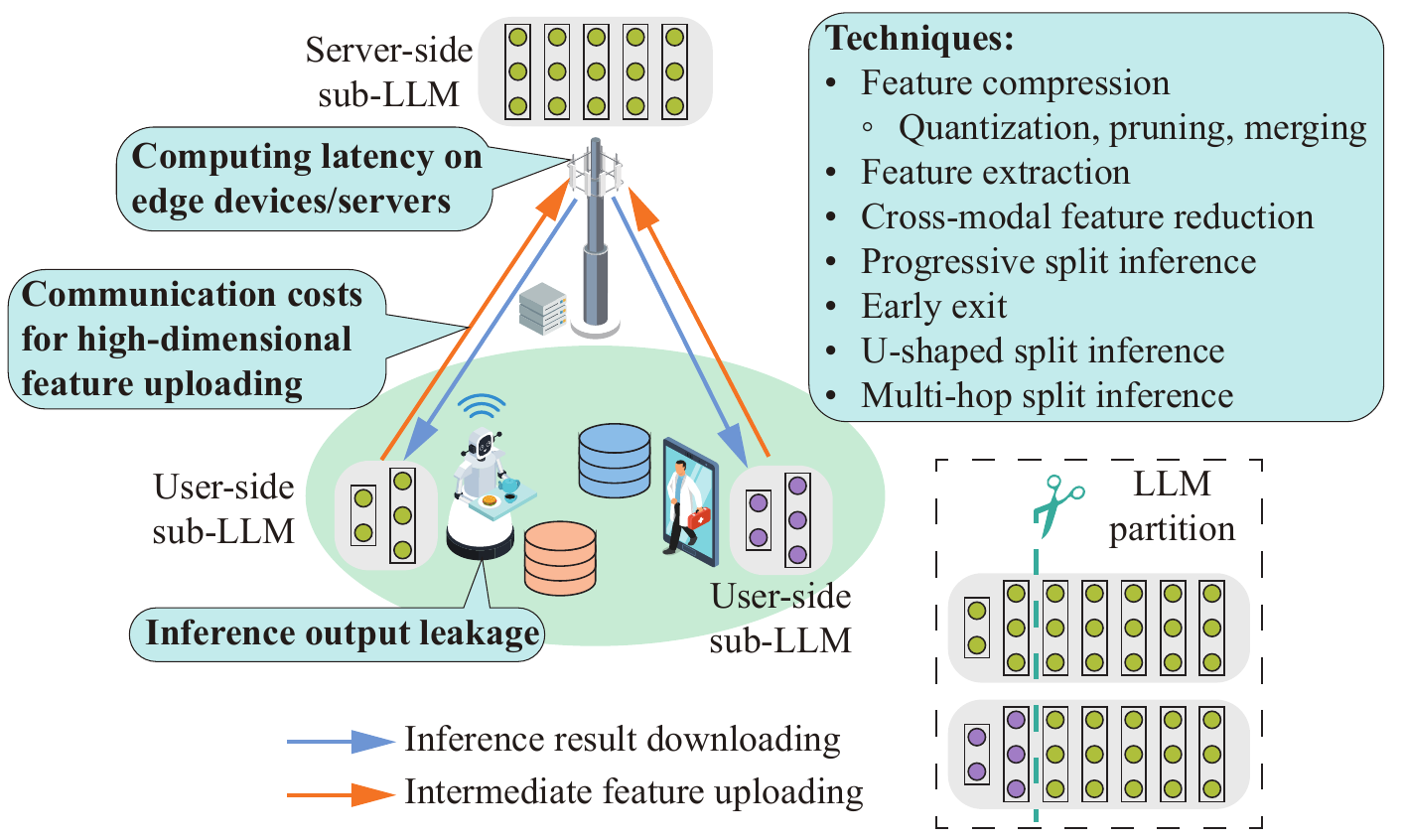}\label{fig_split_inference}}
	\quad
	\subfigure[Collaborative inference.]{\includegraphics[height=1.6in]{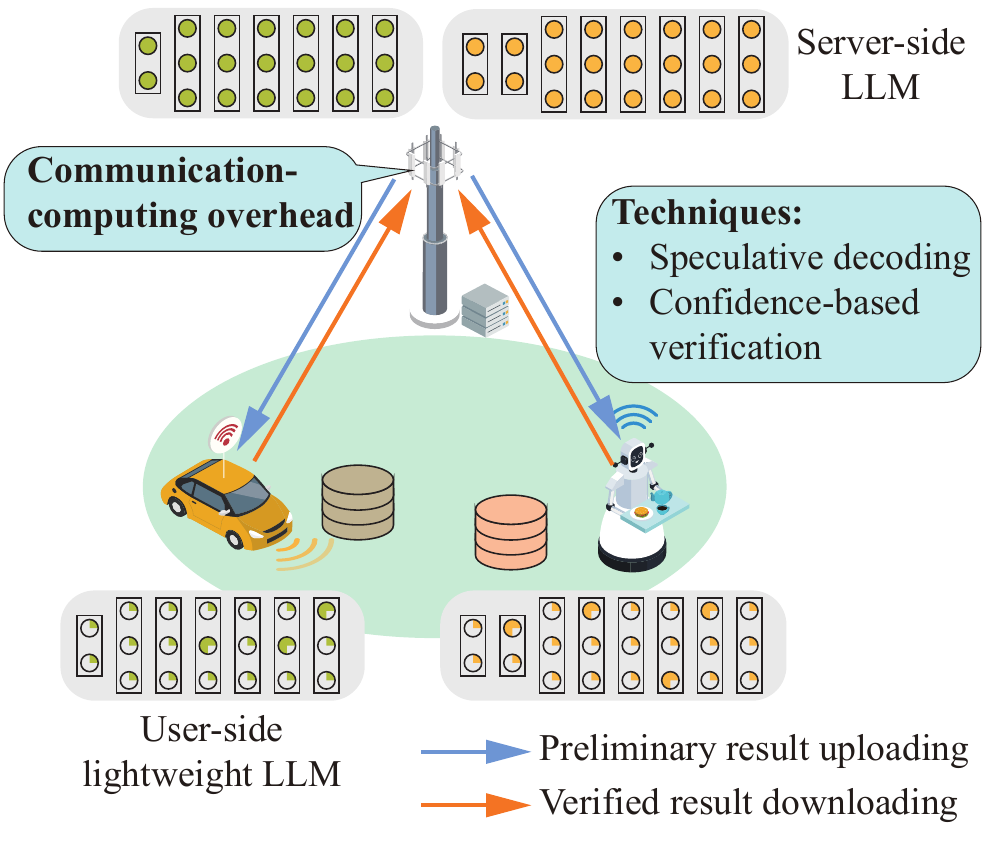}\label{fig_collaborative_inference}}
	\caption{Edge inference frameworks for LLMs. This figure presents the associated challenges and corresponding techniques in each sub-figure. In Fig. \ref{fig_centralized_inference}, edge devices centralized the data to the edge server with LLMs for centralized edge inference. In Fig. \ref{fig_split_inference}, edge devices feed the data into the user-side sub-LLM and upload the intermediate features to the edge server. Subsequently, the edge server conducts the inference with the server-side sub-LLM and returns the inference results to edge devices. In Fig. \ref{fig_collaborative_inference}, edge devices first generate preliminary inference results with the user-side lightweight LLMs and upload the results to the edge server with the larger LLM for verification. }\label{fig:edge_inference}
\end{figure*}

Edge LLM inference leverages edge computing to provide inference services to end users based on well-trained LLMs. Inference services in edge networks involve processing input data, such as text, images, and audio, to generate results like classifications, predictions, and responses with AI models for edge devices, including mobile users and IoT devices. One example is LLM-empowered robotic applications, where a robot uploads captured audio, video, and radar/lidar data across different modalities for feeding into a well-trained LLM on an edge server, thus facilitating the robot to understand the environments and execute actionable policies.
The major difference between edge inference for LLMs and conventional edge inference framework lies in \textit{the exploitation of LLM properties, such as multimodality, parameter-shareability, and the autoregressive process, to accelerate the inference process}. As illustrated in Fig. \ref{fig:edge_inference}, this section elaborates edge LLM inference through three categories: centralized edge inference, split inference, and collaborative inference, with the related works summarized in Table \ref{table_inference}.

\begin{table*}[!t]
\centering
\caption{Summary of related works on edge LLM inference.}
\label{table_inference}
\renewcommand{\arraystretch}{1.4}
\setlength{\tabcolsep}{2mm}
\begin{tabular}{|c|c|c|m{0.64\textwidth}|}
\hline
\textbf{Scenarios} & \textbf{Techniques} & \textbf{Ref.} & \makecell[c]{\textbf{Objectives}} \\ 
\hline
\multirow{9}{*}{\makecell[c]{Centralized\\edge inference}}
               & \multirow{2}{*}{\makecell[c]{LLM inference\\with cross-modal\\input token reduction}} 
                   & \cite{liang2022evit} & Removes some image patches unrelated to the visual contents of the images that input into the ViT, which can be used to reduce communication overhead in centralized edge LLM inference.  \\ \cline{3-4}
                   &  & \cite{jiang2023llmlingua} & Illustrates that input token pruning does not degrade inference accuracy significantly for LLMs. \\ \cline{2-4}
                & \multirow{6}{*}{\makecell[c]{Parameter-sharing\\service\\placement/migration\\for LLM inference}} 
                   & \cite{xu2024cached} & Optimizes the service placement and inference task offloading strategies in a cloud-edge collaborative inference framework.\\ \cline{3-4}
                   &  & \cite{ding2024hybrid} & The requests of end users for the same LLM are offloaded to the same edge server that loads the request LLM, improving memory/storage efficiency. \\ \cline{3-4}
                   &  & \cite{fu2024serverlessllm} & The inference requests of users are forwarded to an edge server with the lowest estimated startup time, thereby providing inference services with low latency. \\ \cline{3-4}
                   &  & \cite{fang2023large} & Users offload LLM inference tasks to an edge server under the jointly optimized communication and computing resource allocation policies. \\ \hline
\multirow{24}{*}{\makecell[c]{Split\\inference}}
               & \multirow{12}{*}{\makecell[c]{Split inference\\for LLMs with\\token representation\\reduction}} 
                   & \cite{caoBTRBinaryToken2024} & Inserts a binarization module after layer norm layers in Transformers to quantize the token representations with 1-bit vectors.  \\ \cline{3-4}
                   &  & \cite{goyal2020powerbert} & Prunes the redundant vectors of intermediate outputs of encoders progressively. \\ \cline{3-4}
                   &  & \cite{bolya2023token} & Proposes ToMe and merges the representations of similar tokens in the outputs of MHAs. \\ \cline{3-4}
                   &  & \cite{shao2021learning} & Uses information bottleneck to enable edge devices to extract and upload informative features to edge servers while keeping comparable inference accuracy in split inference. \\ \cline{3-4}
                   &  & \cite{shi2023taskoriented} & Combines the transmission rate maximization and the information bottleneck together in split inference to improve the robustness of the received features and ensure satisfactory inference performance without incurring too much communication overhead.\\ \cline{3-4}
                   &  & \cite{cao2023pumer} & Removes image token representations irrelevant to the input texts and unimportant to the tasks for ViTs and merges similar text and image token representations independently.\\ \cline{3-4}
                   &  & \cite{10233481} & Adopts deep joint source-channel coding to train encoders and decoders of Transformers by considering the effects of physical channel noises and interference, making split inference for Transformers robust.\\ \cline{2-4}
                & \multirow{1}{*}{\makecell[c]{Progressive\\split inference}} 
                   & \cite{9955582} & Users schedule the offloading sequence of features and upload the features with higher importance to edge servers first until edge servers determine that the uploaded features reach a target confidence level for inference.\\ \cline{2-4}
                & \multirow{1}{*}{\makecell[c]{Split inference\\with early exit}} 
                   & \cite{schuster2022confident} & Demonstrates that the hidden states of the current token need to be forwarded to the subsequent layers for KV cache computation in LLM inference using early exit, illustrating the necessity of designing the uploading strategy of hidden states at exiting layers for split LLM inference with early exit.  \\ \cline{2-4}
                & \multirow{3.5}{*}{\makecell[c]{Multi-hop/U-shaped\\split inference\\for LLMs}} 
                   & \cite{ma2023poster} & An LLM is partitioned into multiple sub-models to be placed on multiple edge devices/servers for multi-hop split inference via inter-device (server) communication.  \\ \cline{3-4}
                   &  & \cite{ohta2023lambdasplit} & Adopts the U-shaped split LLM inference, where only the body sub-LLM with the hidden layers in decoder blocks is placed on the edge server, and only the outputs of intermediate Transformer blocks are exchanged between edge devices and edge servers. \\ \hline
\multirow{3}{*}{\makecell[c]{Collaborative\\inference}}
               & \multirow{3}{*}{\makecell[c]{Speculative\\decoding}} 
                   & \cite{leviathan2023fast} & Speculative decoding enables an edge device to run a smaller on-device LLM while asking the edge server to run a larger LLM to verify and correct the output tokens uploaded by the edge device for reducing E2E latency.  \\ \cline{3-4}
                   &  & \cite{wang2023tabi} & Edge devices can decide whether to upload the generated results to edge servers with powerful LLMs for verification based on calibrated confidence scores. \\ \hline
\end{tabular}
\end{table*}
\subsection{Centralized Edge Inference}
In centralized edge LLM inference, edge devices offload their input token to edge servers for AI inference~\cite{fang2023large}. Note that efficient LLM inference methods, mentioned in Section \ref{sec:on-deviceinference}, can be implemented on edge servers to accelerate edge LLM inference. Centralized edge inference involves raw data uploading from edge devices to edge servers, which shares similar challenges with centralized edge learning in Section \ref{sec:edgelearning_central}, i.e., \textit{communication latency for uploading data} and \textit{computing burden (including excessive memory usage) on edge servers}. In what follows, we will elaborate on the techniques to overcome these challenges.


\subsubsection{LLM inference with cross-modal input token reduction} Analogous to centralized LLM training, input token reduction decreases the volume of data transmissions for fast edge inference. By removing some unimportant tokens from input text sequences and images before inference, the sizes of data that need to be offloaded to edge servers are reduced \cite{liang2022evit,huang2024fewer,li2023compressing}. It has been shown that input token pruning does not introduce significant inference accuracy loss for LLMs \cite{jiang2023llmlingua}. For example, before inputting images into the ViT, removing some image patches unrelated to the visual contents of the images may not degrade the prediction results~\cite{liang2022evit}. Since the significant communication overhead of input tokens comes from visual tokens, an effective approach to reducing communication overhead is to exploit text/audio data tokens to decrease the unrelated visual tokens in multimodal LLM inference. In other words, cross-modality can be fully explored to eliminate communication-computing redundancy in LLM inference.

\subsubsection{Parameter-sharing service placement/migration for LLM inference} For service provisioning in multi-user systems, the scarcity of computing resources, say, memory space, on an edge server is a major concern. This characteristic severely limits the number of edge LLM inference services to be supported by an edge server. For this reason, properly designing service placement in centralized edge inference becomes crucial to accommodate large-sized models. Moreover, considering the mobility of users, service migration, which migrates the AI models from one place to another according to varying user locations, can also be studied. In \cite{xu2024cached}, the authors optimize the service placement and inference task offloading strategies in a cloud-edge collaborative inference framework, where the cloud and edge servers provide inference services with the cached LLMs. The framework aims to minimize the total inference cost of edge servers and the cloud within the GPU memory and computing capacity constraints. Here, the total inference cost of edge servers includes model loading/evicting costs, transmission costs of input prompts and inference results, edge computing costs, and accuracy costs. Another potential method for determining service placement strategies is jointly considering model requests and required computing resources. The queries of end users requesting the same LLM can be offloaded and forwarded to the same edge server \cite{ding2024hybrid}, thereby improving memory/storage efficiency since the LLM can remain in the memory of the edge server. The authors in \cite{fu2024serverlessllm} design a serverless inference system for LLMs, where inference requests can be forwarded to an edge server with the lowest estimated startup time, including LLM loading and migration time. For example, the inference requests can be forwarded to an edge server with the request LLM already in its memory by eliminating additional latency for loading the LLM from the storage disk. 

Although the aforementioned excellent works have been done on service placement for inference, these works have not exploited the shared parameters among models/tasks for edge inference. In fact, by exploiting this feature, multiple LLMs with substantially shared parameters can be loaded into the memory of a server for concurrent inference, thereby significantly improving inference throughput by serving more user requests simultaneously. This requires the design of parameter-sharing service placement schemes for LLM inference by jointly considering the memory, storage, computing, and spectrum constraints of edge networks.


\subsubsection{Resource management for centralized LLM inference} To cater to more users with heterogeneous communication-computing resources and service requirements, integrated communication-computing resource allocation needs to be investigated. A key principle in centralized edge inference involves assigning users with delay-sensitive tasks with more spectrum bandwidth and computing resources to satisfy their QoS requirements. In \cite{fang2023large}, Fang et al. propose an edge LLM inference scheme where users can offload LLM inference tasks to an edge server under the designed communication and computing resource allocation policies. By jointly allocating spectrum bandwidth and computing resources for inference tasks, the average E2E latency and inference accuracy can be optimized. In the context of LLM inference, radio resources can be jointly optimized with input token pruning and parameter-sharing service placement. On the one hand, wireless channels can be allocated more to users with high-dimensional input tokens to upload in order to achieve high inference accuracy. The optimization problem can, therefore, be formulated as an accuracy maximization (or fairness) problem subject to computing-communication latency constraints by jointly optimizing users' token pruning ratios and radio resource allocation. On the other hand, when considering parameter-sharing service placement, a trade-off exists between communication latency and storage/memory efficiency. Specifically, although offloading multiple inference requests requiring the same LLM to a server enhances storage/memory efficiency, as demonstrated above, data transmissions may be subject to longer communication latency, adversely affecting QoS for users. This motivates us to study the parameter-sharing service placement/migration problem under latency constraints while considering channel allocation and data routing over wireless networks~\cite{ding2018beef,chen2022end}.

Furthermore, one particular challenge in edge LLM inference lies in the autoregressive process of LLMs. This characteristic introduces two kinds of difficulty: 1) the task computing delay can hardly be predicted since the length of output tokens during the decoding process may not be known at the beginning, and 2) the memory usage is relevant to the length of output tokens according to KV cache, implying that we must reserve sufficient memory and computing resources for task scheduling. Based on the above observations, we can explore new mathematical modeling of edge computing to address resource optimization for edge LLM inference.


\subsection{Split Inference}
Split inference is a technique that offloads part of the computing workload from edge devices to edge servers by placing partitioned models on edge devices and edge servers, respectively~\cite{lin2023pushing,lin2023split}. The most widely adopted split inference paradigm is the bi-partition paradigm. Edge devices execute on-device inference with the user-side sub-model based on the raw data and upload intermediate features to edge servers for going through the remaining neural networks~\cite{shi2023taskoriented}. Split inference is particularly well-suited for LLM inference. On the one hand, compared with on-device LLM inference, split LLM inference offloads a major part of computation to the edge server, thus reducing the device-side workload, which is of paramount importance for computation-intensive LLM inference. On the other hand, given that LLM applications within edge networks, such as mobile health and autonomous driving, often involve highly sensitive personal data, split LLM inference effectively mitigates privacy issues as edge devices do not need to share private raw data with edge servers.

There are two prominent bi-partition methods for split LLM inference. First, the most resource-intensive modules of LLMs can be placed on the server to fully harness its computing power. In Transformers, the decoder module often requires more computing resources than the encoder module \cite{leviathan2023fast}. Therefore, for LLMs with the encoder-decoder architecture, such as BART \cite{lewis-etal-2020-bart}, the encoder module can often be cached on edge devices while the decoder module can be placed on edge servers. For decoder-only based LLMs, such as GPT-series models, the text and position embeddings as well as several bottom Transformer blocks can be placed on edge devices due to their relatively light workload, whereas the remaining Transformer blocks can be placed on edge servers. Second, the sub-model occupying large storage capacity can be placed on edge servers. For example, the FFNs account for approximately $2/3$ of parameters in LLMs, while only a small portion of parameters in FFNs have a significant contribution to the final inference results~\cite{liu2024ffsplit}. Therefore, these parameters in FFNs, with smaller data sizes yet critical for inference, can be placed on edge devices to save storage/memory resources.


Split inference for LLMs encounters similar challenges to SL, as discussed in Section \ref{sec:edgelearning_split}, i.e., \textit{communication costs for high-dimensional feature uploading}, \textit{computing latency on edge devices/servers}, and \textit{inference output leakage}.  In what follows, we will introduce the techniques to address these challenges.

\subsubsection{Split inference for LLMs with token representation reduction} Considering split inference, there are various ways to reduce the volume of token representations (latent representation)  at the cut layer in LLMs for uploading to edge servers. First, the outputs of intermediate layers in Transformers can be compressed via quantization \cite{liu2024qllm,yang2023dynamic,xiao2023smoothquant}, pruning \cite{goyal2020powerbert,cao2023pumer}, and merging \cite{jin2024unified,jin2024chatunivi}. Uploading the compressed intermediate layer outputs can substantially reduce the communication overhead, thereby achieving low-latency split inference. 1) Quantization enables converting intermediate outputs of hidden layers in Transformers from high bitwidth to low bitwidth. In \cite{caoBTRBinaryToken2024}, the authors insert a binarization module after the layer norm layer inside the Transformer encoder blocks to represent the token representations with 1-bit vectors instead of float-value vectors. 2) Pruning involves removing redundant vectors of token representations, which can apply to both images and text inputs based on a lightweight module \cite{cao2023pumer}. For example, the authors in \cite{goyal2020powerbert} eliminate the redundant vectors of the intermediate encoder outputs, where the remaining intermediate encoder outputs of the last encoder only account for 2\% of the original intermediate encoder outputs. With this method, edge devices can eliminate unimportant vectors of intermediate layer outputs before uploading them to edge servers in split LLM inference. As the size of remaining representation vectors decreases as the encoder goes deeper, which indicates that more encoder blocks remain on edge devices for local computing, it is worthwhile investigating the optimal cut layer selection involving the trade-off between communication efficiency and computing burden on edge devices. 3) Merging can compress the information of intermediate outputs by merging redundant token representations with similar semantic meanings. The authors in \cite{bolya2023token} propose ToMe inserted into the Transformer blocks, which merges the representations of similar tokens in the outputs of MHAs. It is promising to employ this method in split LLM inference so that edge devices merge the representations in intermediate results with the lightweight merging module before uploading to edge servers. 


Feature extraction can also be adopted here to map the original token representation into a new dimension, which is often a lower dimension, while maintaining informative features sufficient for accurate split inference~\cite{huang2023joint}. A popular approach for this is the information bottleneck method, which reduces the size of intermediate features while keeping comparable inference accuracy in split inference~\cite{shao2021learning}. The principle of information bottleneck lies in maximizing the mutual information between the inference results and the extracted features while minimizing the mutual information between the input data and the extracted features, which can be adopted in split LLM inference. This allows edge devices to extract informative features and upload the extracted features to edge servers. Furthermore, the authors in \cite{shi2023taskoriented} propose the robust information bottleneck by combining the transmission rate maximization and the information bottleneck together. By considering the robustness of information distortion in the received features, this principle can improve the robustness of the received features and ensure that edge servers receive sufficient informative features without incurring too much communication overhead. By using the feature extraction method, split LLM inference can alleviate the communication burden of transferring high-dimensional features by uploading informative features to edge servers. 

For multimodal LLMs, one can exploit the correlation across different modalities when reducing the token representation of different modalities, thereby reducing the communication latency. A number of works have discussed how to reduce multimodal token representation in LLMs. In \cite{cao2023pumer}, the authors propose a cross-modal representation reduction method for ViTs. The ViTs use image text cross-modal encoders to integrate input texts and images for vision language tasks, such as visual question answering. The proposed framework aims to reduce the computing overhead in cross-modal interaction. Specifically, it first removes image token representations irrelevant to the input texts and unimportant to the tasks. Then, it independently merges similar text and image token representations to reduce the computing overhead. By applying this method to split inference, edge devices can select and upload the informative representations to edge servers based on the correlation across multiple modalities. Additionally, the semantic information of the selected cross-modal features can be extracted using approaches such as cross-attention mechanisms \cite{10462495} and the feature similarity-based semantic fusion \cite{10233481}. Following the principle of deep joint source-channel coding (JSCC)~\cite{10233481,huang2024d}, the encoders and decoders can be trained by considering the effects of physical channel noises and interference, making split inference robust to such adverse channel effects. Since this field is still in its infancy, cross-modal JSCC for split LLMs can be a significant research direction for 6G edge intelligence.


\subsubsection{Progressive split inference} The progressive split inference mechanism \cite{9955582} can be adopted here to eliminate the unnecessary transmissions of intermediate token representations as long as a desired inference accuracy can be satisfied. In \cite{9955582}, users can schedule the offloading sequence of features before forwarding them to edge servers. The features with higher importance, such as features with higher discriminant gain, are uploaded to edge servers first. The progressive feature offloading will terminate until edge servers determine that the uploaded features reach a target confidence level sufficient for edge servers to conduct inference above the accuracy threshold. By using this paradigm in split LLM inference, communication overhead between end users and edge servers can be reduced, thus saving communication resources. 

\subsubsection{Split inference with early exit} Early exit techniques, as introduced in Section \ref{sec:on-deviceinference}, have been widely used in LLM inference for latency reduction. This approach can be adopted in split inference to reduce computing latency on edge devices/servers. In this context, when adding early exit modules into user-side sub-LLMs, the inference computation in later layers in the user(server)-side sub-LLMs can be skipped, implying that there is no need for uploading intermediate features for latency reduction. However, in LLM inference with the early exit approach, the hidden states of the corresponding token in the later layers are missing when the token outputs the results in the early exit layer. Therefore, in split LLM inference with the early exit technique, if the early exit layers reside at user-side sub-LLMs, the hidden states still need to be uploaded to edge servers for KV cache computation in later layers. As such, we need to design the uploading strategy of hidden states at exiting layers, e.g., opportunistically uploading the hidden states by considering channel conditions.


\subsubsection{Other variants of split inference} Although split inference enhances user privacy by remaining raw data on local devices, the bi-partition paradigm still allows edge servers to obtain the inference results, which can be privacy-sensitive. To address this problem, the U-shaped or $\Lambda$-shaped split LLM inference paradigm can be adopted. In \cite{ohta2023lambdasplit}, an LLM is divided into three sub-models at the decoder blocks. The three sub-models are the head sub-LLM with the text input module, the body sub-LLM with the hidden layers in decoder blocks, and the tail sub-LLM with the text output module. Most decoder blocks, which demand more computing resources, can be placed on edge servers to fully leverage edge servers' computing resources. During the inference process, only the outputs of intermediate Transformer blocks in LLMs are exchanged between the end user and the edge server. These outputs are high-dimensional vectors, which are difficult for edge servers to interpret, thereby effectively preventing edge servers from recovering either the raw data or inference results.

Moreover, split inference can be extended into multi-edge scenarios to unleash the power of distributed edge servers in wireless edge networks, resulting in multi-hop split inference. In these scenarios, an LLM can be partitioned into multiple sub-models to be placed on multiple edge devices/servers according to their computation abilities and the inter-device (server) communication conditions~\cite{ma2023poster}. In this way, edge inference can be carried out sequentially with a mesh of edge servers by transmitting sub-model layer outputs to the next server \cite{borzunov2023distributed}. Obviously, developing efficient algorithms to obtain the optimal splitting decisions for such a scenario becomes much more challenging.


\subsection{Collaborative Inference}
Split inference exchanges high-dimensional intermediate features, resulting in excessive communication overhead. To overcome this limitation, edge devices and edge servers can cooperate in some other modes with the smaller size of information exchange. For instance, speculative decoding~\cite{leviathan2023fast}, which has been introduced in Section \ref{sec:on-deviceinference}, can be adopted in device-server collaborative inference for LLMs. Speculative decoding enables an edge device to run a smaller on-device LLM, called an approximation model, while asking the edge server to run a larger LLM to verify and correct the output tokens uploaded by the edge device. The main advantages of this approach are threefold: First, it empowers edge devices to generate preliminary results/decisions, which can be used for low-latency inference. Second, the parallel decoding process accelerates the inference process on the edge server. Since LLMs generate tokens with the autoregressive decoding technique, verifying a long sequence of tokens is much faster than decoding serially by the server-side LLM alone. Third, concerning communication overhead, the output token can be smaller than cut-layer intermediate features in many cases. To further save communication-computing resources, edge devices can decide whether to upload the tokens generated by on-device LLMs to edge servers for verification~\cite{wang2023tabi} based on calibrated confidence scores, because highly confident outputs might not require an edge server to consume resources for verification.

\subsection{Lessons Learned}
Resource-efficient LLM inference techniques can be adapted to wireless edge environments, enabling effective cooperation between edge devices and edge servers. Thus, edge LLM inference must be jointly optimized with various LLM techniques, such as KV cache optimization, multimodal feature extraction, and autoregressive decoding, all of which significantly impact memory usage, communication overhead, and computing latency for LLM inference. For this reason, the design of edge LLM inference, such as model splitting, early exiting, and radio-computing resource allocation, naturally differs from conventional edge inference systems, creating a rich set of research problems in this area.

\section{Further Research Opportunities\label{sec:further}}
MEI for LLMs remains a largely uncharted direction. This section will discuss the pressing concerns and potential solutions for MEI4LLM.

\subsection{Green Edge LLM\label{subsec:green}}
Energy consumption is a primary public concern over LLMs. For instance, it is estimated that the energy consumption of training GPT-4 is equivalent to the consumption over 5 to 6 years of 1,000 average US households. Moreover, the energy costs of model inference can be even higher due to the frequent service requests from users worldwide. The estimated energy footprint of GPT-4 training and inference are detailed in Table \ref{tab:llm_energy}. These facts underscore the urgent need for the development of energy-efficient LLM training and inference techniques.

\begin{table}[!ht]
	\centering
 \caption{Approximate energy usage of GPT models.}\label{tab:llm_energy}
        \begin{tabular}{|c|c|c|}
         \hline
          \makecell[c]{Name} & \makecell[c]{Total energy consumption\\in training (MWh)} & \makecell[c]{Daily energy consumption\\of responses (MWh)} \\ \hline
          GPT-3  &  1,287 \cite{patterson2021carbon} & 3 \cite{chatgptenergy}\\ \hline
          GPT-4  & $\ge$50,000 \cite{li2024unseen} & 5 \cite{chatgptenergy}\\ \hline 
        \end{tabular}
        \label{tab:llm_training_energy}
\end{table}

By pushing the intelligence to the network edge, MEI4LLM can reduce the energy consumption of LLMs in three aspects. First, making model fine-tuning or inference services available at the network edge eliminates the need to transmit large volumes of data to cloud centers, thereby reducing the energy costs in data transportation. Second, integrated communication-computing design can be jointly optimized to enhance energy efficiency for training/inference. For instance, less data communication can be achieved by data compression or parameter freezing to reduce total transmit power as long as the required training/inference accuracy can be achieved. At last, edge LLM can utilize a small-scale LLM to obtain initial inference results and harness the power of the cloud-based large-scale LLM only when the inference confidence is low. This potentially reduces the energy consumption at cloud centers resulting from invoking large-scale LLMs for every user request.

Regarding the research problems, green edge LLM features the integrated design of wireless communications and computing, which must account for the overall transmission and computing energy. With this in mind, there are two main design goals, i.e., reducing energy consumption on edge devices or decreasing the overall energy consumption for green AI. The first goal is to benefit battery-constrained IoT and mobile devices, making AI services or training more accessible to customers. To achieve this goal, AI training/inference can be offloaded to edge servers as long as it saves energy for end users. For instance, when adopting PEFT and SL, the client-side models can be frozen as much as possible, even entirely frozen, to minimize energy consumption on edge devices. Although this approach might lead to slower learning convergence or increase the energy consumption on edge servers for achieving a target training/inference accuracy, this cost might be less concerning since edge servers are usually more powerful and connected to powerlines. The second one aims to decrease the overall system costs, particularly from the perspective of mobile operators. By minimizing the total energy consumption or maximizing the overall energy efficiency, network operators can improve AI-centric metrics with limited or lower energy costs. A meaningful metric can be ``AI energy efficiency''~\cite{mao2023green}, i.e., the amount of intelligence (AI accuracy) achieved per energy cost. The optimization of AI by considering energy consumption can eliminate the case where the system uses a great amount of energy to merely achieve minor improvements, which is not justifiable from both operators' and society's viewpoints.

\subsection{Secure Edge LLM\label{subsec:secure}}
Secure edge LLM is another important branch of research. Although federated learning and split learning, as alluded to earlier, can serve as privacy-enhancing training frameworks for LLMs to avoid direct access to personal data, there is still privacy risk as malicious servers may launch attacks to recover the raw data based on received models or intermediate features~\cite{wucong2024tifs,yang2023}. However, while LLM security has been extensively studied, edge LLM security has received much less attention. At the mobile edge, security components should be integrated to ensure secure and trustworthy LLMs.

Let us consider two aspects of secure edge LLM. The first aspect is defending against inference attacks to protect user privacy leakage from honest-but-curious edge servers or other edge devices. It has been demonstrated that prompting LLMs may reveal the private information of other users, such as credit card information, in the training process by only inserting a few benign-appearing sentences into the training dataset~\cite{panda2024teach}. For FL and SL, similar problems may still exist. From the user perspective, one potential defense mechanism is to add noise to protect highly sensitive personal data, such as credit card information, based on the theory of differential privacy. The research question is where to insert noise and the amount of noise to insert. 




The second aspect is defending against data poisoning or backdoor attacks by filtering out malicious users who aim to alter the training process, thereby maintaining training effectiveness. These attacks can lead to severe consequences for LLMs, i.e., outputting harmful health guidance if considering a healthcare LLM. Although such attacks have been studied for LLMs~\cite{wan2023poisoning,kandpal2023backdoor}, there is still a lack of research works on designing effective attack/defense mechanisms by considering distributed learning for LLMs, particularly FL and SL, over edge devices. Considering PEFT, the attacker may only be able to change a small proportion of the models, e.g., adaptors or prompts, to impact the training process rather than modifying the full set of parameters. This leads to new challenges/opportunities in designing the attack/defense schemes.

\subsection{Quality-aware Edge LLM Training}
Unlike cloud-based training, data quality at the network edge cannot be well controlled due to the lack of human annotations and monitoring. Unfortunately, training over massive and low-quality data/labels can be detrimental to model performance rather than beneficial. To effectively deploy LLMs at the network edge, it is essential to enable LLM training with automated quality control.

There are two key research directions for improving the quality of edge LLM training.
First, improving the quality of training datasets can directly improve edge LLM training outcomes. On the one hand, edge servers can increase the diversity of training datasets by prioritizing varied data sources. Higher diversity in training datasets increases the likelihood of achieving better performance on corresponding tasks \cite{lee2023beyond}. For example, in federated learning, edge servers can allocate additional training rounds to edge devices with more diverse data. Additionally, in cloud-edge-end FL, the cloud center can assign varying weights to edge LLMs during aggregation based on the regional diversity associated with different edge servers. On the other hand, to ensure high accuracy and reliability in LLM responses, data cleaning for training datasets can be adopted for edge LLM training. Considering LLMs’ sensitivity to input data, data cleaning is essential to detect and remove errors, irrelevant information, duplicates, and inconsistencies before LLM training \cite{hofer2024construction}. For example, in LLM-empowered AI chatbots, data cleaning can identify and remove useless information and grammatical errors in raw data inputs, thereby enhancing the quality of the training datasets and ensuring LLMs' performance \cite{wondercaht}. However, since data cleaning introduces additional computational overhead to edge servers/devices, a trade-off exists between computational efficiency and the quality of training data.

Second, edge LLM training can harness a large volume of unlabeled data or data with noisy labels captured by edge devices. While the training quality of LLMs significantly depends on high-quality training datasets, acquiring enough high-quality labeled data for edge devices requires considerable effort and costs, particularly for the vast number of resource-limited IoT devices at the network edge. Therefore, edge servers must leverage the large volume of unlabeled data collected from distributed edge networks to train LLMs with unsupervised learning \cite{zhang2024ucl} and semi-supervised learning. 
For LLMs, edge servers can fully leverage the relationships between limited labeled data and large volumes of unlabeled data across different modalities. For example, on-device LLM can continuously capture data from multiple modalities, automatically generating labels for data of specific modalities. By utilizing the inherent patterns and correlations in these data, edge servers can reconstruct the labels of unlabeled data, thereby enhancing the training quality and adaptability across various tasks.

\section{Conclusions\label{conclusion}}
In recent years, language models have experienced exponential growth in size, giving birth to numerous LLMs with billions of parameters. This trend urges us to think about how edge intelligence can accommodate these giant models. In this article, we advocated the paradigm shift from cloud computing to 6G MEI for LLM deployment. We highlighted killer applications to motivate this paradigm shift, arguing that cloud computing can hardly fulfill the latency, bandwidth, and privacy requirements. Meanwhile, we identified the key challenges that mainly arise from the resource limitations at the network edge. To address these challenges, we first proposed a 6G MEI architecture for LLMs and then elaborated on several methods to enable efficient edge caching and delivery, edge training, and edge inference for LLMs under the resource-constrained mobile edge. We hope this article can inspire more researchers in the wireless community to explore the deployment of LLMs at the mobile edge and further advance this emerging field.
\bibliographystyle{IEEEtran}
\bibliography{IEEEabrv,NEWmybib}

\end{document}